\documentclass[12pt,a4paper]{iopart}
\usepackage{hyperref}
\usepackage{iopams}
\usepackage{setstack}
\usepackage[english]{babel}
\usepackage{graphicx}
\usepackage{epsfig}
\usepackage{color}
\usepackage{latexsym}
\usepackage{bm}

\usepackage[left=2cm,right=2cm,top=2cm,bottom=2cm]{geometry}

\usepackage{etoolbox}

\makeatletter
\newcommand{\mainmatter}{%
  \setcounter{footnote}{0}%
  \patchcmd{\@makefntext}{\fnsymbol}{\arabic}{}{}%
  \patchcmd{\@thefnmark}{\fnsymbol}{\arabic}{}{}%
  \def\@makefnmark{\textsuperscript{\arabic{footnote}}}%
}
\makeatother

\newcommand{\R}{{\mathbb{R}}}

\newcommand{\C}{{\mathbb{C}}}

\newcommand{\One}{\mathbf{1}}
\newcommand{\QED}{\rule{1.5mm}{3mm}} 

\newcommand{\A}{{\mathcal{A}}}
\newcommand{\D}{{\mathcal{D}}}
\newcommand{\V}{{\mathcal{V}}}

\newcommand{\M}{{\mathcal{M}}}

\newcommand{\cL}{{\mathcal{L}}}

\newcommand{\lact}{\triangleright}

\newcommand{\di}{{\partial}}
\newcommand{\cg}{{\gamma}}
\newcommand{\dg}{{\delta}}
\newcommand{\eg}{{\epsilon}}
\newcommand{\sg}{{\sigma}}
\newcommand{\vareg}{{\varepsilon}}
\newcommand{\Jh}{\hat{J}}
\newcommand{\jh}{\hat{j}}
\newcommand{\Vh}{\hat{\cal V}}
\newcommand{\nuh}{\hat{\nu}}

\newcommand{\cP}{{\cal P}}
\newcommand{\cH}{{\cal H}}

\newcommand{\beq}{\begin{eqnarray}}
\newcommand{\eeq}{\end{eqnarray}}
\newcommand{\ei}{\end{itemize}}
\newcommand{\be}{\begin{enumerate}}
\newcommand{\ee}{\end{enumerate}}
\newcommand{\bq}{\begin{equation}}
\newcommand{\eq}{\end{equation}}

\def\diag{\mathop{\mathrm{diag}}}

\newtheorem{thm}{Theorem}
\newtheorem{prop}[thm]{Proposition}
\newtheorem{clly}[thm]{Corollary}
\newtheorem{lemma}[thm]{Lemma}

\begin{document}

\title{On the classical and quantum Geroch group}
\author{Javier Peraza$^{1,2}$, Miguel Paternain$^2$, and Michael Reisenberger$^1$} 
\address{Instituto de F\a'{\i}sica${}^1$ and Centro de Matem\'atica${}^2$,\protect\\ 
Facultad de Ciencias, Universidad de la Rep\a'ublica Oriental del Uruguay,\protect\\
Igu\a'a 4225, esq. Mataojo, Montevideo, Uruguay\\
\ }
\ead{\mailto{jperaza@cmat.edu.uy},\mailto{miguel@cmat.edu.uy},\mailto{miguel@fisica.edu.uy}}

\date{Jan 31, 2019}

\begin{abstract} 
The Geroch group is an infinite dimensional transitive group of symmetries of classical cylindrically symmetric gravitational
waves which acts by non-canonical transformations on the phase space of these waves. Here this symmetry is rederived and the 
unique Poisson bracket on the Geroch group which makes its action on the gravitational phase space Lie-Poisson is obtained. 
Two possible notions of asymptotic flatness are proposed that are compatible with the Poisson bracket on the phase space, and 
corresponding asymptotic flatness preserving subgroups of the Geroch group are defined which turn out to be compatible with 
the Poisson bracket on the group.

A quantization of the Geroch group is proposed that is similar to, but distinct from, the $\mathfrak{sl}_2$ Yangian, and a 
certain action of this quantum Geroch group on gravitational observables is shown to preserve the commutation relations of 
Korotkin and Samtleben's quantization of asymptotically flat cylindrically symmetric gravitational waves. The action also 
preserves three of the additional conditions that define their quantization. It is conjectured that the action preserves 
the remaining two conditions (asymptotic flatness and a unit determinant condition on a certain basic field) 
as well and is, in fact, a symmetry of their model. Our results on the quantum theory are formal, but a possible rigorous formulation based on algebraic quantum theory is outlined.
\end{abstract}

\mainmatter

\section{Introduction}

Cylindrically symmetric vacuum general relativity is an important truncation of full general relativity (GR) which 
has infinitely many degrees of freedom, yet is tractable because it is integrable \cite{Belinskii_Zakharov}\cite{Maison}. 
It is useful as a toy model for studying both classical and quantum gravitational phenomena, and as a test bed for 
methods of quantizing full vacuum GR \cite{Ashtekar, Varadarajan, Gambini_Pullin, Dominguez_Tiglio, Angulo, 
Microcausality, Isomonodromic, Algebraic_Bootstrap, Asymptotically_safe, Free_Field, Husain_Smolin, Husain}.
It is, for instance, highly relevant to the quantization of null initial data for vacuum GR, since the Poisson 
algebra for these in the general case is almost identical to that of the cylindrically symmetric case \cite{ReiAnd}.

In 1971 Kuchar \cite{Kuchar}\cite{Allen}\cite{Ashtekar_Pierri} quantized asymptotically flat cylindrically symmetric 
vacuum GR subject to a further restriction on the polarization of gravitational waves, and in 1997 Korotkin and Samtleben 
\cite{KS2}\cite{KS} obtained an almost complete quantization without this restriction. Specifically, Korotkin and 
Samtleben obtained a quantization of the Poisson algebra of classical observables (phase space functions) of full 
asymptotically flat cylindrically symmetric vacuum GR, but not a representation of the resulting algebra of 
quantum observables on a Hilbert space that respects the reality conditions. 

Classical cylindrically symmetric vacuum GR possesses a transitive group of symmetries called the {\em Geroch group} 
\cite{Geroch, Kinnersley, Kinnersley_Chitre1, Kinnersley_Chitre2, Kinnersley_Chitre3, Julia85}. The elements 
of this group can be thought of as transformations of the spacetime metric field which are symmetries in 
the sense that they map solutions to solutions. Since the action of these transformations can in some cases 
be worked out explicitly they were initially studied as a means of generating new solutions 
from old. But there is more to the Geroch group. Although its elements are not strictly canonical transformations
they are so in a generalized sense, they preserve the Poisson brackets in a generalized sense that will be
described below. Furthermore, the Geroch group is generated, again in a generalized sense, 
by an infinite set of conserved charges \cite{KS}. 
Here we will explore the possibility that a version of this symmetry is present in the quantum theory. 
Our results suggest that this is the case in the quantum theory of Korotkin 
and Samtleben: We obtain a quantization of the Geroch group and of its action on the cylindrically symmetric 
gravitational field which appears to be a symmetry of the algebra of quantum observables of this field. We 
verify that it preserves the commutation relations and all other conditions defining this algebra, save two 
which we are unable to check.  

This is not entirely surprising since the quantization of \cite{KS} is based on the integrability of 
the model, which in turn is closely related to the symmetry under the Geroch group. Indeed, in \cite{KS}
and \cite{KS3} Korotkin and Samtleben show that the phase space coordinates they quantize to obtain their
quantization of cylindrically symmetric gravity, the matrix valued functions $T_\pm$ defined in section \ref{quantum}, 
are precisely the charges that generate the Geroch group.\footnote{
Despite the conservation of these charges, there is non-trivial dynamics because the charges are explicitly 
time dependent, like the conserved quantity $x - t p/m$ for a free, mass $m$ particle of momentum $p$ and 
position $x$ is.}
They present an explicit map from the Lie algebra of the Geroch group to the vector fields that realize
the corresponding infinitesimal Geroch group transformations of the gravitational phase space, a map which 
may be interpreted as the specificiation of a ``non-Abelian Hamiltonian'' \cite{Babelon_Bernard}, or moment map 
\cite{Lu}, for the Geroch group which generates these vector fields as generalized Hamiltonian flows. 
In addition, they conjecture in \cite{KS3} that the quantization 
of the Geroch group is the $\mathfrak{sl}_2$ Yangian, a well known quantum group introduced by Drinfel'd in 
\cite{Drinfeld3}, and Samtleben \cite{Samtleben_thesis} even makes a partial proposal for the non-Abelian 
Hamiltonian in the quantum theory. 

Our results are complementary to theirs. We do not focus on the realization of the Geroch group as 
generalized Hamiltonian flows generated by certain phase space functions. Instead, we determine the Poisson 
bracket on the Geroch group itself from the requierment that the action of the group on the phase space of 
cylindrically symmetric GR be a Lie-Poisson map, 
which is to say, that it satisfies (\ref{Lie_Poisson}). (See \cite{Babelon_Bernard}\cite{Semenov-Tian-Shansky}.) 
With this bracket the group has a natural quantization, a quantum group similar to an $\mathfrak{sl}_2$ Yangian. 
We then make an ansatz for the form of the action of the quantum Geroch group on the quantum gravitational 
observables, a natural two parameter generalization of the classical action, and fix the parameters by requiring 
the action to be a symmetry. 

The charges, and in particular the quantized non-Abelian Hamiltonian of the Geroch group, are however crucial
for using the Geroch group as a spectrum generating quantum group to organize the Hilbert space of the quantum theory
of cylindrically symmetric gravity. A spectrum generating group of a quantum system is a group that is represented 
irreducibly by unitary operators on the Hilbert space, like the Poincare group on the Hilbert space of a single, free,
relativistic particle.\footnote{
Here we follow \cite{Majid} and use the term ``spectrum generating group'' in a somewhat more ample sense than 
that used in the original applications of the concept to nuclear and particle physics \cite{Bohm_Neeman}.} 
It is especially useful if observables of interest, such as the Hamiltonian for time evolution, can be 
expressed simply in terms of the representations of the generators of the group, because the matrix elements of 
the latter are completely determined by the representation theory of the group. 

Morally, spectrum generating groups of quantum models correspond to transitive groups of canonical transformations
of the corresponding classical model. A rough argument that suggests that irreducibility of the quantum action implies
transitivity of the corresponding action on the classical limit phase space is the following: Phase points correspond 
to nearly classical states, like coherent states, that approximate the evaluation map at the phase point with increasing precision as $\hbar \rightarrow 0$. Such nearly classical states corresponding to distinct phase points become orthogonal in the $\hbar \rightarrow 0$ limit. Now, a unitary transformation can map a nearly classical state to a superposition of such states corresponding to different phase points, which is not a nearly classical state.
But let us suppose that the unitary group action under consideration maps nearly classical states to nearly classical states, and thus descends to a group action on the classical phase space. Since the action is irreducible the orbit of a given nearly classical state must span the Hilbert space. This requires the corresponding orbit in phase space to be dense.

The converse claim, that a group that acts transitively on the phase space, and acts unitarily on the Hilbert space of the quantization, should act irreducibly on this Hilbert space, is a basic principle of quantization. (See \cite{Woodhouse}, section 8.1.)
For instance, the Hilbert space of a single free relativistic particle is irreducible under the action of 
the Poincare group or its double covering \cite{Wigner}. (See \cite{Woodhouse}, sections 6.5, 6.6, and 9.4 for the 
connection to classical theory.) Another example is a pure spin $s$. It is a quantization of the two sphere with 
symplectic form equal to the area form and total area $4\pi \hbar (s + 1/2)$. $SU(2)$ acts transitively on this phase space and irreducibly on the corresponding Hilbert space. (See \cite{Woodhouse} sections 9.2 and 10.4).
In fact, this principle is the basis of the orbit method of group representation theory, in which the irreducible unitary
representations of a group are obtained by quantizing its coadjoint orbits \cite{Kirillov}.

Now let us consider the consequences of unitarity. The action of a unitary representation $U(g)$ of a group element $g$ on state vectors is equivalent to the action
\begin{equation}\label{adjoint_action}
 X \mapsto U(g)^\dagger X U(g)  
\end{equation} 
on operators. But because $U U^\dag = \One$
\begin{equation}\label{commutation_relation_preservation}
 U^\dag [A,B] U = [U^\dag A U, U^\dag B U]
\end{equation}
for any pair of observables $A$ and $B$, so the transformation (\ref{adjoint_action}) preserves commutation relations of observables in the quantum theory. If a unitary group action descends to an action on the classical phase space this action should therefore preserve the Poisson algebra of phase space functions.

The Geroch group acts transitively on the phase space of cylindrically symmetric gravity, but it does not preserve Poisson brackets; Geroch group actions are not canonical transformations. So the Geroch group does not quite fit into the framework
of spectrum generating groups we have outlined. 
However, it does fit if the framework is extended by allowing $G$ to become a {\em quantum} 
group in the quantum theory. In this case $G$ is still a true group in the classical limit, but it acquires 
a non-trivial Poisson bracket which turns it into a phase space, and it is quantized along with the phase 
space $\Gamma$ of the physical system in the quantum theory. The matrix elements $\langle a|U b\rangle$ of 
the symmetry transformations in the Hilbert space $\cH$ of the physical system, instead of being $\C$ valued 
functions on the classical group manifold, become (complex linear combinations of) observables of the quantum 
group. Instead of taking values on each element of the classical group $G$ they assume expectation values on 
each state of the quantum group.  

The notion of a spectrum generating symmetry generalizes directly to quantum groups. See \cite{Macfarlane_Majid} 
and also \cite{Timmermann} Chapter 3. In particular, $U$ is still unitary, but in the sense of quantum group actions:
\begin{equation}
 U^\dagger U = U U^\dagger = \mathbf{1}_\cH \otimes \mathbf{1}_G
\end{equation}
where $\mathbf{1}_\cH$ is the unit operator in the Hilbert space of the physical system, and $\mathbf{1}_G$
is the unit element in the algebra of observables of the quantum group. ($\mathbf{1}_G$ is the quantization 
of the constant function of value $1$ on the group manifold and would be represented by the unit operator in 
a Hilbert space representation of the quantum group.) Such a transformation preserves inner
products in ${\cal H}$ in the sense that the expectation value of $\langle U a| U b\rangle$ in any 
state of the group is equal to the untransformed inner product $\langle a|b\rangle$. It also preserves the
commutation relations of observables of the physical system, because once again equation 
(\ref{commutation_relation_preservation}) 
holds. But note that now the commutator $[U^\dag A U, U^\dag B U]$ receives a contribution from the 
non-trivial commutators between the matrix elements of $U$. As a consequence the Poisson bracket is still 
preserved in the classical limit, but only when a contribution to the transformed bracket from the Poisson 
bracket on $G$ is included: 
\begin{eqnarray}
 \fl\{A,B\}_\Gamma (g \lact \xi) & \equiv \{A,B\}^G_\Gamma(g,\xi) & = 
 \{A^G(g,\cdot), B^G(g,\cdot\}_\Gamma (\xi) + \{A^G(\cdot, \xi), B^G(\cdot, \xi\}_G (g) \label{Lie_Poisson}\\
 && \equiv \{A^G, B^G\}_{G \times \Gamma} (g,\xi)
\end{eqnarray}
Here $g \lact \xi$ denotes the action of a group element $g \in G$ on a phase space point $\xi \in \Gamma$, and
$f^G(g,\xi) = f(g \lact \xi)$ is the corresponding action of $G$ on a function $f$ of $\Gamma$. The right side 
of (\ref{Lie_Poisson}) is simply the Poisson bracket on $\Gamma$ of the functions $A^G$ and $B^G$ with 
$g \in G$ held fixed, plus the bracket on $G$ of these two functions but with $\xi$ held fixed. This is just 
the standard Poisson bracket of $A^G$ and $B^G$ on the phase space $G \times \Gamma$, the phase space of a pair 
of subsystems that are independent in the sense that their phase coordinates Poisson commute. Following 
\cite{Babelon_Bernard} a Lie group action satisfying (\ref{Lie_Poisson}) will be called a {\em Lie-Poisson} action.  

In the present work we will not directly continue the approach of \cite{KS3},\cite{KS}, and \cite{Samtleben_thesis}
by quantizing the non-Abelian Hamiltonian generating Geroch group transformations of gravitational observables.
Rather, we will use the condition (\ref{Lie_Poisson}) to determine the quantization of the Geroch group itself,
and also the associated deformation of its action on the gravitational field (expressed directly, without using the Poisson bracket and generators). Specifically, we will make the classical action of the Geroch group very simple by parametrizing the phase space of cylindrically symmetric vacuum GR by the 
so called ``deformed 2-metric'' $\M$, which is a $2 \times 2$ matrix valued function of one real parameter $w$, the so called spectral parameter. Then we will use the Poisson brackets of $\M$, which are also quite 
nice, to determine the unique Poisson bracket on the Geroch group that ensures that (\ref{Lie_Poisson}) is 
satisfied. The Geroch group with this Poisson bracket has a natural quantization similar to the $\mathfrak{sl}_2$ 
Yangian in the sense that the algebraic relations that define our quantum Geroch group are the same as those of the 
Yangian in the RTT presentation (see \cite{FRT} and \cite{Molev}).
Finally, we obtain the action (\ref{q_Geroch_action}) of the quantized Geroch group on $\M$, a small modification 
of the classical action, which appears to define an automorphism of the quantum algebra of observables (which is 
generated by the quantized $\M$ function) in the quantization of cylindrically symmetric vacuum GR of Korotkin and Samtleben \cite{KS}.
We show that this action preserves the commutation relation (exchange relation) of $\M$, as well as its symmetry, 
reality and positive semi-definiteness. We have not been able to show that the action preserves the Korotkin and Samtleben's 
quantization of the condition ${\rm det}\M = 1$, but suspect that it does. We have also not verified that 
asymptotic flatness is preserved. 

\newcommand{\botheta}{\boldsymbol{\theta}}

The roles of our results and those of Korotkin and Samtleben in developing the Geroch group as a spectrum 
generating symmetry of cylindrically symmetric GR can be illustrated using a simple, partly analogous, partly 
contrasting system:
Consider the rotation group $SO(3)$ in the non-relativistic classical and quantum mechanics of a spinless particle 
in three dimensional space. It acts unitarily, though not reducibly on the Hilbert space of the quantum theory, so
it is not a true spectrum generating group, but it is close enough for the purpose of our analogy.
Classically a rotation by a vectorial angle $\botheta$ transforms the Cartesian coordinate position $\mathbf{q}$ and 
its conjugate momentum $\mathbf{p}$, or indeed any vector $\mathbf{z}$, according to 
\begin{equation}
 \mathbf{z} \mapsto e^{\botheta\times} \mathbf{z}.
\end{equation}
where $\times$ denotes the 3-vector cross product operation.
Quantum mechanically the rotation of any observable is obtained by conjugation with $U(e^{\botheta\times}) 
= e^{\frac{i}{\hbar}\botheta\cdot \mathbf{J}}$ where $\mathbf{J} = \mathbf{q}\times \mathbf{p}$. 
For vector observables
\begin{equation}\label{rotation}
e^{-\frac{i}{\hbar}\botheta\cdot \mathbf{J}}\, \mathbf{z}\, e^{\frac{i}{\hbar}\botheta\cdot \mathbf{J}} 
= e^{\botheta\times} \mathbf{z}.
\end{equation} 
That is, the action of a rotation on a vector observable maps each component to a linear combination of
components in precisely the same way as in the classical theory. But this action can also be obtained 
by conjugating each component with a unitary operator generated by $\mathbf{J}$, or, equivalently, by acting with
this unitary operator on the state vectors of the system. 

The work of Korotkin and Samtleben to obtain the generators of the Geroch group, 
\cite{KS3}\cite{KS}\cite{Samtleben_thesis}, is analogous to identifying the angular momentum 
$\mathbf{J}$ as the generator of rotations.  Our results, on the other hand, deal with the
analogs of the right side of (\ref{rotation}), and of the coordinates $\botheta$ on the group manifold.
In cylindrically symmetric GR the deformed metric $\M$ plays the role of the Cartesian position $\mathbf{q}$ 
and momentum $\mathbf{p}$ as phase space coordinates that transform in a simple way under the symmetry group
in the classical theory. (See (\ref{Geroch_action}) for the action of the classical Geroch group on $\M$.)
But, while the action of a rotation on $\mathbf{q}$ and $\mathbf{p}$ in quantum theory is identical in form 
to the classical action, we find that the action of the Geroch group on $\M$ aquires corrections of order 
$\hbar$ in the quantum theory. (See (\ref{q_Geroch_action}).) Furthermore, the coordinates $\botheta$ on 
the $SO(3)$ manifold are the same in both quantum and classical theory. They do not aquire non-zero commutators 
upon quantization, because rotations are canonical transformations of $\mathbf{q}$ and $\mathbf{p}$. 
But the Geroch group is quantized in quantum cylindrically symmetric GR in the sense that coordinates on the 
Geroch group do have non-zero commutators in the quantum theory. This is relevant also for the realization of 
the group action in terms of the generators obtained by Korotkin and Samtleben, because just as an arbitary 
infinitesimal rotation, the adjoint action of $\botheta\cdot \mathbf{J}$, is a linear combination of the 
adjoint actions of the angular momentum components weighted by the corresponding group (and Lie algebra) 
coordinates $\theta^i$ of the rotation in question, so an arbitary infinitesimal Geroch group transformation 
is a linear combination of the actions of the generators found by Korotkin and Samtleben weighted by 
corresponding coordinates on the Geroch group \cite{KS}. The non-trivial commutators of the group coordinates 
therefore contribute to the commutators of the generators of the Geroch group.

The issue of asymptotic flatness in the quantum theory is intimately connected with the difference between 
our quantum Geroch group and the $\mathfrak{sl}_2$ Yangian.
Although Korotkin and Samtleben's theory of cylindrically symmetric gravitational waves assumes asymptotic flatness,
it includes no conditions specifying the asymptotic falloff of the gravitational variables, either in the classical
or quantum cases. We propose two possible falloff conditions for the classical theory, each of which is compatible 
with the Poisson bracket on the gravitational variables and is invariant under a suitable subgroup of the Geroch group which,
moreover, acts transitively on the solutions satisfying the falloff conditions. 
These conditions are expressed in terms of $\M(w)$, which must tend to an asymptotic value $e_\infty$ as $w \rightarrow \pm\infty$
if the solution is asymptotically flat. One condition, which we consider the more natural one, requires $\M - e_\infty$
to be the Fourier transform of an absolutely integrable function, the other requires simply that $\M - e_\infty$ falls off
as $1/w$, or faster, as $|w| \rightarrow \infty$. We have not settled on a quantization of the first, prefered, falloff 
condition, and so cannot check whether it is preserved by the quantum Geroch group. The second falloff condition
is easily implemented in the quantum theory by requiering that $\M$ be a power series in $1/w$ with order zero term $e_\infty$.
Moreover, it is preserved by the quantum Geroch group action if the quantized matrix elements of the fundamental 
representation of the Geroch group, $s(w)_a{}^b$, are also such power series, with order zero term $\dg_a{}^b$.
This hypothesys on the nature of the quantized matrix elements $s(w)_a{}^b$ together with the algebraic relations, including
commutation relations, that $s(w)_a{}^b$ must satisfy in the quantum Geroch group imply that this quantum group is exactly
the $\mathfrak{sl}_2$ Yangian (see \cite{Molev}). Perhaps the fact that the Yangian comes already equipped with a natural falloff 
condition is the reason Korotkin and Samtleben did not postulate one, although they eventually reject defining their
variables as power series in $1/w$ for much the same reason we do \cite{KS}\cite{Samtleben_thesis}.

What is wrong with postulating that $\M$ and $s$ are power series in $1/w$? Basically, that it places strong and strange 
analyticity requirements on $\M$ and $s$ which are absent in the classical theory. It might be possible to overcome
this limitation by defining the sum of the power series via a suitable resummation technique, but it remains unnatural
to represent essentially arbitrary functions on the real line as power series about $\infty$.
The $\mathfrak{sl}_2$ Yangian emerges naturally as a quantization of (a sector of) an $SL(2,\C)$ loop group, 
consisting of $SL(2,\C)$ valued functions on a circle. If the circle is identified with $|w| = 1$ in the complex $w$
plane, then a Fourier series is an expansion in powers of $w$, and the positive frecuency part (with the angle as time) 
is a series in non-negative powers of $1/w$.
The Geroch group, in contrast, is an $SL(2,\R)$ ``line group'', consisting of $SL(2,\R)$ valued functions on the real $w$ line.\footnote{
There exists a Moebius tranformation mapping the line to the unit circle, but this transformation does not preserve
the Poisson bracket. It transforms the Geroch group into a loop group with a Poisson bracket differing from that of the classical
limit of the Yangian. (See \cite{Babelon_Bernard} for discussion of the classical limit of the Yangian.)}
The natural analogue of a Fourier series in this case is a Fourier integral of modes $e^{ikw}$ for all $k \in \R$. 
We will outline a formulation of the quantum Geroch group in terms of such Fourier integrals within the framework
of algebraic quantum theory. If fully worked out, this would constitute a mathematically rigorous formulation.
A formalism of this type might also provide suitable mathematical underpinnings for Korotkin and Samtleben's 
quantization of cylindrically symmetric vacuum GR. 
% A similar formalism ought to provide suitable definitions for the basic magnitudes in Korotkin and Samtleben's quantization of cylindrically symmetric vacuum GR.

Our main result, the preservation of at least most of the relations defining Korotkin and Samtleben's theory by 
the action of our quantum Geroch group, derive almost exclusively from the algebraic relations defining the gravitational
theory and the quantum group. The only exception is the preservation of the positive semidefiniteness of the matrix
$\M(w)_{ab}$ which invokes the positivity of quantum states (density matrices) on the combined system formed from the group and 
gravitational degrees of freedom. (Positivity of states in the sense used here is one of the basic axioms of algebraic 
quantum theory. See \cite{Khavkine_Morreti}.)

Our results on the quantum theory are thus ultimately still formal, but also largely independent of the detailed 
definitions of the objects involved. Moreover, the structure we find is quite rigid, it admits virtually no modificiation 
without spoiling its consistency, so it would seem to have a good chance of being present unchanged in a completely 
worked out theory.

The remainder of the paper is organized as follows: In Section \ref{cyl_grav} the definition of cylindrically symmetric
vacuum GR is recalled, as well as the structures associated with its integrability, in particular the
auxiliary linear problem and the deformed metric $\M_{ab}$. This leads to a precise description of the
space of solutions, in terms of which the classical Geroch group is defined in a very simple way. 
This section is based on ideas of \cite{Belinskii_Zakharov, Hauser_Ernst, Breit, Nicolai, KS, And} and others. 
But, although the underlying ideas are not new, the specific development given here, which is both 
mathematically rigorous and relatively short, is new as far as the authors know. 
Also in this section two possible definitions of asymptotic flatness are presented and subgroups of
of the Geroch group that preserve the asymptotically flat solutions in the two senses are defined and shown to act
transitively.
In Section \ref{Poisson} the Poisson brackets on the space of solutions is recalled, and from this the
Poisson bracket on the Geroch group that makes the action of this group Lie-Poisson is obtained.
Section \ref{quantum} is dedicated to the quantum theory. After an introductory overview a quantization of 
the Geroch group as a quantum group closely analogous to an $\mathfrak{sl}_2$ Yangian is proposed
in the first subsection. This is formulated as a series of conditions, including commutation relations, on the algebra 
of observables of the Geroch group. A fairly detailed outline of a mathematical definition of this algebra of 
observables is provided. 
In the second subsection Korotkin and Samtleben's quantization of cylindrically symmetric vacuum GR is presented
and an action of the quantum Geroch group on the observables of this model is found which preserves the commutators 
(exchange relations) and three of the further conditions defining the algebra of these observables. The paper closes 
with a discussion of open problems and directions for further research. 

\section{Cylindrically symmetric vacuum gravity and the Geroch group}\label{cyl_grav} 
  
We will consider smooth ($C^\infty$) solutions to the vacuum Einstein field equations 
with cylindrical symmetry. Cylindrically symmetric spacetime geometries have two commuting 
spacelike Killing fields that generate cylindrical symmetry orbits. Furthermore, the 
Killing orbits should be orthogonal to a family of 2-surfaces, a requirement called 
''orthogonal transitivity''. The line element then takes the form %\cite{Kompaneets}
\begin{equation}\label{line_element}
 ds^2 = \Omega^2 (-dt^2 + d\rho^2) + \rho e_{ab}d\theta^a\theta^b,
\end{equation}
where $\theta^a = (\phi, z)$ are coordinates on the symmetry cylinders such that 
$\di_z$ and $\di_\phi$ are Killing vectors, with $\phi \in (0, 2\pi)$ the angle 
around the the constant $z$ sections, which are circles. $h_{ab} = \rho e_{ab}$ 
is the induced metric on the symmetry orbits in these coordinates, with 
$\rho = \sqrt{\det{h}}$ the corresponding area density. The ``conformal metric'' 
of the orbits, $e_{ab}$, is thus a unit determinant, $2\times 2$, symmetric matrix. 
Cylindrical symmetry requires that neither $\rho$, nor $e$, nor the conformal factor 
$\Omega$ that appears in (\ref{line_element}), depend on the coordinates $\theta^a$. 
They are all functions only of the reduced spacetime $\cal S$, the quotient of
the full spacetime by the symmetry orbits, which is coordinatized by $\rho$ and $t$. 

Note that the time coordinate $t$ is determined up to an additive constant by the 
requirement that $t\pm \rho$ be null coordinates of the metric. In fact, since the field equations
require that $\Box\rho = 0$ on the reduced spacetime \cite{NKS} 
\begin{equation}
\rho = \frac{1}{2}(\rho^+ + \rho^-), 
\end{equation}
with $\rho^\pm$ constant on left/right moving null geodesics, making them null coordinates, 
and the time coordinate is 
\begin{equation}
 t = \frac{1}{2}(\rho^+ - \rho^-).
\end{equation}
The coordinates $(t,\rho)$ are good on all of the reduced spacetime provided $d\rho$ is 
spacelike throughout $\cal S$.

Orthogonal transitivity need not be included in the concept of cylindrical symmetry 
\cite{Carot} but traditionally it has been assumed in formulating the cylindrically 
symmetric reduction of GR, and it is not known whether the model is still integrable 
if this assumption is dropped. It is not as stringent a condition as it might appear, 
since it is actually enforced by the vacuum field equations provided only two numbers, 
the so called {\em twist constants}, vanish. See \cite{Wald} Theorem 7.1.1. and 
\cite{Chrusciel}.   

The model quantized by Korotkin and Samtleben is subject to two further restrictions: 
regularity of the spacetime geometry at the symmetry axis, and asymptotic flatness at 
spatial infinity, in a sense to be described below. Regularity at the axis ensures that 
the twist constants vanish, and it eliminates $\Omega$ as an independent field: Absence 
of a conical singularity at the axis requires that 
$\Omega^2 = \lim_{\rho \rightarrow 0} e_{\phi\phi}/\rho$, and the field equations then 
determine $\Omega$ as a functional of $e_{ab}$ on the rest of the spacetime \cite{NKS}. 
The field $e_{ab}$ thus contains all the degrees of freedom of the model.  

Korotkin and Samtleben do not work with these regular, asymptotically flat spacetimes 
directly, but rather with their Kramer-Neugebauer duals. The Kramer-Neugebauer 
transformation \cite{Kramer_Neugebauer}\cite{Breit} is an invertible map from solutions 
to solutions of cylindrically symmetric GR which takes 4-metrics that are regular on the 
axis to geometries for which $e_{ab}$ is regular there. Moreover it maps flat spacetime 
to a solution in which $e_{ab}$ is constant everywhere. This suggests (part of) a
definition of asymptotic flatness: As spatial infinity is approached, that 
is in the limit $\rho \rightarrow \infty$, $e_{ab}$ of the Kramer-Neugebauer dual is 
required to tend to a constant matrix $e_{\infty\,ab}$ (in the basis $\di_\phi$, $\di_z$ 
of Killing vectors). We shall return to the issue of asymptotic flatness in Subsection 
\ref{asymptotic_flatness}, where it will be defined in terms of
the deformed metric $\M$. However, many of our results will be quite independent of 
the hypothesys of asymptotic flatness. 

Thus reduced, vacuum general relativity becomes a non-linear sigma model coupled to a fixed 
(background) dilaton $\rho$. Evaluating the Hilbert action on the class of cylindrically 
symmetric field configurations under consideration one obtains the action 
\cite{Breit}\cite{Nicolai}\cite{KS} (see also \cite{Henneaux_non_lin_sigma})
\begin{equation}\label{action}
 I = -\frac{1}{8}\int_{\cal S} \rho (\di_t e^{ab} \di_t e_{ab} - \di_\rho e^{ab} \di_\rho e_{ab}) d\rho dt 
 = \frac{1}{2}\int_{\cal S} \rho \tr[P_t^2 - P_\rho^2] d\rho dt,
\end{equation}
where $e^{ab}$ is the inverse of $e_{ab}$.\footnote{The Hilbert action usually includes a 
normalization factor depending on Newton's constant. This factor is important for the quantum 
theory. It scales the action in the Feynman path integral, and thus the size of quantum effects.
In the absence of measurements of quantum gravitational effects the factor is determined by the 
measured strength of the gravitational fields of matter fields coupled to gravity, together 
with the measured size of quantum effects in the matter fields, such as the energy of a light 
quantum of given frequency. We have not included such a factor in the action for cylindrically 
symmetric fields because it cannot be obtained from experiment, since the world is not 
cylindrically symmetric, and it is not determined by the action of full GR 
without cylindrical symmetry. Since the symmetry orbits are non-compact the action of 
a complete cylindrically symmetric spacetime is, in general, infinite. One can obtain a finite 
action by integrating over only a finite range of values of the coordinates $(\phi,z)$ parametrising 
the symmetry orbits, and the result is proportional to the action (\ref{action}), but the factor 
of proportionality depends on the range of $(\phi,z)$ chosen. Ultimately this ambiguity is not 
so strange, since quantum gravity probably has no states in which the field fluctuations 
respect exactly cylindrical symmetry, meaning that the cylindrically symmetric theory is not a 
sector of the full theory at the quantum level.}

In the second form the action is a functional, via $P_\mu$, of $\V$, a real, positively oriented, 
zweibein for $e$:
\begin{equation}
\V_a{}^i \V_b{}^j \delta_{ij} = e_{ab}, \qquad \qquad \det
\V \equiv \frac{1}{2} \vareg^{ab}\vareg_{ij}\V_a{}^i \V_b{}^j = 1.
\end{equation}
$P_\mu$ is an $\mathfrak{sl}_2$ valued 1-form on the reduced spacetime $\cal S$, defined as the 
symmetric component of the flat $\mathfrak{sl}_2$ connection
\begin{equation}
 J_{\mu\,i}{}^j = \V^{-1}{}_i{}^a \di_\mu \V_a{}^j. 
\end{equation}
That is, $P_\mu = \frac{1}{2}(J_\mu + J_\mu^t)$, where the superscript $t$ indicates transposition, or, 
more explicitly, $P_{\mu\,i}{}^j = \frac{1}{2}(J_{\mu\,i}{}^j + \dg_{il}J_{\mu\,k}{}^l \dg^{kj})$.

Indices from the beginning of the Latin alphabet, $a,b,c,...$, correspond 
to the tangent spaces of the cylindrical symmetry orbits, while letters 
$i, j, ...$ from the middle of the alphabet denote {\em internal indices}, 
which label the elements of the zweibein viewed as a basis of the space $F$ 
of constant 1-forms of density weight $-\frac{1}{2}$ on the symmetry orbits. 
$\vareg^{ab}$ and $\vareg_{ij}$ are antisymmetric symbols, with 
$\vareg^{\phi z} =1 =\vareg_{12}$; and $\dg_{ij}$ is the Kronecker 
delta. $\V$ may also be viewed as a linear map from an internal vector space 
to $F$. Then $\dg$ is a Euclidean metric on the internal space, and the 
internal indices $i, j, ...$ refer to an orthonormal basis in this space.
Finally, Greek indices $\mu,\nu, ...$ label the reduced spacetime 
coordinates $x^0 \equiv t$ and $x^1 \equiv \rho$.

Notice that the conformal 2-metric $e$ determines $\V$ only up to internal 
rotations at each point, which form the group $SO(2)$. The action is also 
invariant under these rotations, so they constitute a gauge invariance of the model. 

The reduced spacetime field equation for $e$ is the Ernst equation \cite{Ernst}, \cite{KS}
\begin{equation}\label{e_field_eq}
 \di^\mu[\rho e^{ab}\di_\mu e_{bc}] = 0.
\end{equation}
Equivalently
\begin{equation}\label{V_field_eq}
\di_\mu (\rho P^\mu) - [Q_\mu, \rho P^\mu] = 0.
\end{equation}
Here spacetime indices are raised and lowered with the flat metric 
$\eta = - dt\otimes dt + d\rho \otimes d\rho$, and $Q_\mu$ is the antisymmetric component of $J_\mu$. 
$Q$ transforms as a connection under rotations of the zweibein $\V$ and the presence of the commutator 
term makes (\ref{V_field_eq}) invariant under these gauge transformations.   

Note that if a real, unit determinant reference zweibein $Z$ is chosen, then $\V_a{}^j$ may be 
parameterized by an $SL(2,\R)$ group valued scalar field $\V_i{}^j$: $\V_a{}^j = Z_a{}^i \V_i{}^j$. The 
choice of a reference zweibein is not necessary for any of our constructions, but it allows us to 
describe them in the language of Lie groups. Since the conformal 2-metric $e$ determines $\V$ 
only up to local internal $SO(2)$ rotations the space of possible $e$s at each point is the coset 
$SL(2,\R)/SO(2)$. The action (\ref{action}) shows that the model is in fact an $SL(2,\R)/SO(2)$ coset 
sigma model with a fixed, non-dynamical, dilaton $\rho$. 

The cylindrically symmetric truncations of several theories of gravity coupled to matter, 
such as GR with electromagnetism, and supergravity, can be accommodated in the same framework of coset
sigma models with dilaton, by replacing $SL(2)$ by another semi-simple Lie group $G$, and 
$SO(2)$ by the maximal compact subgroup $Q$ of $G$. The quantization scheme of Korotkin and
Samtleben extends to these models \cite{KS}\cite{Samtleben_thesis}\cite{Koepsell}, and we expect that our 
results on the Geroch group do also.

\subsection{Integrability, the auxiliary linear problem, and the deformed metric}

The integrability of the model manifests itself through the existence of an auxiliary linear problem:
From the symmetric ($P$) and antisymmetric ($Q$) components of the flat connection $J$ a new 
$\mathfrak{sl}_2$ connection, the Lax connection $\hat{J}$, is constructed which is flat if 
and only if the field equation (\ref{V_field_eq}) on $\V$ holds. The Lax connection will be 
defined as \cite{Breit}\cite{Nicolai}\cite{NKS}
\begin{equation}\label{deformed_J}
 \hat{J}_\mu = Q_\mu + \frac{1+\cg^2}{1-\cg^2} P_\mu - \frac{2\cg}{1-\cg^2}\vareg_{\mu\nu}P^\nu,
\end{equation}
with 
\begin{equation}\label{gamma_def}
 \cg = \frac{\sqrt{w + \rho^+} - \sqrt{w - \rho^-}}{\sqrt{w + \rho^+} + \sqrt{w - \rho^-}},
\end{equation}
and $w \in \C$ a spacetime independent parameter called the {\em constant spectral parameter}.
($\cg$ is called the {\em variable spectral parameter}.) $\vareg$ is the antisymmetric symbol of 
the reduced spacetime, with $\vareg_{t\rho} = 1$. 

Note that the square roots in the definition (\ref{gamma_def}) of $\cg$ are  principal roots, 
defined by $\sqrt{r e^{i\varphi}} = \sqrt{r}e^{i\varphi/2}$ for all $r \in [0,\infty), \varphi\in (-\pi,\pi]$. 
From this, and the fact that $\Im (w + \rho^+) = \Im (w - \rho^-)$,
it follows that $|\cg| \leq 1$, and if $|\cg| = 1$ then $\Im(\cg) \leq 0$.
 
The Lax connection may also be expressed in terms of the null coordinates $(\rho^+,\rho^-)$:
\begin{equation}\label{null_Lax_connection}
\fl  \hat{J} = Q - \frac{\cg - 1}{\cg + 1} P_+ d\rho^+ - \frac{\cg + 1}{\cg - 1} P_- d\rho^-
          = Q + \frac{\sqrt{w - \rho^-}}{\sqrt{w + \rho^+}} P_+ d\rho^+ + \frac{\sqrt{w + \rho^+}}{\sqrt{w - \rho^-}} P_- d\rho^-. 
\end{equation}

A direct calculation yields the curvature of the Lax connection. Taking into account that $J$ is flat,
that is, $[\di_\mu - J_\mu, \di_\nu - J_\nu] = 0$, one obtains for the curvature of $\hat{J}$
\begin{equation}
 [\di_\mu - \Jh_\mu, \di_\nu - \Jh_\nu] = - \frac{1}{\sqrt{w + \rho^+}\sqrt{w - \rho^-}} \vareg_{\mu\nu}
 (\di_\sigma (\rho P^\sigma) - [Q_\sigma, \rho P^\sigma]), 
\end{equation}
which vanishes iff (\ref{V_field_eq}) holds.

Since $\hat{J}$ is flat on solutions there exists on these a zweibein $\hat{\V}$, the ``deformed
zweibein'', that is covariantly constant with respect to $\hat{J}$:
\begin{equation}\label{auxlin}
 d \hat{\V} = \hat{\V}\hat{J},
\end{equation}
or equivalently $\hat{J} = \hat{\V}^{-1}d\hat{\V}$. That is, $\hat{\V}$ bears the same relationship to 
$\hat{J}$ as $\V$ does to $J$. 

Equation (\ref{auxlin}) is the auxiliary linear problem. A particularly useful solution, $\Vh_0$, is
obtained by setting $\Vh = \V$ at a reference point $\mathbf{0}$ on the worldline in $\cal S$ of the
symmetry axis. Then
\begin{equation}\label{integral_for_Vh}
\Vh_0(x;w) = \V(\mathbf{0})\: {\cP} e^{\int_{\mathbf{0}}^x \Jh(\cdot; w)}\ \ \ \forall x \in {\cal S},
\end{equation}
with $\cP$ indicating a path ordered exponential. Notice that $\Jh = J$ when $\cg = 0$. This is the case 
on the axis, where $\rho^+ = - \rho^-$, so the location of $\mathbf{0}$ within the axis doesn't matter, 
since $\Vh_0 = \V$ on the whole axis. It also occurs on the whole reduced spacetime in the limit that 
$w \rightarrow \infty$. $\Vh_0$ therefore equals $\V$ everywhere in this limit. It follows that if 
$\Vh_0$ at a given point $x \in {\cal S}$ is expressed as a function of $\cg$, via 
\begin{equation}\label{w_of_gamma}
 w = \frac{\rho}{2}(\cg + 1/\cg) - t = \frac{1}{4\cg}[(1 - \cg)^2 \rho^+ + (1 + \cg)^2 \rho^-],
\end{equation}
then $\V(x)$ is just $\Vh_0(x;\cg)$ evaluated at $\cg = 0$. 

A solution $\V$ is thus easily recovered from the deformed zweibein $\Vh_0$ it defines. This provides 
a means to solve the field equation because $\Vh_0$ can be obtained as the solution of a 
``factorization problem''. Just as the zweibein $\V$ determines the conformal metric $e = \V\V^t$ the 
deformed zweibein $\Vh$ defines the deformed (conformal) metric mentioned in the Introduction:
\begin{equation}\label{M_def0}
 \M(x;\cg) = \Vh(x;\cg)\Vh^t(x;\frac{1}{\cg}).
\end{equation}
As we shall see shortly, $\Vh_0$ can be recovered from $\M$, and $\M$, in turn, can be computed by 
integration from initial data. In fact, $\M$ is altogether a useful parametrization of the solutions 
of the theory and plays a central role in the present paper.\footnote{
$\M$ is sometimes called a ``scattering matrix'' \cite{Breit}, although that word is also used for 
the quantum $R$ matrix on occasion, or a ``monodromy matrix'' in \cite{Breit}\cite{NKS} and \cite{KS}, but
that word is more often used for another object, the holonomy of the Lax connection along the entire 
length of a one dimensional space \cite{Fadeev_Takhtajan}\cite{Babelon_Bernard}.}

\begin{figure}
 \begin{center}

\includegraphics[height=8cm]{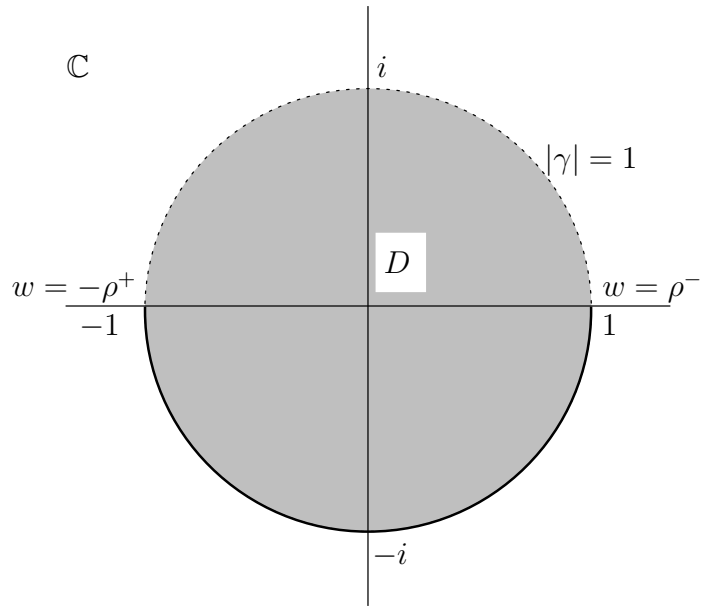}
\caption{The unit circle in the $\cg$ plane, shown in the figure, corresponds to the branch cut of 
$\cg(w)$ along the real segment $[-\rho^+, \rho^-]$ in the $w$ plane, with $\cg = 1$ and $\cg = -1$ 
at the branch points $w = \rho^-$ and $w = -\rho^+$ respectively. The domain $D$, consisting of the 
open disk and the lower unit semi-circle (including $1$ and $-1$), is the range of the function 
$\cg(w)$ defined by (\ref{gamma_def}), and the $\cg$ domain of $\Vh_0(x;\cg)$.}
\label{gamma_plane}
\end{center}
\end{figure}

But before explaining this we must clarify the definition (\ref{M_def0}) of $\M$. For $\cg$ in the set 
$D$ consisting of the open unit disk, $|\cg| < 1$, and the lower half of the unit circle, 
$|\cg|=1, \Im \cg \leq 0$, $\Vh(x;\cg)$ will be set to $\Vh_0(x;\cg)$. See Fig. \ref{gamma_plane}. One 
could equate $\Vh$ with $\Vh_0$ also in the complement of $D$, using the fact that $w(\cg) = w(1/\cg)$ 
according to (\ref{w_of_gamma}), but this would in general lead to a $\Vh$ which is discontinuous on 
the unit circle. It is more useful to define $\Vh$ on the unit circle to be the limit of $\Vh_0$ inside 
the unit disk.

Note that $D$ is precisely the range of $\cg(w)$ as defined by (\ref{gamma_def}), and that $\cg(w)$
is holomorphic in $w$ on all of the Riemann sphere except at a branch cut on the segment 
$-\rho^+ \leq w \leq \rho^-$ of the real axis. On the branch cut $\cg(w)$ is continuous from above, 
that is, $\cg(w) = \lim_{\eg \rightarrow 0, \eg>0} \cg(w + i\eg)$. From the definition (\ref{auxlin}, 
\ref{integral_for_Vh}) it follows that $\Vh_0(x;w)$ shares these analyticity and continuity properties 
in $w$: The right side of (\ref{auxlin}) is holomorphic in $\Vh$ and $w$ away from the lines 
$\rho^- = w$ and $\rho^+ = - w$ in spacetime where $\cg = \pm 1$, implying (by \cite{Lefschetz} 
proposition 10.3, Chapter II, section 5) that $\Vh_0$ is holomorphic in $w$ provided the path of 
integration in (\ref{integral_for_Vh}) may be chosen to avoid these lines, which is the case for all 
$w \in \mathbb{C}$ except the branch cut $-\rho^+ \leq w \leq \rho^-$. Since the singularities of $\Jh$ 
at the singular lines are integrable a dominated convergence argument adapted to path ordered exponentials 
(see prop. 4 of appendix A of \cite{ReiAnd}) shows that $\Vh_0$ is continuous from above in $w$ at the 
branch cut. At any point $x$ of the reduced spacetime $\cg(w)$ maps the complement of the branch cut to 
the interior of the unit disk, so $\Vh_0(x;\cg)$ is analytic in $\cg$ in $|\cg|<1$, and approaching the 
branch cut from above in $w$ corresponds to approaching the lower half of the $\cg$ unit circle 
$\{|\cg| = 1, \Im \cg \leq 0\}$ from the inside of the unit disk, so $\Vh_0(x;\cg)$ on the lower half 
of the unit circle is the limit of $\Vh_0(x;\cg)$ inside the unit disk. 

It remains to analyze the implications of setting $\Vh(x;\cg)$ on the upper half of the $\cg$ unit circle equal to the limit of $\Vh_0(x;\cg)$ inside the unit disk. Off the branch cut 
$\overline{\Vh_0(x;w)} = \Vh_0(x;\bar{w})$ and thus $\overline{\Vh_0(x;\cg)} = \Vh_0(x;\bar{\cg})$
for $|\cg|<1$, which by continuity holds for $\Vh(x;\cg)$ on the entire closed unit disk, including 
the circle $|\cg|=1$. Notice that $\cg(w)$ can be extended to a double valued analytic function of $w$, namely the inverse of $w(\cg)$ defined by (\ref{w_of_gamma}) which takes values $\cg(w)$ and $1/\cg(w)$ at each $w \in \mathbb{C}$. To define $\Vh(x;\cg)$ on the closed $\cg$ unit disk we have made
also $\Vh(x;w)$ double valued on the branch cut $-\rho^+ \leq w \leq \rho^-$, taking the values
$\Vh^{(1)}(x;w) = \Vh_0(x;w)$ and $\Vh^{(2)}(x;w) = \overline{\Vh_0(x;w)}$ corresponding to $\cg(w)$
and $\overline{\cg(w)} = 1/\cg(w)$ respectively. $\Vh^{(2)}$ satisfies the auxiliary linear problem
with connection $\Jh^{(2)} = \Vh^{(2)\,-1} d \Vh^{(2)} = \overline{\Vh^{(1)\,-1} d \Vh^{(1)}} = \overline{\Jh^{(1)}}$. This is just $\Jh$ evaluated using $\overline{\cg(w)} = 1/\cg(w)$ in place of $\cg(w)$, which by (\ref{deformed_J}) yields $-\Jh^{(1)\,t}$. Thus 
\begin{equation}\label{second_branch_connection}
\Jh^{(2)} = -\Jh^{(1)\,t}. 
\end{equation}

Having defined $\Vh(x;\cg)$ on the closed unit disk we have defined $\M(x;\cg)$ only on the unit circle, since for any $\cg$ that lies in the interior of the unit disk $1/\cg$ lies outside the closed unit disk. But this will suffice for our purposes. Equation (\ref{M_def0}) provides a factorization
of $\M(x;\cg)$ on the $\cg$ unit circle into a product of an $SL(2,\mathbb{C})$ valued function 
$\Vh(x;\cg)$ holomorphic inside the unit circle having a continuous limit on the circle itself, and an $SL(2,\mathbb{C})$ valued function $\Vh(x;1/\cg)$ holomorphic outside the unit circle in the Riemann sphere, having a continuous limit on the circle. This factorization is essentially unique, for suppose 
$\Vh'(\cg)$ defines another such factorization, then $\Vh'(\cg)\Vh'^t(1/\cg) = \Vh(\cg)\Vh^t(1/\cg)$ on $|\cg| = 1$, which implies that $f_+(\cg) = \Vh^{-1}(\cg)\Vh'(\cg)$ and $f_-(\cg) = \Vh^t(1/\cg)\Vh'^{t\,-1}(1/\cg) = f_+^{t\,-1}(1/\cg)$ are 
functions that are holomorphic inside and outside the unit circle respectively and match on the circle. Together they therefore define a continuous function on the Riemann sphere which is holomorphic off the unit circle. By Morera's theorem this function is actually holomorphic on the whole Riemann sphere and thus constant. Indeed it must be a constant orthogonal matrix since it must equal $f_+(1) 
= f_+^{t\,-1}(1)$. $\M$ therefore determines $\Vh(x;\cg)$, and thus also $\V(x)$, up to an $SO(2)$ gauge transformation.

Since we will reuse this holomorphy argument several times let us enshrine it in a lemma:
\begin{lemma}\label{analyticity}
Suppose a continuous function $f$ on the Riemann sphere is holomorphic everywhere except possibly on the unit circle
$|z| = 1$, then it is a constant.
\end{lemma}

\noindent {\em Proof:} By Cauchy's integral theorem the integral of $f$ around a closed, 
rectifiable, piecewise $C^1$ curve entirely inside or entirely outside the unit circle 
vanishes. Because $f$ is continuous a uniform convergence argument shows that the integral 
also vanishes on curves that contain segments of the unit circle but do not cross this 
circle. This shows that the integral around any closed, rectifiable, piecewise $C^1$ 
curve at all vanishes. By Morera's theorem $f$ is therefore holomorphic on the entire 
Riemann sphere, and thus, by Liouville's theorem, constant. \hfill\QED

A solution may therefore be recovered, up to gauge, from the deformed metric $\M$ it defines. 
We will now show a remarkable property of $\M$ expressed as a function of $x$ and $w$ instead 
of $x$ and $\cg$. Namely, that $\M(x;w)_{ab}$ corresponding to a solution of the field equation 
depends only on $w$, and in fact is equal to the conformal metric $e_{ab}$ on the axis worldline at time 
$t = -w$. In the next subsection we will see that {\em any} $\C^\infty$ function $\M(w)$ sharing the 
algebraic properties of a conformal metric (that it be a real, positive definite, symmetric matrix 
of unit determinant) determines a solution uniquely up to gauge, so $\M(w)$ provides a natural 
parametrization of the space of solutions (including ones that are not asymptotically flat). 

\begin{figure}
 \begin{center}

\includegraphics[height=8cm]{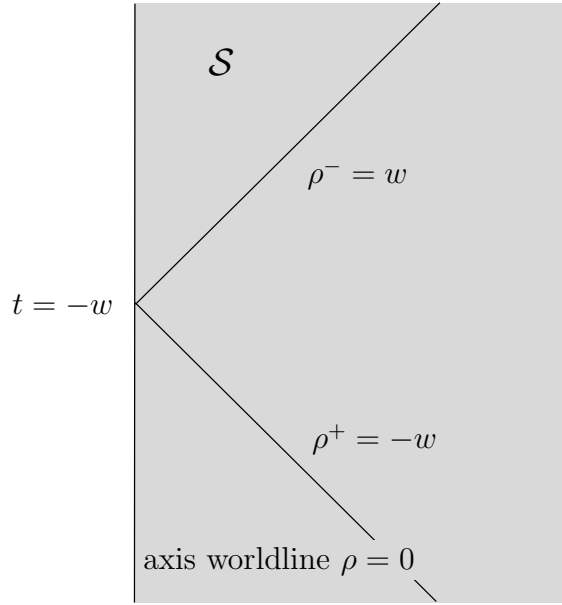}
\caption{The reduced spacetime $\cal S$ is indicated by the shaded area. 
It is bounded by the worldline of the cylindrical symmetry axis, shown as 
a vertical line on the left. For a given value of $w$ the functions 
$\cg(x;w)$ and $\V_0(x;w)$ are singular (non-holomorphic) on the diagonal 
lines $\rho^- = w$ and $\rho^+ = -w$. The deformed metric $\M(w)$ is equal 
to the conformal metric $e$ on the symmetry axis at the time $t = -w$ where 
the two singular lines meet.}
\label{singularities}
\end{center}
\end{figure}

Let us explore the properties of $\M(x;w)$. To each $w$ lying on the branch cut $[-\rho^+,\rho^-]$
there corresponds a point $\cg(w)$ on the lower half of the $\cg$ unit circle, and another point 
$1/\cg(w) = \overline{\cg(w)}$ lying on the upper half of the unit circle. {\em A priori} these
define two values of $\M$, 
\begin{equation}
\M^{(1)}(x;w) \equiv \M(x;\cg(w)) = \Vh^{(1)}(x;w) \Vh^{(2)\,t}(x;w) = \Vh_0(x;w)\overline{\Vh^t_0(x;w)}
\end{equation}
and 
\begin{equation}
\M^{(2)}(x;w) \equiv \M(x;\overline{\cg(w)}) = \Vh^{(2)}(x;w) \Vh^{(1)\,t}(x;w) = \overline{\M^{(1)}(x;w)} 
\end{equation}
respectively. 
Now notice that by (\ref{second_branch_connection}) the spacetime gradient of $\M^{(1)}$ at constant $w$ vanishes! 
\begin{equation}\label{dM1}
 d\M^{(1)} = \Vh^{(1)} \Jh^{(1)} \Vh^{(2)\,t} + \Vh^{(1)}\Jh^{(2)\,t}\, \Vh^{(2)\,t} = 0. 
\end{equation}
If $w$ is real it lies in the branch cut of the function $\cg(x;w)$ whenever the spacetime point 
$x$ is spacelike separated from the point $t = -w$ on the symmetry axis worldline. These spacetime 
points form a wedge bounded to the past by the null line $\rho^+ = -w$ and to the future by the null
line $\rho^- = w$, lines at which $w$ lies on the edge of the branch cut and $\cg = -1$ and 
$\cg = 1$ respectively. See Fig. \ref{singularities}. $\M^{(1)}(x;w)$ is clearly independent of position on the interior of this
wedge. On the boundary of the wedge $\Jh^{(1)}(x;w)$ is singular in $x$, but the singularity is integrable so $\Vh_0(x;w)$ defined by the path ordered exponential (\ref{integral_for_Vh}) is continuous in $x$ throughout $\cal S$. $\M^{(1)}$ thus has the same constant value on the boundaries of the wedge as in its interior. It may therefore be evaluated at the vertex of the wedge, the point $t = -w$ on the symmetry axis worldline, where $\Vh_0 = \V$ and hence 
$\M^{(1)} = \V\V^t = e$. Since this is real it also equals $\M^{(2)}$: 
\begin{equation}\label{M_e_relation}
 \M(x;w) \equiv \M^{(1)}(x;w) = \M^{(2)}(x;w) = e(\rho = 0, t = -w).
\end{equation}
One may conclude that $\M(x;w)_{ab}$ is real, positive definite, of unit determinant, symmetric in its indices, and independent of $x$, being a smooth function of $w$ only. As a consequence of the last property 
\begin{equation}\label{M_conjugation_invariance}
 \M(x;\bar{\cg}) = \M(x;\cg)
\end{equation}
on the $\cg$ unit circle. (Note that while the $x$ independence of $\M(x;w)$ has not been demonstrated for all $x$ in spacetime it has been demonstrated for all $x$ such that $|\cg(x;w)| = 1$, which is the subset of spacetime on which $\M(\cdot;w)$ has been defined.) 

\begin{figure}
 \begin{center}

\includegraphics[height=8cm]{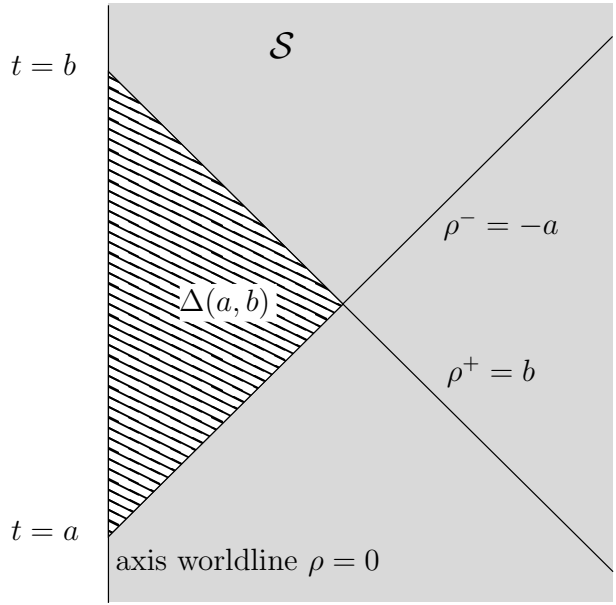}
\caption{The conformal metric $e$ on the segment $a \leq t \leq b$ of the worldline of the symmetry axis determines the solution in the reduced spacetime domain $\Delta(a,b)$, the hatched region in the figure.}
\label{diamond}
\end{center}
\end{figure}

Equation (\ref{M_e_relation}) provides an interpretation of $\M$ as the conformal metric on the axis
worldline in the reduced spacetime. Given $e$ on a segment $a \leq t \leq b$ of the axis worldline
this determines $\M(w)$ for $a \leq -w \leq b$, which in turn determines $\M(x;\cg)$ on the whole 
$\cg$ unit circle at all points $x$ within the ``causal diamond'', $\Delta(a,b)$, of $a \leq t \leq b$, that is, within the intersection of the past light cone of the point $t = b$ and the future light cone of the point $t = a$ on the axis. See Fig. \ref{diamond}. As we have seen this allows the solution $\V$ to be recovered within the
causal diamond up to $SO(2)$ gauge. $\M$ may also be calculated from $\V$ and its normal derivative
along any Cauchy surface of this diamond by first calculating $\Vh_0$ on this surface (really a line
in $\cal S$). For instance, it can be calculated from $\V$ on one of the null boundaries of the
diamond, as discussed in detail in \cite{ReiAnd}.

\subsection{The deformed metric and the space of smooth solutions to the field equation}\label{M_solution_space}

We have seen that the deformed metric $\M$, or equivalently the conformal metric on the axis, determines
the solution uniquely up to $SO(2)$ gauge. In this subsection it will be shown that {\em any} smooth
function $\M$ from a real interval $a \leq -w \leq b$ to real, symmetric, positive definite, unit
determinant $2 \times 2$ matrices defines a smooth solution $\V(\rho,t)$ on the causal diamond 
$\Delta(a,b)$ such that $e(t = -w)_{cd} = \M(w)_{cd}$ on the axis worldline.   

The space $\D_n$ of smooth functions from a given interval in $\mathbb{R}$ to the group 
$GL(n,\mathbb{C})$ of invertible $n \times n$ complex matrices naturally has the structure of a group,
under pointwise multiplication, and of an infinite dimensional manifold modeled on the topological 
vector space $E$ formed by smooth functions taking values in the set of all (not necessarily invertible)
$n \times n$ complex matrices with the topology of uniform convergence of the functions and all their partial derivatives on compact sets. (See \cite{Pressley_Segal} Chapter 3.)
It follows that the space of solutions in the causal diamond of the axis worldline segment 
$a \leq t \leq b$ is precisely the submanifold of $\D_2$ for the interval $a \leq -w \leq b$ in which the 
matrices are real, symmetric, positive definite, and of unit determinant. 

The demonstration of the existence of solutions corresponding to all $\M$ relies on the existence of
factorizations of ``loops'', which are smooth functions from the unit circle to $GL(n,\mathbb{C})$. The
set $\cL_n$ of all such functions admits a natural group structure and manifold structure exactly
analogous to that of $\D_n$, and is called the ``$GL(n,\mathbb{C})$ loop group'' \cite{Pressley_Segal}.  

The loop in our proof will be the deformed metric $\M$ ``expressed as function of $\cg$ at
a given reduced spacetime point $x$''. More precisely, at each point $x$ in the causal diamond 
$\Delta(a,b)$ the given function $\M(w)$ on $a \leq -w \leq b$ defines the function
\begin{equation}\label{Mc_def}
 \M_C(x;\cg) \equiv \M\circ w (x;\cg) = \M(\frac{\rho}{2}(\cg + 1/\cg) - t)                                    
\end{equation}
for all $\cg$ on the unit circle in $\mathbb{C}$. This is ``$\M$ as a function of $x$ and 
$\cg$'', which we denote $\M_C$ here to avoid confusion because it is, after all, distinct
as a function from the function $\M$ of $w$. (The subscript $C$ is meant to suggest 
``circle'' or ``$\cg$''.) 
$\M_C$, like $\M$, takes values in the real, symmetric, positive definite, unit determinant
$2 \times 2$ matrices. The definition (\ref{Mc_def}) implies, furthermore, that it is a
smooth function of $\cg$ (and $\rho$ and $t$), and that
\begin{equation}
 \overline{\M_C(\cg)} = \M_C(\bar{\cg})
\end{equation}
on the $\cg$ unit circle, since $\bar{\cg} = 1/\cg$ there. (The $x$ dependence of the two sides 
of this equation has been left implicit, as it will be in other equations in which it plays no role.)

We shall see that these properties of $\M_C$, imply that $\M_C$ admits a factorization of the form
\begin{equation}\label{factorization}
 \M_C(\cg) = \nuh(\cg)\nuh^t(\cg^{-1}), 
\end{equation}
which defines a deformed zweibein $\nuh$ and a zweibein 
$\nu(x) = \nuh(x;\cg = 0)$. This zweibein will be the solution defined by $\M$. 

The main theorem underlying this result is (\cite{Pressley_Segal} Theorem 8.1.2):

\begin{thm}[Birkhoff factorization theorem]\label{Birkhoff}
 Let $M$ be a smooth function from the unit circle $|z| = 1$ in the complex plane to
 the group $GL(n,\mathbb{C})$ of invertible $n \times n$ complex matrices, that is $M \in \cL_n$, then there exist 
 smooth functions $A$ and $B$ from the unit disk $|z|\leq 1$ to $GL(n,\mathbb{C})$, which are
 holomorphic on the interior $|z|<1$ of the disk, such that
 \begin{equation}
  M(z) = A(z)D(z)B(z^{-1})
 \end{equation}
 on $|z| = 1$, where $D(z) = \diag[z^{k_1}, z^{k_2}, ..., z^{k_n}]$ with 
 $k_1\geq k_2\geq ...\geq k_n$ integers called partial indices.
 
 The partial indices are uniquely determined by $M$. For a dense open subset 
 of $\cL_n$ all the partial indices vanish, and then $A$ and $B$ are also 
 uniquely determined by $M$, provided the value of $B(0)$ is fixed. Indeed on this
 open subset of $\cL_n$, once $B(0)$ is fixed, the map 
 $\cL_n \rightarrow \cL_n \times \cL_n$ defined by $M \mapsto (A, B)$ is a diffeomorphism.  
\end{thm}
(In the statement of the theorem in \cite{Pressley_Segal} $A$ and $B$ are smooth functions 
on the circle $|z| = 1$ which are the boundary values of functions holomorphic
in $|z| < 1$. But it is easy to show that this implies that these functions are smooth on
$|z| \leq 1$: The smoothness of such a function along the boundary $|z| = 1$ implies that
the power series in $z$ for the function and for each of its derivatives converge uniformly 
on $|z| \leq 1$, ensuring that each derivative is continuous in this domain.
Seeley's theorem \cite{Seeley} then implies that the function has a $C^\infty$ extension
to the whole $z$ plane.)

We will also use Theorem 1.13 of \cite{Faro}: 

\begin{thm}\label{Positive}
 If $M \in \cL_n$ is Hermitian and positive definite, then all partial indices of $M$ vanish.
\end{thm}

Now we are ready to use the properties of $\M_C$ to demonstrate (\ref{factorization}): 

\begin{prop}\label{Mvv}
 Suppose $\M_C \in \cL_2$ is real, symmetric, positive definite and of unit determinant, 
 and that $\overline{\M_C(\cg)} = \M_C(\bar{\cg})$. Then there exists a smooth, 
 $2 \times 2$ matrix valued function $\nuh$ on $|\cg| \leq 1$ which is holomorphic on 
 $|\cg| < 1$ and satisfies  
 \begin{equation}\label{Factorization}
  \M_C(\cg) = \nuh(\cg)\nuh^\dagger(\cg) = \nuh(\cg)\nuh^t(\cg^{-1})
 \end{equation}
 on $|\cg| = 1$, 
 \begin{equation}\label{det1}
  \det \nuh = 1,  
 \end{equation}
 and 
 \begin{equation}\label{reality_condition}
  \overline{\nuh(\cg)} = \nuh(\bar{\cg}).  
 \end{equation}
 These conditions determine $\nuh$ up to right multiplication by a constant real
 $SO(2)$ matrix $K$. Thus $\nuh' = \nuh K$ also satisfies the conditions. 
\end{prop}  
  
\noindent{\em Proof}:
By Theorem \ref{Birkhoff} there exist smooth functions $A_0$ and $B_0$ from the closed 
$\cg$ unit disk to $GL(2,\mathbb{C})$, holomorphic on $|\cg| < 1$ such that 
\begin{equation}\label{factorization0}
 \M_C(\cg) = A_0(\cg)B_0(\cg^{-1})
\end{equation}
with $B_0(0) = \One$. The diagonal factor in the factorization is absent by Theorem \ref{Positive}. 
Define $A_1(\cg) = A_0(\cg)B_0(1)$ and $B_1(\cg) = B_0^{-1}(1)B_0(\cg)$, so that
\begin{equation}\label{factorization1}
 \M_C(\cg) = A_1(\cg)B_1(\cg^{-1})
\end{equation}
on $|\cg| = 1$, with $B_1(1) = \One$ and $A_1(1) = \M_C(1)$.

Because $\M_C$ is Hermitian  
\begin{equation}
 A_1(\cg)B_1(\bar{\cg}) = \M_C(\cg) = \M_C^\dagger(\cg) = B_1^\dagger(\bar{\cg})A_1^\dagger(\cg), 
\end{equation}
so 
\begin{equation}
 A_1^{-1}(\cg) B_1^\dagger(\bar{\cg}) = B_1(\bar{\cg})A_1^{\dagger\,-1}(\cg) = [A_1^{-1}(\cg) B_1^\dagger(\bar{\cg})]^\dagger 
\end{equation}
on $|\cg| = 1$. Note that the left side is holomorphic on $|\cg|<1$ while the right side is
holomorphic on $|\cg| > 1$, and both sides are continuous at $|\cg| = 1$. Thus by Lemma 
\ref{analyticity} both sides are equal to a constant Hermitian matrix $K_1$: 
\begin{equation}
 A_1^{-1}(\cg) B_1^\dagger(\bar{\cg}) = K_1 = K_1^\dagger.
\end{equation}
Evaluating at $\cg = 1$ shows that $K_1 = \M_C^{-1}(1)$. 

Since $\M_C(1)$ is real, symmetric, and positive definite there exists a real zweibein $C$ for $\M_C(1)$ such that $\M_C(1) = C C^t$. This zweibein will be made unique by requiring it to be a lower 
triangular matrix with positive diagonal entries, so that $C$ defines what is called a  
Cholesky factorization of $\M_C(1)$ \cite{Cholesky}. For $2 \times 2$ matrices the Cholesky factors are easily written down explicitly: For an arbitrary real, symmetric, positive definite $2 \times 2$ matrix $S$ the Cholesky factorization $L L^t = S$ is given by  
\begin{equation} \label{explicit_Cholesky}
 L = \frac{1}{\sqrt{S_{11}}}\left[\begin{array}{cc} S_{11} & 0 \\ S_{21} & \sqrt{\det S} \end{array}\right]. 
\end{equation}

With this definition $B_1(\bar{\cg}) = C^{t\,-1} C^{-1} A_1^\dagger(\cg)$,
so 
\begin{equation}\label{factorization2}
 \M_C(\cg) = \nuh(\cg)\nuh^\dagger(\cg)
\end{equation}
on $|\cg| = 1$, with $\nuh(\cg) = A_1(\cg)C^{t\,-1}$. As a consequence $\nuh(1) = C$.

Equation (\ref{factorization2}) and the condition $\overline{\M_C(\cg)} = \M_C(\bar{\cg})$ implies that
$\overline{\nuh(\cg)}\nuh^t(\cg) = \nuh(\bar{\cg})\overline{\nuh^t(\bar{\cg})}$, and thus
\begin{equation}
 \overline{\nuh^t(\bar{\cg})}\nuh^{t\,-1}(\cg) = \nuh^{-1}(\bar{\cg})\overline{\nuh(\cg)} =
 \overline{[\overline{\nuh^t(\bar{\cg})}\nuh^{t\,-1}(\cg)]}^{t\,-1}
\end{equation}
on $|\cg| = 1$.
Once again Lemma \ref{analyticity} shows that the two sides must be equal to a constant matrix.
Evaluating at $\cg = 1$ shows that this constant is $\bar{C}^t C^{t\,-1} = \One$, since $C$ is real. Thus
\begin{equation}\label{reality_condition2}
 \overline{\nuh(\cg)} = \nuh(\bar{\cg}).
\end{equation}  

The fact that $\det \nuh = 1$ is demonstrated similarly: $\det \M_C = 1$ implies that 
$\det \nuh(\cg) = 1/\overline{\det \nuh(\cg)}$ on $|\cg| = 1$, so Lemma \ref{analyticity}
shows that $\det \nuh(\cg)$ is independent of $\cg$. Thus
\begin{equation}
 \det \nuh(\cg) = \det \nuh(1) = \det C = 1.
\end{equation}
Here we have taken into account that $1 = \det\M_C(1) = (\det C)^2$ and $\det C > 0$ because 
$C$ is lower triangular with positive diagonal elements.

The function $\nuh$ which has been found satisfies conditions (\ref{Factorization}), (\ref{det1}), 
and (\ref{reality_condition}), and $\nuh(1) = C$ is lower triangular with positive diagonal elements. 

It is evident the zweibein $\nuh' = \nuh K$ also satisfies (\ref{Factorization}), (\ref{det1}), 
and (\ref{reality_condition}) provided $K$ is a constant real $SO(2)$ matrix. This is so
{\em only} if $K$ is a constant real $SO(2)$ matrix because if (\ref{Factorization}) holds for
both $\nuh$ and $\nuh'$ then $\nuh^{-1}(\cg)\nuh'(\cg) = \nuh^t(1/\cg)\nuh'^{-1\,t}(1/\cg)$ on 
$|\cg| = 1$. Lemma \ref{analyticity} then implies that $K = \nuh^{-1}(\cg)\nuh'(\cg)$ is  an 
orthogonal matrix independent of $\cg$.\hfill\QED

An important issue is how $\nuh(x;\cg)$ depends on the spacetime position $x$.

\begin{prop}\label{smoothness}
 If the $SO(2)$ freedom in $\nuh$ is fixed by requiring $\nuh(1)$ to be lower triangular 
 with positive diagonal elements then $\nuh$ is a smooth function of $\cg$ and spacetime 
 position $x$ on $|\cg| = 1$. 
\end{prop}  

\noindent{\em Proof}:
Fix an arbitrary smooth curve, parameterized by $\lambda \in \R$, in the reduced spacetime. It 
defines a curve $\M_C(x(\lambda), \cg)$ in $\cL_2$. Our first task will be to show that
this curve is $C^\infty$ in $\cL_2$. Then, adopting the notation of (\ref{factorization0}) we know that 
$\M_C(x;\cg) = A_0(x;\cg)B_0(x;\cg^{-1})$ at each spacetime point, with $A_0$ and $B_0$
smooth on $|\cg|\leq 1$ and holomorphic on $|\cg|< 1$, and $B_0(x;0) = \One$. Theorem
\ref{Birkhoff} tells us, in addition, that $A_0(x(\lambda),\cg)$ and
$B_0(x(\lambda),\cg)$ viewed as curves in $\cL_2$ parameterized by $\lambda$ are smooth.  
The second step of the proof consists in demonstrating that this implies
that $A_0(x(\lambda),\cg)$ and $B_0(x(\lambda),\cg)$ are $C^\infty$ functions of $\lambda$ and $\cg$ on $|\cg| = 1$.
From this the claim of the proposition is then easily deduced.

Let us turn to the first task, showing that $\M_C(x(\lambda), \cg)$ is a $C^\infty$ curve in $\cL_2$.
Let $e^{i\theta} = \cg$ on $|\cg| = 1$, then
\begin{equation}\label{curve}
 \M_C(x(\lambda), e^{i\theta}) = \M(\rho(\lambda)\cos(\theta) - t(\lambda)).
\end{equation}
We will work with the coordinates of this curve in the atlas that Pressley and Segal 
\cite{Pressley_Segal} use to define the manifold structure of $\cL_2$. This atlas is 
composed of right translates (under $\theta$-pointwise multiplication) of a chart 
$\Phi_1$ on a neighborhood of the identity in $\cL_2$, the $\Phi_1$ coordinates of a
point $X \in \cL_2$ in this neighborhood being the matrix elements $[\ln X]_{ab}(\theta)$ of the
logarithm (inverse exponential map) of $X$. Note that since $GL(2,\C)$ is a finite
dimensional Lie group the logarithm is a diffeomorphism in a neighborhood of the identity 
in this group.

Let $\lambda_0$ be the value of $\lambda$ at some particular point of the curve 
(\ref{curve}), then $F(\lambda, \theta) = \M_C(x(\lambda),\cg)\M^{-1}_C(x(\lambda_0),\cg)$
is a right translate of this curve that passes through the identity of $\cL_2$ at
$\lambda = \lambda_0$, and $\ln F(\lambda, \theta)$ is the matrix of coordinates of this
curve in the chart $\Phi_1$. From the definition (\ref{curve}), and the fact that $\M$ is 
$C^\infty$, it follows that $\ln F$ is $C^\infty$ in its arguments, and vanishes at 
$\lambda = \lambda_0$. As a consequence $\ln F$ and all its $\theta$ derivatives converge
to $0$ uniformly in $\theta$ at $\lambda = \lambda_0$, that is, they converge uniformly to
$\ln F(\lambda_0,\theta)$ and its $\theta$ derivatives. $F$ is thus a continuous curve
in $\cL_2$ at $\lambda = \lambda_0$, and so is $\M_C(x(\lambda), e^{i\theta})$.

Now let us consider the first derivative in $\lambda$. The difference
\begin{equation}
 [\ln F(\lambda,\theta) - \ln F(\lambda_0,\theta)]/(\lambda - \lambda_0) - \di_\lambda \ln F(\lambda_0,\theta)
\end{equation}
is $C^\infty$ in its arguments and vanishes at $\lambda =\lambda_0$. (Here $\di_\lambda$ is
the partial derivative in $\lambda$ at constant $\theta$.) Thus the ratio
$[\ln F(\lambda,\theta) - \ln F(\lambda_0,\theta)]/(\lambda - \lambda_0)$ converges to 
$\di_\lambda \ln F(\lambda_0,\theta)$ uniformly in $\theta$, and all $\theta$ derivatives
of the ratio converge uniformly to the corresponding $\theta$ derivatives of 
$\di_\lambda \ln F(\lambda_0,\theta)$. Thus the first derivative in $\lambda$ of the curve 
$F$ in $\cL_2$ exists, and its components in the chart $\Phi_1$ are simply the matrix elements of 
$\di_\lambda \ln F$. This has been established for $\lambda = \lambda_0$ but, 
because right translations act as diffeomorphisms, it clearly holds for all $\lambda$ such that $F(\lambda, \cdot)$ 
lies in the domain of $\Phi_1$. 
Now the preceding argument can be applied to $\di_\lambda \ln F$, instead of $\ln F$, to establish that the 
second derivative also exists, and finally, that the curve $F$, and also $\M_C(x(\lambda), e^{i\theta})$, is $C^\infty$ 
in $\cL_2$.

From this result it follows, by Theorem \ref{Birkhoff}, that the factors $A_0$ and $B_0$ in the factorization 
$\M_C(x;\cg) = A_0(x;\cg)B_0(x;\cg^{-1})$ also define smooth curves, $A_0(x(\lambda),\cdot)$ and
$B_0(x(\lambda),\cdot)$, in $\cL_2$. Our aim is now to show that this implies 
that $A_0(x(\lambda),e^{i\theta})$ and
$B_0(x(\lambda),e^{i\theta})$ are $C^\infty$ functions of $\lambda$ and $\theta$. 

It is sufficient to consider $A_0$. The coordinates in the chart $\Phi_1$ of the right translate of $A_0(x(\lambda),\cdot)$ 
that passes through the origin at $\lambda = \lambda_0$ are the matrix elements of
$\ln H$ where $H(\lambda, \theta) = \ln(A_0(x(\lambda),e^{i\theta})A_0^{-1}(x(\lambda_0),e^{i\theta}) )$. 
The definition of $\cL_2$ implies that $A_0(x(\lambda),e^{i\theta})$, and thus also $\ln H$, is $C^\infty$ in $\theta$.

Since $A_0(x(\lambda),\cdot)$ is a differentiable curve in $\cL_2$ the ratio
$[\ln H(\lambda) - \ln H(\lambda_0)]/(\lambda - \lambda_0)$ and all its $\theta$ derivatives converge uniformly in $\theta$
as $\lambda \rightarrow \lambda_0$. Therefore $\di_\lambda \ln H$, as well as $\di_\lambda\di^n_\theta \ln H$ exist at
$\lambda = \lambda_0$ for all integers $n > 0$, and, because the convergence is uniform, the multi derivatives obtained by
changing the order of differentiation in any one of the above expressions are defined and equal to the original expression.
This argument is equally valid at any other value of $\lambda$ such that $\ln H(\lambda)$ lies in the domain of $\Phi_1$, so 
the derivatives are defined at these values of $\lambda$ as well.

Of course $A_0(x(\lambda),\cdot)$ is actually a $C^\infty$ curve, so the argument can be applied to $\di_\lambda \ln F$ in place of $\ln F$ to demonstrate the existence of the second $\lambda$ derivatives, and, iterating, the $\lambda$ derivatives of all orders. 
$\ln H$ is therefore a $C^\infty$ function of $\lambda$ and $\theta$. It follows that $H = 
A_0(x(\lambda),e^{i\theta})A_0^{-1}(x(\lambda_0),e^{i\theta})$ is also. Thus, since $A_0(x(\lambda_0), e^{i\theta})$ is a
$C^\infty$ function of $\theta$, $A_0(x(\lambda), e^{i\theta})$ is $C^\infty$ in $\lambda$ and $\theta$. 
Because $x(\lambda)$ is an arbitrary smooth curve in the reduced spacetime, it follows that $A_0(x,e^{i\theta})$
is a $C^\infty$ function of $x$ and $\theta$.

The definitions in the proof of Theorem \ref{Mvv} imply that
$\nuh(x;e^{i\theta}) = A_0(x;e^{i\theta})B_0(1)C(x)^{-1\,t} = A_0(x;e^{i\theta})A^{-1}_0(x;1)C(x)$ where $C(x)$ is the lower triangular matrix with positive definite diagonal elements appearing in the Cholesky factorization $\M_C(x;1) = C(x) C(x)^t$.
It follows that $\nuh(x;e^{i\theta})$ is $C^\infty$ in its arguments, because 
$A_0$ and $\M_C$ are $C^\infty$ in their arguments, with $\det A_0 = 1$, and $C(x)$ is holomorphic in the components of 
$\M_C(x;1)$.

\hfill\QED

Now consider the limit of $\nuh(x;\cg)$ as $x$ approaches the worldline of the symmetry 
axis. $\M_C(x;\cg) = \M(\frac{\rho}{2}(\cg + 1/\cg) - t)$ approaches the constant, that is, 
$\cg$ independent, value $\M(w = -t)$ in the topology of the manifold $\cL_2$. $\nuh$ must 
therefore approach a factorization of this constant. Since a factorization by constant 
matrices exists, and the factorization is unique up to a constant $SO(2)$ transformation, 
$\nuh$ must be $\cg$ independent at the symmetry axis. It follows that 
$\nuh(x;\cg) = \nuh(x;0) = \nu(x)$ there.

So far we have shown that any smooth function $\M(w)$ from the interval $a \leq -w \leq b$
to $2 \times 2$ matrices that are real, symmetric, positive definite, and of unit 
determinant defines a zweibein $\nu(x) \equiv \nuh(x;\cg = 0)$ at each point $x$ in the 
causal diamond $\Delta(a,b)$ of the segment $a \leq t \leq b$ of the axis worldline in the 
reduced spacetime. But, is this zweibein a solution to the field equations? Does its 
deformed metric reproduce the function $\M$ we started with? We will now show that the 
answer to both questions is ``yes''.

The key point is that $\M$ depends only on $w$. Thus the spacetime differential of 
$\M_C(x;\cg) = \M(w(x;\cg))$ at constant $w$ vanishes:
\begin{equation}
 0 = [d\M_C]_w. 
\end{equation}
This forces $\jh \equiv \nuh^{-1}[d\nuh]_w$ to be exactly the Lax connection 
(\ref{deformed_J}) 
constructed from the zweibein $\nu$. As a consequence, since $\nuh = \nu$ on the axis, 
$\nuh$ is the deformed zweibein $\nuh_0$ constructed from $\nu$ via the integral 
(\ref{integral_for_Vh}), and the given function $\M$ is the corresponding deformed metric. 
On the other hand the equation $\jh = \nuh^{-1}[d\nuh]_w$ also implies that $\jh$ is flat. 
Since the Lax connection is flat iff the zweibein satisfies the field equation $\nu$ is a 
solution, which moreover satisfies the boundary condition $e(t) = \M(w=-t)$ on the 
symmetry axis worldline.

Let us carry out this argument in detail. Specifically, let us demonstrate the key 
claim that $\jh$ equals the Lax connection defined by $\nu$: 
$[d\M_C]_w(\cg) =  [d\nuh(\cg)]_w \nuh^t(\cg^{-1})+ \nuh(\cg)[d\nuh^t(\cg^{-1})]_w$, so
\begin{equation}\label{dM_w_expansion}
\fl 0 = \nuh^{-1}(\cg)[d\M_C]_w(\cg) \nuh^{-1\,t}(\cg^{-1})
 = \nuh^{-1}(\cg)[d\nuh(\cg)]_w + \left[\nuh^{-1}(\cg^{-1})[d(\nuh(\cg^{-1})]_w \right]^t.
\end{equation}
The differential of $\nuh$ at constant $w$ is 
\begin{equation}
 [d\nuh]_w = [d\nuh]_\cg + \frac{\di \nuh}{\di\cg}[d\cg]_w,
\end{equation}
where 
\begin{equation}\label{d_gamma_w}
 [d\cg]_w = -\frac{\cg}{2\rho}\left[ \frac{\cg - 1}{\cg + 1} d\rho^+ + \frac{\cg + 1}{\cg - 1} d\rho^-\right], 
\end{equation}
as follows from the differential of the expression (\ref{w_of_gamma}) for $w$ as a 
function of $\cg$, $\rho^+$, and $\rho^-$, 
\begin{equation}
 dw = \frac{1}{4\cg} [ (1 - \cg)^2 d\rho^+ + (1 + \cg)^2 d\rho^-] + \frac{\rho}{2}(1 - \frac{1}{\cg^2}) d\cg.
\end{equation}
(The null coordinates $\rho^+, \rho^-$ on the reduced spacetime have been used here only 
because they are slightly more convenient than the coordinates $t,\rho$.)

The variable spectral parameter $\cg$ is of course double valued as a function of $w$ 
(at a given spacetime position). This does not introduce any ambiguity in $[d\cg]_w$ 
expressed as a function $\cg$ since the value of $\cg$ identifies the branch. 
The fact that $\cg$ and $\cg^{-1}$ correspond to the same $w$ does lead to the identity
\begin{equation}
 [d\cg^{-1}]_w(g) = -\cg^{-2}[d\cg]_w(g) = [d\cg]_w(g^{-1}), 
\end{equation}
which may easily be verified directly from (\ref{d_gamma_w}). Here the argument of 
each expression is the value of $\cg$ at which it is evaluated. It follows that
\begin{eqnarray}
 [d\nuh(\cg^{-1})]_w(g) & = [d\nuh]_\cg(g^{-1}) + \frac{\di \nuh(\cg^{-1})}{\di\cg^{-1}}(g)[d\cg^{-1}]_w(g)\\
 & = [d\nuh]_\cg(g^{-1}) + \frac{\di \nuh(\cg)}{\di\cg}(g^{-1})[d\cg]_w(g^{-1}) 
 = [d\nuh]_w(g^{-1}) 
\end{eqnarray}
Defining
\begin{equation}\label{Khat_expanded}
 \jh \equiv \nuh^{-1}[d\nuh]_w = \nuh^{-1}[d\nuh]_\cg 
 - \nuh^{-1}\frac{\di \nuh}{\di\cg}
\frac{\cg}{2\rho}\left[\frac{\cg - 1}{\cg + 1} d\rho^+ + \frac{\cg + 1}{\cg - 1} d\rho^-\right], 
\end{equation}
the condition (\ref{dM_w_expansion}) may be expressed as
\begin{equation}\label{K_match}
 \jh(\cg) = -\jh^t(\cg^{-1})\ \ \ \mathrm{on}\ |\cg| = 1.
\end{equation}

The connection $\jh$ is holomorphic inside the $\cg$ unit disk. This follows from 
(\ref{Khat_expanded}), the definition of $\nuh$ and the fact that $[d\nuh]_\cg$ is 
holomorphic in this domain. The latter follows from Theorem \ref{smoothness}: The 
holomorphy of $\nuh$ inside the unit $\cg$ disk implies that it can expressed 
in terms of its boundary values on the unit circle via Cauchy's integral formula. 
The spacetime partial derivative at constant $\cg$, $\di_\mu$, may be taken inside 
this integral because by Theorem \ref{smoothness} $[d\nuh]_\cg$ is jointly 
continuous in spacetime position and $\cg$ on $|\cg| = 1$. Therefore
\begin{equation}
 \di_\mu \nuh(\cg) = \frac{1}{2\pi i} \int_{|\cg'| = 1} \frac{\di_\mu \nuh(\cg')}{\cg' - \cg} d\cg'
\end{equation}
at all $\cg$ inside the unit disk, which implies that $[d\nuh]_\cg$ is indeed holomorphic there.

Equation (\ref{K_match}) is thus yet another condition that requires a function holomorphic inside the unit circle and a function that is holomorphic outside this circle to match on the circle itself. But this time the limiting 
functions on the circle are not bounded. In fact
\begin{equation}\label{Khat_poles}
 \jh = q - \frac{\cg - 1}{\cg + 1} p_+ d\rho^+ - \frac{\cg + 1}{\cg - 1} p_- d\rho^-, 
\end{equation}
where $q$ is holomorphic inside the unit disk and continuous on its boundary, and 
$p_\pm = \mp \left[\nuh^{-1}\frac{\di \nuh}{\di\cg}\right]_{\cg = \mp 1}\frac{1}{2\rho}$.
This should match
\begin{equation}
 -\jh^t(\cg^{-1}) = -q^t(\cg^{-1}) - \frac{\cg - 1}{\cg + 1} p_+^t d\rho^+ - \frac{\cg + 1}{\cg - 1} p_-^t d\rho^-. 
\end{equation}
(Note that $\frac{\cg^{-1} - 1}{\cg^{-1} + 1} = - \frac{\cg - 1}{\cg + 1}$.) Equality requires
firstly that the coefficients of the singular terms match:
\begin{equation}
 p_\pm = p_\pm^t,
\end{equation}
and, secondly, that the regular remainder also matches, which requires that
\begin{equation}
 q(x;\cg) = -q^t(x;\cg^{-1})
\end{equation}
on $|\cg| = 1$. Because $q$ is continuous on $|\cg| = 1$ Lemma \ref{analyticity} applies, so $q$ is in fact independent of $\cg$, 
though it can still depend on $x$, and it is antisymmetric: $q(x) = -q^t(x)$.

Now evaluate $\jh(x;\cg)$ at $\cg = 0$. From (\ref{Khat_poles}) it follows that
\begin{equation}
 \jh(x;0) = q + p_+ d\rho^+ + p_- d\rho^-,
\end{equation}
while from (\ref{Khat_expanded}) it follows that $\jh(x;0) = j(x) \equiv \nu^{-1}d\nu (x)$, because the second term in this expression, proportional to $\cg$, vanishes at $\cg = 0$.
$p = p_+ d\rho^+ + p_- d\rho^-$ and $q$ are thus the symmetric and antisymmetric components, respectively, of $j$, and $\jh$ is the Lax connection (\ref{null_Lax_connection}) calculated from the zweibein field $\nu$. 

\subsection{Asymptotically flat solutions}\label{asymptotic_flatness}

Let us consider the Kramer-Neugebauer duals of infinite asymptotically flat solutions
treated in Korotkin and Samtleben's theory of cylindrically symmetric gravitational waves.
The solution spacetimes of Korotkin and Samtleben's theory are both spatially and temporally infinite
and are asymptotically flat at spatial infinity, although they do not 
specify precise asymptotic falloff conditions. Furthermore, instead of treating these solutions directly 
the theory treats their Kramer-Neugebauer duals.
The space of solutions treated thus corresponds to smooth conformal metrics $e_{cd}$ on the
whole real $t$ axis, or equivalently smooth $\M$ on the whole real $w$ axis. Asymptotic
flatness of the Kramer-Neugebauer dual requires that $e_{cd}(x)$ tends to an asymptotic
value $e_{\infty\,cd}$ at spacelike infinity, that is, as $\rho$ tends to $\infty$ 
while $t$ is held constant. (Of course $e_{\infty\,cd} = \dg_{cd}$ in a suitable basis.) 
Thus at spacelike infinity the solution is the Kramer-Neugebauer dual of Minkowski space. 

The following proposition shows that this is ensured by requiring 
that $e$ tends to $e_\infty$ as $t \rightarrow \pm \infty$ along the symmetry axis 
worldline. Equivalently, one requires that $\M(w)$ tends to $e_\infty$ as $w \rightarrow \mp \infty$. 

\begin{prop}
If $\M(w) \rightarrow e_\infty$ as $|w| \rightarrow \infty$ then $e(x)\rightarrow e_\infty$ as 
$\rho \rightarrow \infty$ with $t$ held constant. 
\end{prop}

\noindent {\em Proof}: 
Let us consider the limit $\rho \rightarrow \infty$ with $t$ held constant. In this limit 
$\M_C(x; e^{i\theta}) = \M(\rho \cos \theta - t)$ tends to $e_\infty$ for all $\theta$ except $\pm \pi/2$, 
while it also remains bounded since $\M$ is bounded: $||\M_C(x,e^{i\theta})|| \equiv \max_{ab}|\M_C(x,e^{i\theta})_{ab}| \leq  
\sup_w ||\M(w)|| < \infty$.  
It follows that 
\begin{equation}\label{M_avg_lim}
 \langle \M_C\rangle \rightarrow e_\infty
\end{equation}
where
\begin{equation}
 \langle \M_C\rangle(x) = \frac{1}{2\pi} \int_{-\pi}^\pi \M_C(x; e^{i\theta}) d\theta.
\end{equation}
Now recall that $\M_C(e^{i\theta}) = \nuh(e^{i\theta})\nuh^t(e^{-i\theta}) = \nuh(e^{i\theta})\nuh^\dagger(e^{i\theta})$ with
\begin{equation}
 \nuh(e^{i\theta}) = \nu + \sum_{n = 1}^\infty \alpha_n e^{in\theta},
\end{equation}
since $\nuh(\cg)$ is holomorphic in $|\cg|<1$ and $\nuh(0) = \nu$. Thus 
\begin{equation}\label{M_avg_Fourier_sum}
 \langle \M_C\rangle = e + \sum_{n = 1}^\infty \alpha_n \alpha^\dagger_n.
\end{equation}
It turns out that this and (\ref{M_avg_lim}) implies that $e \rightarrow e_\infty$ as $\rho \rightarrow \infty$.
Each of the matrices $e$ and $\alpha_n \alpha^\dagger_n$ are Hermitian and positive semi-definite. 
($e$ of course also has unit determinant and is real by Theorem \ref{Mvv}.)
Hermitian $2\times 2$ matrices are the real span of the unit matrix $\One$ and the Pauli matrices. 
A Hermitian matrix $A$ may therefore be represented by a real 4-vector $a$ defined by
\begin{equation}
 A = \left[\begin{array}{cc} a^0 + a^3 & a^1 - i a^2 \\ a^1 + i a^2 & a^0 - a^3 \end{array}\right] 
= a^0 \One + a^{\rm i} \sigma_{\rm i} 
\end{equation}
The determinant of $A$ is the Lorentzian norm squared $[a^0]^2 - [a^1]^2 - [a^2]^2 - [a^3]^2$ of $a$.
The unit determinant Hermitian matrices correspond to the unit norm shell, and the positive semi-definite ones to the closed future light cone. It follows that $e(x)$ lies on the future unit norm shell, and, by (\ref{M_avg_Fourier_sum}), in the closed past lightcone of $\langle \M_C\rangle(x)$. 
But $e_\infty$, being the limit of $\M$, must be real, symmetric, positive definite and have unit determinant, just like $e(x)$, so it too lies on the future unit norm shell. It follows that if $\langle \M_C\rangle \rightarrow e_\infty$ then 
$e \rightarrow e_\infty$ as well. \hfill\QED

How fast should $\M$ approach $e_\infty$? This is partly a question of convenience. One does not
need not worry that overly strong falloff conditions limit the local dynamics, because the asymptotic 
behaviour of $\M$ has no effect at all on the solutions within a given finite causal diamond $\Delta(a,b)$, 
which depends only on $\M(w)$ in the interval $a \leq -w \leq b$. It is, however, strongly limited by the
Poisson bracket on $\M$, which will be presented in the next section. We want the submanifold $\Gamma_\infty$ of 
the phase space $\Gamma$, consisting of $\M$s corresponding to solutions with asymptotically flat Kramer-Neugebauer 
duals and with a common asymptotic value $e_\infty$, to form a phase space in its own right.\footnote{
Note that in the following, when confusion is unlikely, we will often call deformed metrics $\M$ which approach 
an asymptotic value $e_\infty$ as ``asymptotically flat'' even though it is actually the Kramer-Neugebauer dual 
of the corresponding solution that is asymptotically flat. In the same spirit we will call $\Gamma_\infty$ the 
asymptotically flat phase space.}
In order that the Poisson bracket on the full phase space $\Gamma$ restricts to a Poisson bracket 
on $\Gamma_\infty$ it is necessary that the Hamiltonian flow generated by observables defined on 
$\Gamma_\infty$ is everywhere tangent to $\Gamma_\infty$, i.e. does not flow off $\Gamma_\infty$.
We can therefore not require $\M-e_\infty$ to fall off more rapidly as $|w| \rightarrow \infty$ than the 
perturbation of $\M$ generated by an observable does. 

A simple prescription that will turn out to be consistent with this requirement is that $\M - e_\infty$ falls 
off as $1/w$ or faster. 
Another prescription for asymptotic flatness that is consistent with the Poisson bracket and is natural for our 
purposes requires $\M$ to be a Wiener function, that is, the Fourier transform of an $L^1[\R]$ (or ``absolutely 
integrable'') function plus a constant. (See \cite{Reiter_Stegeman}\cite{Faro}\cite{Gohberg_Krein}\cite{Wiener_review}.)
This space of functions contains all differentiable functions such that both the function minus the constant term, and 
its derivative, are square integrable \cite{Beurling}, so on any finite interval they are fairly unrestricted. On the other hand, the 
Riemann-Lebesgue lemma shows that the Fourier transforms of $L^1[\R]$ functions vanish at infinity, so a 
Wiener function reduces asymptotically to its constant term, which in the case of $M$ would be 
$e_\infty$. Furthermore, the condition that $\M$ be a Wiener function is sufficient to guarantee the existence 
and essential uniqueness of the factorization of $\M$ into positive frequency and negative frequency factors from which 
the quantization of Korotkin and Samtleben is constructed. (See subsection \ref{q_Geroch_as_symmetry}.)

\subsection{The classical Geroch group for cylindrically symmetric gravity with and without asymptotic flatness}\label{classical_Geroch}

The Geroch group is a group of maps of the solution space to itself. We saw in Subsection \ref{M_solution_space}
that the space of solutions on the causal diamond $\Delta(a,b) \subset {\cal S}$ 
of a segment $a \leq t \leq b$ of the axis worldline can be identified with the set 
of smooth conformal metrics $e_{cd}(t)$ on this segment. Equivalently, it consists of the 
smooth deformed metrics $\M(w)$ on the range  $a \leq -w \leq b$ of spectral parameters 
$w$. The Geroch group consists of the smooth functions 
$s: [-b, -a] \rightarrow SL(2,\R)$
of $w$, and acts via
\begin{equation}\label{Geroch_action}
 \M(w) \mapsto s(w) \M(w) s^t(w),
\end{equation}
or $\M_{cd} \mapsto s_c{}^{c'}s_d{}^{d'}\M_{c' d'}$ in terms of components \cite{Breit}\cite{KS}. 
This can be thought of as a $w$ dependent change of the basis in which the matrix elements are evaluated. 

The action is a symmetry in the sense that it maps solutions to solutions: The only 
conditions that $\M$ must satisfy is that it be smooth in $w$ and that $\M(w)$ be real,
symmetric positive definite, and of unit determinant. (\ref{Geroch_action}) clearly 
preserves all these conditions.

The action is also transitive. The properties of $\M$ guarantee that there exists a Cholesky
factorization $\M(w) = s(w)s^t(w)$ with $s$ smooth in $w$ and $s(w) \in SL(2, \R)$ (and lower
triangular with positive diagonal elements). Any solution can therefore be reached via a 
Geroch group action from the solution corresponding to $\M = {\bf 1}$ (the Kramer-Neugebauer
dual of flat spacetime), and of course by applying the inverse element $s^{-1}$ of the Geroch
group one returns from the given solution to $\M = {\bf 1}$. It follows that any solution 
$\M'$ can be reached via a Geroch group action from any other $\M$ by mapping
$\M \mapsto {\bf 1} \mapsto \M'$, so the Geroch group acts transitively on the space of
solutions.

Now let us consider the Geroch group for asymptotically flat spacetimes. One definition of asymptotic flatness
that we proposed in the last subsection had $\M - e_\infty$ falling 
off as $1/w$ or faster. Then the corresponding asymptotic flatness preserving Geroch group is naturally defined 
to consist of maps $s:\R \rightarrow SL(2,\R)$ such that $s -\One$ also falls off as $1/w$ or faster. 
This group, acts transitively on the asymptotically flat phase space $\Gamma_\infty$ defined by the $1/w$ falloff.
(Note that this is not the full subgroup of the Geroch group of $\Gamma$ which preserves the asymptotic 
boundary conditions. The group of $w$ independent transformations that are isometries of $e_\infty$ has been 
divided out so that all $s$ tend to the single value $\One$ at infinity.)

In the other definition of asymptotic flatness we proposed in Subsection \ref{asymptotic_flatness} $\M$ is required 
to be a Wiener function with constant term $e_\infty$. There exists a transitive asymptotic flatness preserving 
Geroch group also in this case.  
Products, sums, and multiples of Wiener functions are easily seen to be Wiener functions. Thus imposing the requirement
that $s$ be a Wiener function with constant term $\One$ ensures that the Geroch group maps the asymptotically 
flat phase space, in the Wiener sense, to itself. Moreover the action of these transformations is transitive. 
This can be proved in much the same way that transitivity was demonstrated in the smooth case without boundary 
conditions. In outline: since $\M$ is real, symmetric, positive definite and of unit determinant, and a Wiener 
function of $w$, with constant term $e_\infty$, it admits a (unique) Cholesky factorization $\M = L L^t$ with 
$L$ a real, unit determinant, lower triangular matrix valued Wiener function with positive definite diagonal 
elements. The constant term admits an analogous Cholesky factorization, $e_\infty = L_\infty L^t_\infty$, so
\begin{equation}\label{AF_Cholesky}
 \M(w) = s(w) e_\infty s^t(w),
\end{equation}
with $s = L L^{-1}_\infty$. $s$ is a Wiener function, and because the matrix elements of $L$ are continuous 
functions of those of $\M$, its constant term, equal to its asymptotic value, is $\One$. From 
(\ref{AF_Cholesky}) it follows that any Wiener function deformed metric $\M$ can be reached from any other 
sharing the same asymptotic value $e_\infty$ via the action (\ref{Geroch_action}) of the Geroch group, with 
$s$ an $SL(2,\R)$ valued Wiener function with constant term $\One$.

The non-trivial claim made in this argument is that the Cholesky factor $L$ of $\M$ is a Wiener function. By 
(\ref{explicit_Cholesky}) the matrix elements of $L$ are polynomials in the matrix elements of $\M$ 
multiplied by $(\M_{11})^{-1/2}$. Because $\M$ is real and positive definite, and has a positive definite 
asymptotic value $e_\infty$, $\M_{11}$ is real, positive and bounded away from zero. It then follows from 
the Wiener-Lev\'y theorem (theorem 1.3.4 \cite{Reiter_Stegeman}), and the fact that $\M_{11}$ is a Wiener 
function, that $(\M_{11})^{-1/2}$ is a Wiener function. This then implies that $L$ is also a Wiener function. 

Note that in the discussion of the Wiener function approach we have not required smoothness of the functions 
involved. Indeed, it may be that the whole theory, also of the phase space of asymptotically flat solutions 
of cylindrically symmetric vacuum GR, could be developed in terms of Wiener functions, dispensing with additional 
smoothness assumptions, but this will not be worked out here. It is important to notice, however, that the 
asymptotic flatness preserving Geroch group remains a transitive symmetry if one requires both $\M$ and $s$ 
to be $C^\infty$ in addition to being Wiener functions in $w$, or if one imposes the additional requierment 
of real analyticity.

The Geroch group as defined here is not a loop group because it maps the real line, instead of a
circle, into the group $SL(2,\R)$. It is tempting to turn the real line into a unit circle via the
Moebius transformation $w \mapsto z = \frac{i + w}{i - w}$. Smooth functions on the unit
circle $|z| = 1$ then correspond to smooth functions of $w$ which are also smooth functions 
of $1/w$ on $w \neq 0$. These necessarily fall off to a constant limiting value as 
$|w| \rightarrow \infty$. The Geroch group for 
asymptotically flat spacetimes with a suitable falloff condition on $\M$ is therefore equivalent to 
a loop group. However, for our purposes there is a serious problem with this way of presenting 
the group: The Poisson bracket $\{s(w_1),s(w_2)\}$ between group elements, given 
in (\ref{gerochbracket}), depends only on the difference $w_1 - w_2$ but is not similarly translation 
invariant when expressed in terms of the Moebius transformed spectral parameter $z$. The lack 
of translation invariance complicates the quantization because the close connection to the 
$\mathfrak{sl}_2$ Yangian is lost. We therefore choose to continue with the Geroch group in 
it's ``almost loop group form'' in terms of functions on the real $w$ line. It might aptly be called 
a ``line group''.

\section{Poisson brackets on the space of solutions and on the Geroch group}\label{Poisson}                   

The deformed metric $\M$ given as a function of $w$ characterizes solutions completely, and any smooth, real,
symmetric, unit determinant function $\M$ defines a solution. The function $\M$ can thus be taken as a 
coordinate system on the space of solutions, that is, the phase space $\Gamma$ of (not necessarily asymptotically flat)
cylindrically symmetric gravitational waves. Any gravitational observable is therefore a functional of $\M$. To specify the 
Poisson bracket on $\Gamma$ space it is thus sufficient to give the Poisson brackets between the matrix elements 
$\M(v)_{ab}$ and $\M(w)_{cd}$.

These Poisson brackets have been evaluated from the action (\ref{action}) by Korotkin and Samtleben \cite{KS}, and 
in a different way in \cite{ReiAnd}. The result is
\begin{equation}\label{Monbracket}
\begin{array}{r}
\{ \overset{1}{\M}(v) , \overset{2}{\M}(w) \} = {{{}}} {\rm p.v.} \left(\frac{1}{v-w}\right) \left[ \Omega \overset{1}{\M}(v) \overset{2}{\M}(w) +  \overset{1}{\M}(v)\, {}^t\Omega \overset{2}{\M}(w) + \right. \\
\left. +  \overset{2}{\M}(w) \Omega^t \overset{1}{\M}(v) + \overset{1}{\M}(v) \overset{2}{\M}(w) \,{}^t\Omega^t  \right].
\end{array} 
\end{equation}
Here ``tensor notation'' has been used, in which a tensor product $A \otimes B$ of a tensor $A$ acting on space $1$
and a tensor $B$ acting on space $2$ is denoted $\overset{1}{A} \overset{2}{B}$. $\Omega$ denotes the Casimir 
element of $\mathfrak{sl}_2$: In terms of Pauli's sigma matrices $\Omega = \frac{1}{2} \left( \overset{1}{\sigma_x} \overset{2}{\sigma_x} + \overset{1}{\sigma_y} \overset{2}{\sigma_y} + \overset{1}{\sigma_z} \overset{2}{\sigma_z} \right)$. In terms of Kronecker deltas 
\begin{equation}\label{Omega_components}
 \Omega_a{}^b{}_c{}^d = \dg_a^d \dg^b_c - \frac{1}{2}\dg_a^b \dg_c^d,
\end{equation}
with indices $a$, $b$ corresponding to space $1$ and indices $c$, $d$ corresponding to space $2$. 
$\Omega^t$ denotes $\Omega$ transposed in the indices corresponding to space $2$, that is,
$\Omega^t_a{}^b{}^c{}_d = \Omega_a{}^b{}_d{}^c$, ${}^t\Omega$ is $\Omega$ similarly transposed in the indices corresponding to space $1$, and ${}^t\Omega^t$ is obtained by transposing in both spaces. 
Finally, ${\rm p.v.} \left(\frac{1}{u}\right)$ denotes the {\em Cauchy principal value} of $\frac{1}{u}$, a distribution 
defined by 
\begin{equation}
{\rm p.v.} \left(\frac{1}{u}\right) = \lim_{\epsilon \rightarrow 0^+}\frac{1}{2}\left(\frac{1}{u + i\epsilon} + \frac{1}{u - i\epsilon}\right), 
\end{equation}
where the limit is taken after integration against the test function.

The Poisson bracket (\ref{Monbracket}) is really just a symmetrization, in both the indices associated with space
$1$ and in those associated with space $2$, of a simpler expression:
\begin{equation}
\{ \overset{1}{\M}(v) , \overset{2}{\M}(w) \} = 4\, {\rm p.v.} \left(\frac{1}{v - w}\right){\rm Sym}_1{\rm Sym}_2 
\left[ \Omega \overset{1}{\M}(v) \overset{2}{\M}(w) \right], 
\end{equation}
where ${\rm Sym}X = \frac{1}{2}(X + X^t)$ for any matrix $X$. In index notation 
\begin{equation}\label{index_M_bracket}
\fl \{ \M(v)_{ab} , \M(w)_{cd} \} = 4\, {\rm p.v.} \left(\frac{1}{v - w}\right) {\rm Sym}_{(ab)}{\rm Sym}_{(cd)} \left[\M(v)_{cb}\M(w)_{ad} - \frac{1}{2}
 \M(v)_{ab}\M(w)_{cd}\right],
\end{equation}
where ${\rm Sym}_{(ab)}$ denotes symmetrization with respect to the pair of indices $a,b$.

From (\ref{index_M_bracket}) it follows immediately that
\begin{equation}
 \{ \det\M(v) , \M(w)_{cd} \} = 0,
\end{equation}
and thus that $\{ \det\M(v) , F \} = 0$ for any functionally differentiable functional $F$ of $\M$.
This is important because, as already pointed out, the Hamiltonian flow generated via the Poisson bracket by observables must not flow off the phase space $\Gamma$, and $\det\M(v) = 1$ throughout $\Gamma$. If it did flow off $\Gamma$ then the bracket between observables defined only 
on $\Gamma$ would not be defined. If the definitions of the observables were extended off $\Gamma$ to the space of all smooth matrix valued functions of $w$ then their brackets would be defined but would depend on the extensions.

An analogous issue arises for the smaller asymptotically flat phase space $\Gamma_\infty$ consisting
of deformed metrics $\M$ approaching a common asymptotic value $e_{\infty}$ as $w \rightarrow \pm\infty$
at a rate determined by some given falloff condition. The issue is whether the Hamiltonian flows generated 
by observables preserve asymptotic flatness.
Does the flow generated by $\M(w)$ preserve asymptotic flatness? Although $\M(v)$ approaches
$e_{\infty}$ as $v \rightarrow \pm \infty$, the numerator in (\ref{index_M_bracket}) does not in general
approach zero. The perturbation of $\M(v)$ generated by $\M(w)$ thus falls off as $1/v$ as 
$v \rightarrow \pm \infty$. The same is true for the Hamiltonian flow generated by a functional $F$ of
$M$ with smooth, compactly supported functional derivative $\dg F/\dg\M(w)_{ab}$, and in this case the 
perturbation generated by $F$ is also smooth.
Thus, if the falloff condition that defines $\Gamma_\infty$ is that 
$\M(v) - \e_\infty$ falls off as $1/v$ then the flow generated by localized observables preserves $\Gamma_\infty$. 

This is also the case if the falloff condition is that $\M$ be a Wiener function with constant term 
$\e_\infty$: The right side of (\ref{index_M_bracket}) is linear in $\M(v)$ and can therefore be separated into two terms, one obtained by replacing $\M(v)$ by its asymptotic value $e_\infty$, and the other by replacing $\M(v)$ by $\M(v) - \e_\infty$, a Wiener function with vanishing constant term.
Integrating both terms against $\dg F/\dg\M(w)_{ab}$ to obtain the perturbation generated by 
$F$ one finds that both terms are Wiener functions with vanishing constant term. The first term is because it as well as its derivative is square integrable, the second term because it is a sum of products of Wiener functions with vanishing constant term, and such Wiener functions form an algebra.
With either falloff condition the Poisson bracket is well defined on $\Gamma_\infty$ for a large class of observables which is complete in the sense that it separates points on $\Gamma_\infty$.

Let us turn now to the Geroch group. Recall that the action of a Geroch group element $g$ on the deformed metric $\M$ maps $\M$ to
\begin{equation}\label{Geroch_action2}
\M^G (w) (g, \xi) \equiv \M(w)(g \lact \xi) = s(w)(g) \M(w)(\xi) s^t(w)(g)
\end{equation}
at all points $\xi$ in the phase space $\Gamma$, which defines the action on any gravitational observable 
since these are functionals of $\M$. (For clarity in equation (\ref{Geroch_action2}) we have 
distinguished between the Geroch group element $g$ and its faithful representation $s(g)$, and 
between the point $\xi$ in the phase space $\Gamma$ and $\M(\xi)$. This will usually not be done.)

We will show that there exists a unique Poisson bracket on the Geroch group such that this action 
is a Poisson map $G\times \Gamma \rightarrow \Gamma$. That is, that it preserves the Poisson algebra of 
gravitational observables: For any pair, $A$ and $B$ of such observables $\{A,B\}^G_\Gamma(g,\xi) 
\equiv \{A,B\}_\Gamma (g \lact \xi)$ equals $\{A^G, B^G\}_{G \times \Gamma} (g,\xi) \equiv
\{A^G(g,\cdot), B^G(g,\cdot\}_\Gamma (\xi) + \{A^G(\cdot, \xi), B^G(\cdot, \xi\}_G (g)$.

\begin{thm}\label{propgerochS}
The action of the Geroch group $G$ on the space of solutions $\Gamma$ is a Poisson map $G\times \Gamma \rightarrow \Gamma$ if and only if the Poisson bracket on $G$ is
\begin{equation}\label{gerochbracket}
\{ \overset{1}{s}(v), \overset{2}{s}(w) \}_G = {{{}}} {\rm p.v.} \left(\frac{1}{v-w}\right) \left[ \Omega , \overset{1}{s}(v) \overset{2}{s}(w) \right]
\end{equation}
\end{thm}

\noindent {\em Proof}:
Since all gravitational observables are functionals of $\M$ it is necessary and sufficient to verify that the action (\ref{Geroch_action2}) preserves 
equation (\ref{Monbracket}), that is, that
\begin{equation}\label{eq1}
\{ \overset{1}{s} \overset{1}{\M} \overset{1}{s}{}^t , \overset{2}{s} \overset{2}{\M} \overset{2}{s}{}^t \}_{G\times \Gamma} (g , \xi) 
= \{ \overset{1}{\M} , \overset{2}{\M} \}_{\Gamma} (g \lact \xi).
\end{equation}
When the argument is not given, tensors in space $1$ are evaluated at spectral parameter $v$, and those in space $2$ at $w$.

The left side of equation (\ref{eq1}) is equal to
\beq
\{ \overset{1}{s} \overset{1}{\M} \overset{1}{s}{}^t , \overset{2}{s} \overset{2}{\M} \overset{2}{s}{}^t \}_{G\times \Gamma} &=& \overset{1}{s} \overset{2}{s} \{  \overset{1}{\M} , \overset{2}{\M}\}_{\Gamma} \overset{1}{s}{}^t \overset{2}{s}{}^t  + \{ \overset{1}{s} , \overset{2}{s} \}_{G} \overset{1}{\M} \overset{1}{s}{}^t \overset{2}{\M} \overset{2}{s}{}^t \nonumber \\
&& \overset{2}{s} \overset{2}{\M} \{ \overset{1}{s} , \overset{2}{s}{}^t \}_{G} \overset{1}{\M} \overset{1}{s}{}^t + \overset{1}{s} \overset{1}{\M} \{  \overset{1}{s}{}^t , \overset{2}{s} \}_{G} \overset{2}{\M} \overset{2}{s}{}^t \nonumber \\
&& \overset{1}{s} \overset{1}{\M} \overset{2}{s} \overset{2}{\M} \{\overset{1}{s}{}^t ,  \overset{2}{s}{}^t \}_{G}
\eeq
Substituting the Poisson bracket (\ref{Monbracket}) between deformed metrics into this equation, and setting 
$\mathcal{Y} \equiv \{ \overset{1}{s} , \overset{2}{s} \}_{G} \overset{1}{s}{}^{-1} \overset{2}{s}{}^{-1}$, one obtains
\beq
\fl\{ \overset{1}{s} \overset{1}{\M} \overset{1}{s}{}^t , \overset{2}{s} \overset{2}{\M} \overset{2}{s}{}^t \}_{G\times \Gamma} &=& \overset{1}{s} \overset{2}{s}\: {{{}}} {\rm p.v.} \left(\frac{1}{v-w}\right) \left[ \Omega \overset{1}{\M} \overset{2}{\M} +  \overset{1}{\M} \phantom{}^t {\Omega}{} \overset{2}{\M} + \overset{2}{\M} {\Omega^t} \overset{1}{\M} + \overset{1}{\M} \overset{2}{\M} \phantom{}^t {\Omega}{^t}  \right] \overset{1}{s}{}^t \overset{2}{s}{}^t  \nonumber \\
&& + \mathcal{Y}  \overset{1}{\M}{}^G \overset{2}{\M}{}^G + \overset{2}{\M}{}^G {\mathcal{Y}^t} \overset{1}{\M}{}^G \nonumber \\
&&  + \overset{1}{\M}{}^G \phantom{}^t \mathcal{Y}  \overset{2}{\M}{}^G + \overset{1}{\M}{}^G \overset{2}{\M}{}^G \phantom{}^t\mathcal{Y}^t\\
&=& 4\, {\rm Sym}_1{\rm Sym}_2\left[\left( {{{}}} {\rm p.v.} \left(\frac{1}{v-w}\right) \overset{1}{s} \overset{2}{s} \Omega \overset{1}{s}{}^{-1} \overset{2}{s}{}^{-1} + \mathcal{Y} \right)  \overset{1}{\M}{}^G \overset{2}{\M}{}^G\right] 
\eeq

The right side of (\ref{eq1}) is just the Poisson bracket (\ref{Monbracket}) evaluated with $\M^G$ in place of $\M$:
\begin{equation}
4\, {\rm Sym}_1{\rm Sym}_2\left[ {{{}}} {\rm p.v.} \left(\frac{1}{v-w}\right) \Omega   \overset{1}{\M}{}^G \overset{2}{\M}{}^G\right].
\end{equation}
The action of the Geroch group maps $\Gamma$ to itself, so in order to satisfy (\ref{eq1}) on all of $G \times \Gamma$ it is necessary and sufficient that
\begin{equation} 
\fl{\rm Sym}_1{\rm Sym}_2 \left[ \left( {{{}}} {\rm p.v.} \left(\frac{1}{v-w}\right) \overset{1}{s} \overset{2}{s} \Omega \overset{1}{s}{}^{-1} \overset{2}{s}{}^{-1} + \mathcal{Y} \right) \overset{1}{\M} \overset{2}{\M} \right] 
= {\rm Sym}_1{\rm Sym}_2 \left[ {{{}}} {\rm p.v.} \left(\frac{1}{v-w}\right) \Omega  \overset{1}{\M} \overset{2}{\M} \right] 
\end{equation}
for all $\M$ corresponding to solutions. This equation may be written as
\begin{equation}\label{symops}
{\rm Sym}_1{\rm Sym}_2[\overset{12}{Z} \overset{1}{\M} \overset{2}{\M} ] = 0  
\end{equation}
with
\begin{equation}
\overset{12}{Z} = {{{}}} {\rm p.v.} \left(\frac{1}{v-w}\right) \left[\overset{1}{s} \overset{2}{s} \Omega 
\overset{1}{s}{}^{-1} \overset{2}{s}{}^{-1} - \Omega\right] + \mathcal{Y}. 
\end{equation}

Clearly, setting $Z = 0$ is sufficient to ensure that (\ref{symops}) is satisfied for all $\M$, and thus that the action (\ref{Geroch_action2}) preserves the Poisson bracket (\ref{Monbracket}) on the solution space. 
$Z = 0$ implies that
\begin{equation}
\{ \overset{1}{s} , \overset{2}{s} \}_{G} = \mathcal{Y}\overset{1}{s}\overset{2}{s} 
= {{{}}} {\rm p.v.} \left(\frac{1}{v-w}\right) \left[ \Omega  \overset{1}{s} \overset{2}{s} - \overset{1}{s} \overset{2}{s} \Omega \right],
\end{equation}
which is precisely the claimed result (\ref{gerochbracket}).

In fact $Z = 0$ is also necessary. That is, the validity of (\ref{symops}) for all $\M$ corresponding to solutions 
implies $Z = 0$ (and thus of course (\ref{gerochbracket})).
To show this we begin by simplifying (\ref{symops}). The space of solutions corresponds to smooth functions $\M(w)$ that 
are real, symmetric, positive definite, and of unit determinant. But if (\ref{symops}) holds for $\M$ it also
holds for a multiple of $\M$ by a scalar function of $w$, so if (\ref{symops}) holds for all $\M$ corresponding to
solutions it must still hold if the unit determinant requirement is dropped. In other words, it must hold 
for all smooth matrix valued functions which are real, symmetric, and positive definite. In particular, it must hold for $\M_{ab} = n_a n_b$ where $n$ is a smooth co-vector valued function of the spectral parameter having compact support in $\R$. Thus
\begin{equation}\label{Znnnn}
 Z_{(a}{}^b {}_{(\!(c}{}^d(v,w) n_{a')}(v) n_b(v) n_{c')\!)}(w) n_d(w) = 0,  
\end{equation}
where symmetrization with respect to interchange of the indices $c$ and $c'$ has been indicated by a double parenthesis, $(\!(\:)\!)$ to distinguish it from the symmetrization with respect to $a, a'$, indicated by a single
parenthesis.

The following lemma is useful for extracting the consequences of this condition
\begin{lemma}\label{XYlemma}
Let $X$ and $Y$ be vectors (one dimensional matrices) of commuting components, then 
$X_{(a} Y_{b)} = 0$ implies that either $X = 0$ or $Y = 0$.
\end{lemma}

\noindent{\em Proof}: Suppose $X \neq 0$ then there exists $\psi^a$ so that the contraction $\psi^a X_a \neq 0$.
Thus $0 = \psi^a X_a Y_b + \psi^a Y_a X_b$ implies that $X$ and $Y$ are linearly dependent, so that 
$X_{[a} Y_{b]} = 0$. It follows that $X_{a} Y_{b} = 0$, and from this that $Y = 0$. \hfill\QED

Even more useful will be the following corollary. 
\begin{clly}\label{XYcorollary}
Suppose $X_{C\,a}$ is a multidimensional matrix of commuting components depending on the index $a$ and on possible additional indices subsumed in the multiindex $C$, and $Y_{b\,D}$ is defined analogously, then 
$X_{C\,(a} Y_{b)\,D} = 0$ implies that either $X = 0$ or $Y = 0$.
\end{clly}

This is an immediate consequence of the proof of the lemma, which is unaffected by the presence of multiindices. 

Applying this corollary to (\ref{Znnnn}) with $X_{a'} = n_{a'}(v)$ and $Y_{a c c'}(v,w) = Z_{a}{}^b {}_{(c}{}^d(v,w) n_b(v) n_{c')}(w) n_d(w)$ one obtains
\begin{equation}\label{Znnn}
 Z_{a}{}^b {}_{(c}{}^d(v,w) n_b(v) n_{c')}(w) n_d(w) = 0,  
\end{equation}
and applying it again, one obtains
\begin{equation}\label{Znn}
 Z_{a}{}^b {}_{c}{}^d(v,w) n_b(v) n_d(w) = 0.  
\end{equation}

To solve this equation for $Z$ we use another lemma: 
\begin{lemma}
If $A^{ab\;C}(v,w)$ is a distribution on $\R^2$ with indices $a$, $b$, and possibly others subsumed in the multiindex $C$, then
\begin{equation}
A^{ab\,C}(v,w)n_a(v)n_b(w) = 0 
\end{equation}
(as a distribution) for any $n_a$ a smooth vector valued function of compact support on $\R$ implies that 
$A^{ab\,C}(v,w) = \alpha^{ab\,C}(v)\dg(v - w)$, with $\alpha^{ab\,C}(v)$ a distribution on $\R$ which is antisymmetric under interchange of the indices $a$ and $b$.
\end{lemma}

\noindent{\em Proof}: Let $\varphi$ be a smooth function on $\R^2$, then the hypothesies of the lemma imply that
\begin{equation}
 \int \varphi(v,w) A^{ab\,C}(v,w) n_a(v) n_b(w) dv\, dw = 0.
\end{equation}
The functional derivative of this equation by $n^c(u)$ and $n^d(z)$ yields
\begin{equation}
 \varphi(u,z) A^{cd\,C}(u,z) + \varphi(z,u) A^{dc\,C}(z,u) = 0.
\end{equation}
Putting $\varphi = 1$ this reduces to
\begin{equation}\label{varphi_1_eqn}
 \varphi(u,z) A^{cd\,C}(u,z) = - \varphi(z,u) A^{dc\,C}(z,u),
\end{equation}
while $\varphi(u,z) = u - z$ gives
\begin{equation}
 (u - z)[ A^{cd\,C}(u,z) - \varphi(z,u) A^{dc\,C}(z,u)] = 0.
\end{equation}
Combining these two results we obtain 
\begin{equation}
 (u - z) A^{cd\,C}(u,z) = 0.
\end{equation}
The general solution to this last equation is 
\begin{equation}
 A^{cd\,C}(u,z) = \alpha^{cd\,C}(u)\dg(u - z),
\end{equation}
with $\alpha^{cd\,C}(u)$ a distribution on $\R$. By (\ref{varphi_1_eqn}) $\alpha^{cd\,C}(u)$ must be  antisymmetric under interchange of the indices $c$ and $d$. \hfill\QED

This lemma and (\ref{Znn}) imply that
\begin{equation}
 Z_{a}{}^b {}_{c}{}^d(v,w) = k_{ac}(v) \varepsilon^{bd}\dg(v - w),  
\end{equation}
with $k_{ac}(v)$ a distribution on $\R$ and $\varepsilon^{bd}$ the two dimensional antisymmetric symbol.

Substituting this form of $Z$ into the original condition (\ref{symops}) yields
\begin{equation}
 k_{(a(\!(c}(v) \varepsilon_{a')c')\!)}{\rm det}\M(v)\dg(v - w) = 0.  
\end{equation}
If this is to hold for all real, symmetric, positive definite, and smooth $\M(v)$ then $k_{(a(\!(c}(v) \varepsilon_{a')c')\!)} = 0$. Equivalently
\begin{equation}
 k_{(a |\!c|}(v)r^c \varepsilon_{a')c'} r^{c'}= 0
\end{equation}
for all 2-vectors $r$. Lemma \ref{XYlemma} then implies that $k_{ac}(v) r^c = 0$ for all $r$, so we may conclude that $k = 0$, and thus that $Z = 0$. This completes the proof of the theorem. \hfill\QED

Note that the proof does not depend on boundary conditions at $|w| \rightarrow \infty$, so the theorem applies also
to the asymptotically flat phase spaces $\Gamma_\infty$ we have defined and their corresponding asymptotic flatness
preserving Geroch groups.

The bracket (\ref{gerochbracket}) is the well known Sklyanin bracket on group manifolds 
\cite{Sklyanin}\cite{Babelon_Bernard} for the $SL(2,\R)$ line group - the Geroch group. 

Note that (\ref{gerochbracket}) is actually not singular at $v = w$, because the numerator vanishes there. $[\Omega , \overset{1}{s}(v) \overset{2}{s}(v)] = 0$ 
since $\Omega$ is invariant under simultaneous, equal $SL(2,\C)$ transformations in both space $1$ and space $2$. More explicitly, by (\ref{Omega_components}),
\begin{equation}
 \Omega = \Pi - \frac{1}{2}I,
\end{equation}
where $I = \overset{1}{\One}\overset{2}{\One}$ is the product of the identity matrices in spaces 1 and 2, and $\Pi_a{}^b{}_c{}^d = \dg_a^d \dg^b_c$ exchanges spaces 1 and 2: $\Pi \overset{1}{x} \overset{2}{y} = \overset{2}{x} \overset{1}{y}$ on any pair of 2-vectors
$x$ and $y$. Thus
\begin{eqnarray}
 [\Omega , \overset{1}{s}(v) \overset{2}{s}(w)] 
&= [\Pi , \overset{1}{s}(v) \overset{2}{s}(w)] 
 = [\Pi \overset{1}{s}(v) \overset{2}{s}(w)\Pi, \Pi]
 = - [\Pi , \overset{2}{s}(v) \overset{1}{s}(w)]\nonumber\\
&= - [\Pi , \overset{1}{s}(w) \overset{2}{s}(v)]
 = - [\Omega, \overset{1}{s}(w) \overset{2}{s}(v)],
\end{eqnarray}
which shows that $[\Omega , \overset{1}{s}(v) \overset{2}{s}(w)]$ vanishes when $v = w$.

The Sklyanin bracket (\ref{gerochbracket}) may therefore be written in two alternative forms
\begin{equation}\label{gerochbracket2}
\{ \overset{1}{s}(v), \overset{2}{s}(w) \}_G =  \lim_{\epsilon\rightarrow 0^+}\frac{1}{v-w \pm i\epsilon} 
[ \Omega , \overset{1}{s}(v) \overset{2}{s}(w) ]. 
\end{equation}
The Jacobi relation for the Sklyanin bracket then follows from the classical Yang-Baxter equation
\begin{equation}
 [\overset{12}{r}_\pm, \overset{13}{r}_\pm] + [\overset{12}{r}_\pm, \overset{23}{r}_\pm] + [\overset{13}{r}_\pm, 
 \overset{23}{r}_\pm] = 0,
\end{equation}
which holds for either of the two alternative classical $r$ matrices $\overset{ij}{r}_\pm \equiv \lim_{\epsilon \rightarrow 0^+} 
\left(\frac{1}{w_i-w_j\pm i\epsilon}\right) \overset{ij}{\Omega}$ appearing in the form (\ref{gerochbracket2}) of the bracket. 

Note also that the determinant $\det s(v)$ Poisson commutes with any functional of $s$ under the Sklyanin bracket, because 
\begin{eqnarray}
 \{ \det \overset{1}{s}(v), \overset{2}{s}(w) \}_G & = \det \overset{1}{s}(v)\,\overset{1}{\tr}[\overset{1}{s}{}^{-1}(v) \{ \overset{1}{s}(v), 
 \overset{2}{s}(w) \}_G]\\
 & = {\rm p.v.} \left(\frac{1}{v-w}\right) \det \overset{1}{s}(v)\,
 \overset{1}{\tr}[\Omega \overset{2}{s}(w) - \overset{2}{s}(w) \Omega] = 0.
\end{eqnarray}
This property has to hold for any valid Poisson bracket on an $SL(2,\R)$ line group, because $\det s(v) = 1$ identically on the group manifold. If one tries to impose a Poisson bracket that does not have this property, then the equations $\{ \det \overset{1}{s}(v), \overset{2}{s}(w) \} = 0$ become non-trivial constraints that restrict the phase space of allowed group elements $s$ to a proper subset of the 
$SL(2,\R)$ line group.  

The Sklyanin bracket also preserves the asymptotically flat Geroch group with either of the two 
falloff conditions on $s(w)$ as $|w| \rightarrow \infty$ that we have considered. The proof is virtually identical to that of the analogous result for the asymptotically flat gravitational phase space $\Gamma_\infty$. Although the Sklyanin bracket (\ref{gerochbracket}) differs from the bracket (\ref{Monbracket}) on the deformed metric $\M$ it shares all the features that made the latter proof possible.
Therefore, for the asymptotically flat Geroch group, defined either by the requierment that $s - \One$
tends to zero as $1/w$ or faster as $|w| \rightarrow \infty$, or by the requierment that $s$ be
a Wiener function with constant term $\One$, the Hamiltonian flow generated via the Sklyanin bracket (\ref{gerochbracket}) by any observable $\Phi$ of the group (functionals of $s$) such that $\dg \Phi/\dg s(w)_a{}^b$ is smooth and compactly supported on the $w$ line is tangent to this group.
The Sklyanin bracket therefore defines Poisson brackets between such observables on the asymptotically
flat Geroch group.

It is straightforward to check that the action of the Geroch group on itself via the multiplication map, 
$(g_1, g_2) \mapsto g_1 g_2$, that takes two group elements to their product is a Poisson map: 
\begin{equation}
\fl \{\overset{1}{s}(\cdot)\overset{1}{s}(g_2),\overset{2}{s}(\cdot)\overset{2}{s}(g_2)\}_G(g_1) 
 + \{\overset{1}{s}(g_1)\overset{1}{s}(\cdot),\overset{2}{s}(g_1)\overset{2}{s}(\cdot)\}_G(g_2)
 = \{\overset{1}{s}(\cdot),\overset{2}{s}(\cdot)\}_G(g_1 g_2)
\end{equation}
The Geroch group equipped with the bracket (\ref{gerochbracket}) is thus a Poisson-Lie group, and its action 
(\ref{Geroch_action2}) on the phase space $\Gamma$ is a Lie-Poisson action in the terminology of \cite{Babelon_Bernard}.
The calculation does not depend on boundary conditions at $|w| \rightarrow \infty$ so the result applies 
equally well to the asymptotic flatness preserving Geroch groups.

\section{The quantum Geroch group and its action on quantum cylindrically symmetric gravitational waves}\label{quantum}

The present section treats the quantization of the Geroch group and of its action on cylindrically symmetric 
vacuum gravitational fields. Before entering into a detailed exposition we give a brief outline of the results 
(with forward equation references).  

The quantization will be carried out only at the algebraic level. The algebra 
generated by the quantized matrix components $s(w)_a{}^b$ of the Geroch group elements is specified and the 
action on the algebra generated by the quantized deformed metric components $\M(w)_{ab}$ is given,
but no representations of these algebras on Hilbert spaces are found. This last step, which is non-trivial, is left 
to future investigations. A representation has been proposed in \cite{KS2} for the quantized cylindrically symmetric 
gravitational data of Korotkin and Samtleben, which implies a representation of the components $\M(w)_{ab}$, but it 
has not yet been demonstrated that the inner product is positive definite on the proposed representation space.

Furthermore, our results are formal. They are derived from algebraic relations satisfied by $s(w)_a{}^b$ and $\M(w)_{ab}$,
and general properties of quantum theories, without a full definition of the objects $s$ and $\M$.
However, we do present a fairly detailed outline of such a definition for $s$, which can also be applied {\em mutatis
mutandis} to define $\M$. Note also that the structure encountered appears to be quite rigid, there seems to be little 
or no freedom to alter it without spoiling its consistency, so there seems to be a good chance that our results will 
hold also in a fully worked out theory. 

What we obtain is a quantization (\ref{q_Geroch_exchange}) of the Sklyanin bracket (\ref{gerochbracket}) and 
compatible quantizations of the unit determinant (\ref{qdets_is_1}) and reality (\ref{reality1}) conditions 
that correspond to the requirement $s(w) \in SL(2,\R)$ for 
all $w \in \R$ in the definition of the classical Geroch group. In this way an algebra of quantized 
polynomials in the matrix components of $s$ is obtained which is similar to the $\mathfrak{sl}_2$ Yangian
algebra which quantizes a subgroup of the $SL(2,\C)$ loop group \cite{Drinfeld1}, \cite{Molev}. 
This is our proposal for the quantum Geroch group. It incorporates quantum versions of all the conditions 
that define the classical Geroch group, as presented in subsection \ref{classical_Geroch}), except the smoothness 
requierment which, as is usual in quantization, is not imposed on the quantum theory.
We do not settle on a proposal for a condition that defines the quantized asymptotic flatness preserving 
subgroup of the Geroch group, although some posibilities are discussed. 

In addition to quantizing the Geroch group we also define an action of the quantized Geroch group element 
$s$ on Korotkin and Samtleben's quantization of the deformed metric $\M$. This action is a quantization of 
the classical action of the Geroch group on the classical deformed metric and it appears to be a symmetry
of Korotkin and Samtleben's model of quantum cylindrically symmetric gravity \cite{KS}, because 
it preserves the exchange relation for $\M$ and several of the additional conditions that define
the quantization of $\M$ in their model. However we are not able to check that all such conditions are 
preserved. In particular, we have not been able to verify that the action of the quantum Geroch group 
preserves the quantization of the condition that $\det \M = 1$, and whether or not the action preserves 
asymptotic flatness is also unresolved. 

As we saw in subsection \ref{M_solution_space} the space of solutions of classical $C^\infty$ cylindrically 
symmetric vacuum gravity can be identified with the set of smooth functions $w \mapsto \M(w)$ specifying 
the deformed metric for all $w \in \R$. $\M(w)$ is required to be real, symmetric, of unit determinant, and 
positive semi-definite (which implies positive definite because $\det \M = 1$). This space of solutions is 
turned into a phase space, $\Gamma$, by equipping it with the Poisson bracket (\ref{Monbracket}) derived
from the cylindrically symmetric Hilbert action for vacuum gravity (\ref{action}).
If the phase space is restricted to fields with asymptotically flat Kramer-Neugebauer duals
then $\M(w)$ must tend to an asymptotic value $e_\infty$ as $|w| \rightarrow \infty$, the falloff conditions 
being dictated by the precise definition of asymptotic flatness adopted. 

In the quantum model of Korotkin and Samtleben $\M$ satisfies an exchange relation that quantizes 
(\ref{Monbracket}) and, as we will see, also quantum versions of all the conditions on $\M$ that define 
$\Gamma$ save the smoothness requierment. Note however that their model is not formulated in terms of $\M$, 
but rather in terms of the elements, $T_+$ and $T_-$, of a factorization of $\M$:
\begin{equation}\label{M_product_def}
 \M(w)_{ab} = T_{+}(w)_a{}^c T_{-}(w)_b{}^d e_{\infty\,cd},
\end{equation}
with 
\begin{equation}
 T_-(w)_a{}^b = [T_+(\bar{w})_a{}^b]^*.
\end{equation} 
Classically $T_\pm$ are $SL(2,\C)$ line group elements that are holomorphic on the upper/lower half complex $w$-plane and have Poisson brackets (\ref{T+T+bracket} 
- \ref{T+T-bracket}) similar to the Sklyanin bracket of the Geroch group elements $s$.
These Poisson brackets imply the Poisson bracket (\ref{Monbracket}) for $\M$, and {\em vice versa}.

The model of Korotkin and Samtleben assumes asymptotic flatness of the Kramer-Neugebauer dual spacetime geometry,
which implies a falloff of $T_+$ to a constant matrix as $|w| \rightarrow \infty$, but a precise falloff condition 
is not given. In Subsection \ref{asymptotic_flatness} we have suggested two possible such conditions on $\M$
for the classical theory. One was simply to require $\M$ to fall off to its asymptotic value as $1/w$ or faster. 
The other proposal was to require 
$\M$ to be a Wiener function, which implies that $T_+$ is also a Wiener function, and thus approaches a constant 
asymptotic value as $w\rightarrow \infty$ \cite{Gohberg_Krein}. This slightly more complicated condition is natural 
because it also guarantees the existence of the factorization (\ref{M_product_def}). 

Korotkin and Samtleben quantize the Poisson brackets of $T_+$ and $T_-$ in such a way that $\M$, defined 
by (\ref{M_product_def}), satisfies a closed exchange relation (\ref{RMR'M}), that is, one expressible 
entirely in terms of $\M$ and constants. This exchange relation quantizes the Poisson bracket 
(\ref{Monbracket}) of $\M$. Korotkin and Samtleben 
then show that the condition (\ref{qdetT+}), that a certain quantization of the determinant of $T_+$ is 
$1$, is compatible with the exchange relations, as is the condition that $\M_{ab}$ be symmetric in its 
indices. Indeed, these compatibilities determine the detailed form of the exchange relation between 
$T_+$ and $T_-$ completely. Imposing the unit quantum determinant condition on $T_+$ implies a similar 
condition on $T_-$ and both together imply in turn that $\det \M = 1$ in the classical limit. The unit 
quantum determinant condition on $T_+$ thus indirectly quantizes the clasical unit determinant condition on $\M$.
The condition that $\M_{ab}$ be symmetric implies, via the factorization (\ref{M_product_def}) that
the components $\M(w)_{ab}$ are real for all real $w$, and also that $\M$ is positive 
semi-definite. In this way all the conditions defining the possible values of the matrix $\M(w)$ in the 
classical phase space $\Gamma$ are implemented in their quantization, although the unit determinant condition 
on $\M$ is implemented only indirectly through a condition on $T_+$. Since $\M(w)$ is the natural variable 
describing the classical field, at least in our presentation, it would be desireable to have a presentation 
of the quantum theory exclusively in terms of the quantum magnitudes $\M(w)_{ab}$, but this has so far eluded us.

What about asymptotic flatness in the quantum theory? One could require the expectation 
value of $\M$ to be a Wiener function, or that it falls off as $1/w$ or faster to an asymptotic value. The second 
approach actually fits 
fairly nicely with the definition of the Yangian algebra \cite{Molev}: The fact that $T_+$, like $s$, satisfies 
the exchange relation (\ref{T+T+exchange}) of a Yangian generating function \cite{KS} suggests 
that $T_+$ could be a formal power series in $1/w$ with constant term $\One$, like the Yangian generating function. 
$\M$ would then also be a formal power series in $1/w$. Taking the expectation values of the coefficients in this power 
series in a state of the gravitational field would yield a numerical formal power series for the expectation value of $\M$
which should converge, possibly by means of a non-trivial summation method, to a function $\langle \M\rangle$ of $w$.
If the summation method used in this proposed definition of the quantum theory is at all reasonable this expectation 
value would fall off to an asymptotic value, $e_\infty$, as $1/w$ or faster.
But Korotkin and Samtleben do not adopt this approach, because it seems to clash with the analytic properties 
in $w$ of $T_+$ in the classical limit. See \cite{Samtleben_thesis} equation (3.82) and comments on p. 40 - 41 and p. 61. We are 
inclined to agree with them. We will argue that it is unnatural, though perhaps not impossible, to require $\M$ to 
be such formal power series. 

In subsection \ref{q_Geroch_as_symmetry} we will make an ansatz as to the form of the action of the 
quantum Geroch group on the quantum magnitudes $\M(w)_{ab}$ and show that essentially only one action 
out the proposed family, namely (\ref{q_Geroch_action}), can be a symmetry of Korotkin and Samtleben's 
model of cylindrically symmetric gravity. (More precisely, we show that all the actions from the ansatz 
that can be symmetries are strictly equivalent to (\ref{q_Geroch_action}).) The action (\ref{q_Geroch_action}) 
seems be a symmetry because, to the extent that we are able to check, it preserves all the conditions that 
define Korotkin and Samtleben's model of cylindrically symmetric gravity. It preserves the exchange relation 
(\ref{Monbracket}) and the symmetry, reality, and 
positive semi-definiteness of $\M$. It has not been possible to check whether it preserves the quantization 
of the unit determinant condition on $\M$, because there is as yet no formulation of this condition directly 
in terms of $\M$. Neither can the preservation of the quantized unit determinant condition on
$T_+$ be checked because no action of the quantum Geroch group on $T_+$ has been found. 
For the same reason the preservation of the exchange relations of $T_+$ and $T_-$ could not be checked
directly. Of course, this is not necessary if these are consequences of the exchange relation (\ref{RMR'M}) 
of $\M$, which seems likely but has not been proved.
Finally, whether or not asymptotic flatness is preserved by the quantum Geroch group action depends on how asymptotic 
flatness is defined, and how the asymptotic flatness preserving subgroup of the Geroch group is quantized. 
If asymptotically flat quantum $\M(w)$ are assumed to be a formal power series in $1/w$ 
(with a suitable summation rule for expectation values) and $s$ is also such a power series, with constant term 
$\One$, then it is clear that the action (\ref{q_Geroch_action}) preserves asymptotic flatness.
On the other hand, no asymptotic flatness preserving quantum Geroch group has been developed corresponding
to the other proposed implementation of asymptotic flatness, in which the expectation value of $\M$ is required 
to be a Wiener function. 

In addition to these results on the relations satisfied by $\M$ and $s$ and the action of the quantum Geroch 
group, we also outline definitions of $\M$ and, in more detail, $s$. One possibility that has already been 
mentioned is to define them as formal power series in $1/w$, in analogy with the generating functions of the Yangian
in the ``RTT'' formulation \cite{FRT}\cite{Molev}.
The Yangian can be obtained as the quotient of the free algebra generated by the coefficients of this formal 
power series by the exchange and other relations that the generating functions are required to satisfy \cite{Molev}.
But we consider it more natural to postulate that $\M$ and $s$ are Fourier integrals, that is, 
sums of Fourier modes $e^{ikw}$, with the coefficients of these modes being the generators of a free algebra 
on which the exchange relations and others are then imposed. (More precisely, the free algebra in question is
the tensor algebra of a space of test functions.)
Either of these frameworks, of formal power series or of Fourier integrals, might serve to underpin the results 
of the present work. That is, they could provide definitions of $s$ and of $\M$ for which the formal results of 
the present work hold rigorously. We outline the steps necessary to realize this program in the Fourier integral 
framework.

\subsection{Quantization of the Geroch group}\label{q-Geroch}

The Sklyanin bracket (\ref{gerochbracket}) admits a natural quantization, namely the exchange relation 
\begin{equation}\label{q_Geroch_exchange}
R(v-w) \overset{1}{s}(v) \overset{2}{s}(w) = \overset{2}{s}(w) \overset{1}{s}(v) R(v-w)
\end{equation}
for the quantized Geroch group element $s$, with $R(u) = (u - i \hbar /2)I - i \hbar \Omega$. $s$ is a 
spectral parameter dependent, $2 \times 2$ matrix of non-commuting ``quantum magnitudes'' $s(w)_a{}^b$.
In a state of the quantized Geroch group each component $s(w)_a{}^b$ will have an expectation value, 
so the expectation value of $s$ is a $2 \times 2$ complex matrix valued function (or distribution) of 
the spectral parameter, much like a classical Geroch group element. $R$, the so-called ``R matrix'', is 
a four index tensor which defines a linear map of the tensor product of spaces 1 and 2 to itself. 
In components 
\beq\label{R_matrix}
R(u)_a{}^b{}_c{}^d = u\delta_a^b \delta_c^d - i\hbar \delta_a^d\delta_c^b, 
\eeq
with $a, b$ space 1 indices, and $c, d$ space 2 indices.

The exchange relation (\ref{q_Geroch_exchange}) can be rewritten as the commutation relation
\begin{equation}\label{s_commutator}
 [\overset{1}{s}(v), \overset{2}{s}(w)] = \frac{i\hbar}{v - w}[\Omega, \frac{1}{2}\left(\overset{1}{s}(v) \overset{2}{s}(w) + \overset{2}{s}(w)\overset{1}{s}(v)\right)],
\end{equation}
making the connection with the Sklyanin bracket manifest.
This expression can be found by combining (\ref{q_Geroch_exchange}) with an alternative form of (\ref{q_Geroch_exchange}),
\begin{equation}\label{q_Geroch_exchange2}
R(w-v) \overset{2}{s}(w) \overset{1}{s}(v) = \overset{1}{s}(v) \overset{2}{s}(w) R(w-v),
\end{equation}
obtained by interchanging $v$ and $w$ and the spaces $1$ and $2$, and using the fact that $R$ is invariant under the latter interchange, or alternatively, by multiplying (\ref{q_Geroch_exchange}) by $R(w-v)$ from both the left and the right.
Conversely, the exchange relation (\ref{q_Geroch_exchange}) may be recovered easily from the commutation relation (\ref{s_commutator}) if one first notes that (\ref{s_commutator}) 
implies that $[\overset{1}{s}(v), \overset{2}{s}(w)] = [\overset{1}{s}(w), \overset{2}{s}(v)]$ because the numerator of (\ref{s_commutator}),
$[\Omega, \overset{1}{s}(v) \overset{2}{s}(w) + \overset{2}{s}(w)\overset{1}{s}(v)] 
= [\Pi, \overset{1}{s}(v) \overset{2}{s}(w) + \overset{2}{s}(w)\overset{1}{s}(v)]$, satisfies 
\begin{eqnarray}
[\Pi, \overset{1}{s}(v) \overset{2}{s}(w) + \overset{2}{s}(w)\overset{1}{s}(v)] 
 = [\Pi \left(\overset{1}{s}(v) \overset{2}{s}(w) + \overset{2}{s}(w)\overset{1}{s}(v)\right)\Pi, \Pi]\nonumber \\
 = - [\Pi, \overset{1}{s}(w) \overset{2}{s}(v) + \overset{2}{s}(v)\overset{1}{s}(w)]. \label{numerator_id}
\end{eqnarray}

Notice that the preceding equation implies that the numerator in the commutation relation (\ref{s_commutator}) vanishes when $v = w$. The pole singularity in (\ref{s_commutator}), like the one in the Sklyanin bracket it quantizes, is therefore only apparent.

Note also that the exchange relation (\ref{q_Geroch_exchange}) is essentially the ``RTT'' relation, or ``ternary relation'', of the $\mathfrak{sl}_2$ Yangian algebra \cite{Molev}.  In fact, when (\ref{q_Geroch_exchange}) is expressed in terms of the rescaled spectral parameters $v' = -i v/\hbar$ and $w' = -i w/\hbar$, (\ref{q_Geroch_exchange}) becomes precisely the RTT relation of the 
$\mathfrak{sl}_2$ Yangian (or the $\mathfrak{gl}_2$ Yangian).\footnote{
The definition of the quantum Geroch group will also include a quantization of the unit determinant condition,
(\ref{qdets_is_1}), given further on, which is identical to that in the $\mathfrak{sl}_2$ Yangian. 
Moreover, one may define on the quantum Geroch group all the Hopf algebra structures of the $\mathfrak{sl}_2$ Yangian:
the co-product $\Delta(s(w)_a{}^b) = s(w)_a{}^c \otimes s(w)_c{}^b$, the antipode 
$S(s(w)_a{}^b) = s^{-1}(w)_a{}^b = \vareg_{ac}\vareg^{bd} s(w + i\hbar)_d{}^c$ and the co-unit
$\epsilon(s(w)_a{}^b) = \delta_a{}^b$. (The formula for the inverse of $s$ follows from the restriction 
(\ref{ihbar_dif_exchange}) of the exchange relation and the unit quantum determinant condition
(\ref{qdets_is_1}).)}
The Yangian is a symmetry of the quantum Heisenberg spin chain, and various other physical systems. In the Heisenberg spin chain it arises as the quantization of a subgroup of the $SU(2)$ loop group which is a symmetry of the classical spin chain, the so called ``dressing symmetry'' \cite{Babelon_Bernard}.

The exchange relation (\ref{q_Geroch_exchange}) defines a quantization of the Poisson algebra of polynomials 
in the matrix components of classical Geroch group elements. The quantization of such a polynomial is obtained 
by replacing each classical matrix component by the corresponding non-commuting quantum matrix component and 
then symmetrizing each monomial with respect to the ordering of its factors, yielding a symmetrized polynomial.\footnote{
It would be natural to call the result a symmetric polynomial, but this term already has another meaning. It refers to
a polynomial in several (usually commuting) variables which is symmetric under interchange of the variables, like 
$xy + yz + zx$. Our ``symmetrized polynomials'' are polynomials in several non-commuting variables that are symmetric 
under permutation of the order of the factors in products, like $xy + yx + yz + zy$.}
The algebra of the resulting quantized polynomials is then determined by the exchange relation: Using (\ref{q_Geroch_exchange}) any polynomial in the quantum matrix components $s(w)_a{}^b$  can be expressed as a symmetrized polynomial. In particular, any product of symmetrized polynomials can be expressed as a symmetrized polynomial.

Let us examine this in detail. The exchange relation (\ref{q_Geroch_exchange}) may be written as an expression for the product 
$\overset{2}{s}(w) \overset{1}{s}(v)$ as the image under a linear map of the reversed product, $\overset{1}{s}(v) \overset{2}{s}(w)$: 
\begin{eqnarray}
  \overset{2}{s}(w) \overset{1}{s}(v) & = -\frac{1}{(w-v)^2 + \hbar^2}R(v - w) \overset{1}{s}(v) \overset{2}{s}(w) R(w-v) \label{explicit_exchange}\\
& = \overset{1}{s}(v) \overset{2}{s}(w) - \frac{2i\hbar}{w - v + i\hbar}A\, \overset{1}{s}(v)\overset{2}{s}(w) \, S + \frac{2i\hbar}{w - v - i\hbar}S\, \overset{1}{s}(v)\overset{2}{s}(w)\, A, \label{explicit_exchange2}
\end{eqnarray}
with $A = \frac{1}{2}(I - \Pi)$ the antisymmetrizer and $S = \frac{1}{2}(I + \Pi)$
the symmetrizer with respect to interchange of space 1 and space 2 indices. (In 
components $A_a{}^b{}_c{}^d = \dg_{[a}^b \dg_{c]}^d = \frac{1}{2}\vareg_{ac}\vareg^{bd}$
and $S_a{}^b{}_c{}^d = \dg_{(a}^b \dg_{c)}^d$.) 

Using (\ref{explicit_exchange}) a symmetrized product may be expressed as a sum of linearly ordered products: a product of $n$ matrix components $s(w_i)_{a_i}{}^{b_i}$ symmetrized with respect to reordering of the factors can be written as a linear combination of products of (possibly different) components of the matrices $s(w_i)$ ordered according to the index $i$ of the spectral parameter argument $w_i$, that is, with the factors appearing in the order $s(w_1)$, $s(w_2)$, ... $s(w_n)$. For instance,
\begin{eqnarray}
\fl\frac{1}{2}\left(\overset{1}{s}(v) \overset{2}{s}(w) + \overset{2}{s}(w) \overset{1}{s}(v)\right)\nonumber \\ 
= \overset{1}{s}(v) \overset{2}{s}(w) - \frac{i\hbar}{w - v + i\hbar}A\, \overset{1}{s}(v) \overset{2}{s}(w)\, S + \frac{i\hbar}{w - v - i\hbar}S\, \overset{1}{s}(v) \overset{2}{s}(w)\, A, 
\end{eqnarray}
and in general
\begin{equation}
\fl \frac{1}{n!}\sum_{\sg \in S_n} s(w_{\sg(1)})_{a_{\sg(1)}}{}^{b_{\sg(1)}}\cdot ... \cdot s(w_{\sg(n)})_{a_{\sg(n)}}{}^{b_{\sg(n)}} 
 = \sum_{\bf c, d} X_{\bf{a}}{}^{\bf{b}}{}_{\bf{d}}{}^{\bf{c}}\,
  s(w_1)_{c_1}{}^{d_1}\cdot ... \cdot s(w_n)_{c_n}{}^{d_n},
\end{equation}
where $\bf{a}$ denotes the whole sequence of indices $a_1,a_2,...,a_n$ and $\bf{b}$, $\bf{c}$ and $\bf{d}$ are defined similarly, and $S_n$ is the symmetric group.

Inverting $X$ one obtains an expression for the ordered product $s(w_1)_{c_1}{}^{d_1}\cdot ... \cdot s(w_n)_{c_n}{}^{d_n}$ as a linear combination of symmetrized products. For instance, 
\begin{equation}\label{symmetrized_ss_product}
\fl \overset{1}{s}(v)\overset{2}{s}(w) = \frac{1}{2}\left(\overset{1}{s}(v) \overset{2}{s}(w)
 + \overset{2}{s}(w)\overset{1}{s}(v)\right) + \frac{i\hbar/2}{v - w}\,[\Pi, \frac{1}{2}\left(\overset{1}{s}(v) \overset{2}{s}(w) + \overset{2}{s}(w)\overset{1}{s}(v)\right)],
\end{equation}
as can also be seen directly from the commutation relation (\ref{s_commutator}).
Note that $X$ is a rational function of $\hbar$ and the spectral parameters of the factors,
and that it reduces to the identity map when $\hbar = 0$. Its inverse is therefore also a rational 
function, and hence finite except at pole singularities. 
 
As already pointed out, the commutator (\ref{s_commutator}), and thus (\ref{symmetrized_ss_product}), 
does not really have a pole singularity, since the numerator of the apparent pole term vanishes along 
with the denominator when the two spectral parameters coincide. Nevertheless, (\ref{s_commutator}) and 
(\ref{symmetrized_ss_product}) are not, strictly speaking, defined at coinciding spectral parameters. 
What can be defined is the limit as the apparent singularity is approached. As a consequence the algebra 
of polynomials we have described is actually built not only of the components of $s(w)$ at different 
values of the spectral parameter, but also of the derivatives of these matrix components. This can be 
seen immediately in the commutator at coinciding spectral parameters:
\begin{equation}\label{s_coincident_commutator}
 [\overset{1}{s}(v), \overset{2}{s}(v)] = i\frac{\hbar}{2}[\Omega, \frac{d\overset{1}{s}}{dv}(v) \overset{2}{s}(v) + \overset{2}{s}(v)\frac{d\overset{1}{s}}{dv}(v)].
\end{equation}
(The same of course occurs in the corresponding Poisson bracket.) 

Will it always be the case that the residues of the (apparent) poles that appear when reordering
a monomial vanish? If $s$ is a matrix of {\em functions} of $w$, mapping from a domain $U \subset \C$ 
to an algebra over $\C$, then this must be so within $U$. If $B = C/f$ with $B$ and $C$ polynomials 
in $s$ (with any ordering) depending on a set of spectral parameters, and $f$ a linear function of 
these spectral parameters, then $C = f B$, so $C$ must vanish when $f$ vanishes. 
Then, for instance, (\ref{explicit_exchange2}) implies that 
\begin{equation}\label{AssS}
 A\, \overset{1}{s}(v)\overset{2}{s}(w)\, S = 0,
\end{equation}
if $v = w + i\hbar$, and similarly
\begin{equation}\label{SssA}
 S\, \overset{1}{s}(v)\overset{2}{s}(w)\, A = 0
\end{equation}
if $v = w - i\hbar$. 
In fact, under the hypothesys that the product $s_a{}^b(x) s_c{}^d(y)$ is sufficiently regular on the lines 
$x - y \mp i\hbar = 0$ so that $(v - w - i\hbar)\overset{1}{s}(v) \overset{2}{s}(w)$ and 
$\overset{2}{s}(w)\overset{1}{s}(v)(v - w - i\hbar)$ vanish when $(v - w - i\hbar)$ does, (\ref{AssS}) and 
(\ref{SssA}) both follow immediately from the exchange relation (\ref{q_Geroch_exchange}): The $R$ matrix 
(\ref{R_matrix}) evaluated at spectral parameter difference $u = i\hbar$ is 
$R(i\hbar)_a{}^b{}_c{}^d = i\hbar \vareg_{ac}\vareg^{bd}$, so (\ref{q_Geroch_exchange}) implies that
\begin{equation}\label{ihbar_dif_exchange}
 \vareg_{ac}\vareg^{bd} s(z + i\hbar/2)_b{}^e s(z - i\hbar/2)_d{}^f = s(z - i\hbar/2)_c{}^d s(z + i\hbar/2)_a{}^b 
 \vareg_{bd}\vareg^{ef},
\end{equation}
which implies (\ref{AssS}) and (\ref{SssA}). These equations will be important in the present work. 

The assumption that $s(w)$ is a function of $w$ with a definite value at each point of the domain of 
interest seems too strong a hypothesys to impose on the quantum theory without further justification.
What sort of mathematical object should $s$ be? Recall that classically the deformed metric $\M(w)$ 
is just the conformal metric $e$ on the cylindrical symmetry axis at time $t = -w$. 
The action, $\M(w) \mapsto s(w)\M(w)s^t(w)$, of the Geroch group thus provides a quite concrete 
meaning for $s(w)$ as a field on the symmetry axis worldsheet. We therefore expect it to be represented 
in the quantum theory by some sort of quantum field operator. In particular, its expectation value, in 
a state of the Geroch group, should be a distribution in $w$, giving a numerical value when integrated 
against a test function of $w$. Also, some products of $s$ with itself, and with the field $\M$ should 
be defined, such as that ocurring in the quantization (\ref{q_Geroch_action}) of the action 
$\M(w) \mapsto s(w)\M(w)s^t(w)$. 

We shall assume that the quantization of the Geroch group can be realized within the general framework of 
algebraic quantum theory. See \cite{Khavkine_Morreti} for a recent review. The starting point for formulating
our quantization in this framework will be  
the associative, unital algebra $\A_0$ generated by a unit element $\One$ and the quantization $s[f]$ of
the integrals, $\sum_{a,b}\int s(w)_a{}^b f(w)_b{}^a dw$, of the classical Geroch group element $s$ against 
test functions $f$ of real $w$. An involution, $*$, turns $\A_0$ into a $*$-algebra. The involution quantizes 
complex conjugation of classical variables. Thus, for instance, $s[f]^*$ is the quantization of the complex 
conjugate, $\overline{\sum_{a,b}\int s(w)_a{}^b f(w)_b{}^a dw}$, of the classical Geroch group element 
integrated against $f$, and the quantization $r$ of a real classical quantity is $*$ real, that is 
$r = r^*$.
The smeared quantum Geroch group element $s[f]$ will be assumed to be linear in $f$ like it's 
classical counterpart.

A state $\varphi$ of the Geroch group is a linear function which maps elements of this $*$-algebra to their 
expectation values. States are required to be positive and normalized:
\begin{eqnarray}
 \varphi(a^* a) \geq 0 \ \ \mbox{for all elements, $a$, of $\A_0$}\label{state_positivity}\\
 \varphi(\One) = 1.\label{state_normalization}
\end{eqnarray}
(Note that these conditions imply that $\varphi(a^*) = \overline{\varphi(a)}$ for all $a \in {\A_0}$.)
We will often denote the expectation value $\varphi(a)$ by $\langle a \rangle$, leaving the state involved 
implicit.  

States correspond essentially to density matrices in the standard Hilbert space formulation of quantum mechanics: 
Via the Gelfand-Naimark-Segal (GNS) construction (extended to $*$-algebras) a Hilbert space can be found for any 
given state such that that state, and a subset of the other states on the $*$-algebra, are represented by density 
matrices on the Hilbert space, and the elements of the $*$-algebra are represented by operators defined on a 
common dense subspace. The $*$ operation in the algebra corresponds to taking the adjoint in the Hilbert space 
representation, so only $*$ real algebra elements can be represented by self-adjoint operators. For this reason
the quantizations of real classical observables must be $*$ real exactly, and not just to order zero in 
$\hbar$.\footnote{
Only consider real classical and quantum observables will be considered.
By labeling measurement outcomes by complex numbers one can also construct complex classical and
quantum observables. However the real and imaginary parts of such observables must be $∗$ real, so no 
generality is gained. Moreover, they must commute with each other, so it is by no means the case that an 
arbitrary complex algebra element can be interpreted as a complex observable. It seems, therefore, that 
allowing such complex observables would not bring any gain in simplicity either.}

The expectation value of a product $s[f_1]s[f_2]...s[f_n]$ is a linear functional of each test function involved. 
If a state $\varphi$ is such that the dependence of the expectation value on each test function is also continuous
then the expectation value defines a distribution \cite{Khavkine_Morreti}. That is, $\varphi(s[f])$ defines a 
distribution $\langle s(w) \rangle = \varphi(s(w))$ in $w \in \R$, $\varphi(s[f]s[g])$ defines a distribution 
$\langle s(v)s(w) \rangle$ in $(v,w) \in \R^2$, etc.. Following standard practice in quantum field theory we will 
admit only states that are continuous in this sense. In order to work within an established mathematical framework 
we shall, somewhat arbitrarily, take the space of test functions to be ${\mathfrak{S}}^{2 \times 2}$, the space 
of $2$ by $2$ matrices of Schwartz functions, the space $\mathfrak{S}$ of Schwarz functions having the usual 
topology (see for instance \cite{Hormander}). Then the expectation values 
$\langle s(w) \rangle$, $\langle s(v)s(w) \rangle$, ... are tempered distributions. 

The quantized Geroch group element $s$ itself is a linear map ${\mathfrak{S}}^{2\times 2} \rightarrow \A_0$, 
but we cannot call it an $\A_0$ valued distribution, since we have not defined a topology on $\A_0$ with 
which to define continuity. However we may still use the notation $s(w)$ of distribution theory, in which
the ``argument'' $w$ is the argument of the test functions on which $s$ acts. With this understanding products 
of the form $s(w_1)s(w_2)...s(w_n)$, in which the arguments $w_1$, ..., $w_n$ are independent, are defined.

The framework as developed thus far suffices to define the exchange relation (\ref{q_Geroch_exchange}) 
for real spectral parameters. We may also impose a reality condition
\begin{equation}\label{reality_real_axis}
 [s(w)_a{}^b]^* = s(w)_a{}^b \ \ \ \forall w \in \R.
\end{equation}
which quantizes the requierment that $s(w)_a{}^b$ be real in the classical theory.
Since products $s[f_1]s[f_2]...s[f_n]$ are completely characterized by the sequence of test functions
$f_1, f_2, ..., f_n$, in fact are linear images of $f_1 \otimes f_2 \otimes ... \otimes f_n$, $\A_0$
must be homomorphic to the tensor algebra of test functions, 
\begin{equation}
T({\mathfrak{S}}^{2\times 2}) \equiv \C\, \oplus\, {\mathfrak{S}}^{2\times 2}\, \oplus\, {\mathfrak{S}}^{2\times 2}\otimes {\mathfrak{S}}^{2\times 2}\, \oplus\,...
\end{equation}
divided by the two sided ideal $\cal I$ generated by the exchange relations, with the involution $*$ defined by
(\ref{reality_real_axis}). We do not want $\A_0$ to be subject to any constraints {\em apart} from 
the exchange relation and the reality condition, so we take $\A_0$ to be isomorphic to the quotient 
$T({\mathfrak{S}}^{2\times 2})/{\cal I}$ with $*$ defined by (\ref{reality_real_axis}).

Defining $s(w)$ for real $w$ is not enough for our purposes. 
As mentioned earlier, our quantization of the Geroch group requires the quantized group elements $s(w)$ 
to satisfy a quantization of the unit determinant condition:
\begin{equation}\label{qdets_is_1}
 {\rm qdet}\, s(w) = 1,
\end{equation}
where the {\em quantum determinant} of $s(w)$ is defined by
\begin{equation}\label{qdets_def}
 {\rm qdet}\, s(w) \equiv s(w+i\hbar/2)_1{}^1 s(w - i\hbar/2)_2{}^2 - s(w + i\hbar/2)_2{}^1 s(w - i\hbar/2)_1{}^2.
\end{equation}
The quantum determinant is defined in this way so that it commutes with $s$. (For a proof that
it commutes see \cite{Samtleben_thesis} section 5.1. or \cite{Molev} sections 2.6 and 2.7.) 
If this were not so the condition (\ref{qdets_is_1}) would imply additional, spurious,
constraints $0 = [{\rm qdet}\, s(w), s(v)]$ analogous to those that would arise in the classical 
theory if $\det s(w)$ did not Poisson commute with $s(v)$. We will return to this issue at the 
end of this subsection.

But (\ref{qdets_def}) is not a complete definition until the quantum matrix components 
$s(w)_a{}^b$ themselves are defined on complex $w$ with non-zero imaginary components, at least for 
imaginary components $\pm i\hbar/2$. Furthermore products of the form
\begin{equation}\label{shifted_product}
 s(w + i\hbar/2)_a{}^b s(w - i\hbar/2)_c{}^d,
\end{equation}
in which the spectral parameter arguments of the factors are not independent, must also be defined.
The product (\ref{shifted_product}) appears not only in the quantum determinant
(\ref{qdets_def}), but also in the important restriction (\ref{ihbar_dif_exchange}) of the exchange relation,
and in the quantization (\ref{q_Geroch_action}) of the action (\ref{Geroch_action}) of the Geroch group 
on cylindrically symmetric gravitational data. In fact, (\ref{shifted_product}) is the only
product of quantized group elements $s$ at related spectral parameter values that is strictly
necessary for our results. We will make an ansatz for the Geroch group action which uses a more 
general class of products, but the action we finally obtain - the only satisfactory one among the 
candidates of the ansatz - requires only the product (\ref{shifted_product}).

Extending $s$ to complex $w$ could introduce new, spurious, degrees of freedom in the model.
Classically, the components $s_a{}^b$ of a Geroch group element are four real valued functions defined 
only on real values of the spectral parameter $w$. At least near the classical limit, the expectation 
value $\langle s \rangle$ of $s$ in the quantum theory should have the same number of degrees of freedom, 
and indeed the reality condition (\ref{reality_real_axis}) implies that the four components 
$\langle s(w)_a{}^b \rangle$ are real on the real $w$ axis. How can one ensure that $\langle s \rangle$ 
off the real axis has no additional, independent degrees of freedom? The most obvious, straightforward way 
is to require $\langle s \rangle$ off the real $w$ axis to be the analytic continuation of $\langle s \rangle$ 
on the axis.

However, the requierment that there exist an analytic continuation, at least to $\Im w = \pm \hbar/2$,
restricts $\langle s \rangle$ on the axis to a subclass of real analytic functions, which seems a very 
strong restriction. An apparently more general way to achieve the same end is to evaluate the Fourier transform 
\begin{equation}\label{Fourier}
 \tilde{s} (k)_a{}^b = \frac{1}{\sqrt{2\pi}} \int_{-\infty}^{\infty}e^{-ikw}  s(w)_a{}^b  dw,
\end{equation}
of $s$ along the real $w$ axis, and then define $s$ on (a portion of) the complex $w$ plane to be the
inverse Fourier-Laplace transform 
\begin{equation}\label{inv_Fourier_Laplace}
  s(w)_a{}^b  = \frac{1}{\sqrt{2\pi}}\int_{-\infty}^{\infty}e^{ikw} 
 \tilde{ s} (k)_a{}^b dk.
\end{equation}
with the $k$ integral running over the real line. Of course the integral (\ref{Fourier}) of $s$ is 
just a formal expression. Really $\tilde{s}$ is defined by the equation 
\begin{equation}\label{Fourier_def}
\tilde{s}[\bar{\tilde{f}}] = s[\bar{f}]\ \forall f \in {\mathfrak{S}}^{2\times 2}, 
\end{equation}
where the Fourier transform $\tilde{f}$ of the test function $f \in  {\mathfrak{S}}^{2\times 2}$ is 
defined by an integral of the form of (\ref{Fourier}) with $f$ in place of $s$. (Note that the Fourier transform 
defines a topological automorphism of ${\mathfrak{S}}^{2\times 2}$,
so in particular, $\tilde{f}$ can be any element of ${\mathfrak{S}}^{2\times 2}$.)
The inverse transform of $\tilde{s}$ is also defined by (\ref{Fourier_def}), and 
its extension to complex $w = x + i\eta$, the inverse Fourier-Laplace transform, is defined by
$s(x + i\eta) \equiv s_\eta(x)$ with
\begin{equation}\label{inv_Fourier_Laplace_def}
  s_\eta[\bar{f}] = \tilde{s}[e^{-k\eta}\bar{\tilde{f}}].
\end{equation}
Here $\eta \in \R$ is a constant and $x\in \R$ the argument of $f$, so $s(w)$ for any given value, $\eta$, of 
$\Im w$ is a linear map from functions $f$ on the line $\Im w = \eta$ in the complex plane to $\A_0$.  
(See \cite{Hormander} section 7.4.)

The Fourier transform provides a valuable alternative perspective on the field $s$, but apparently not  
an approach to extending $s(w)$ to complex $w$ that is more general than analytic continuation. On the 
contrary, the Fourier transform based approach formulated here amounts to just analytic continuation with an 
extra condition on $s(w)$:
We have already restricted the states we admit by the requierment that the expectation value 
$\langle s(w) \rangle$ on the real $w$ axis be a tempered distribution. This expectation value
therefore has a Fourier transform $\langle \tilde{s}(k) \rangle$ which is also a tempered 
distribution, determined by the relation (\ref{Fourier_def}). The inverse Fourier-Laplace transform
(\ref{inv_Fourier_Laplace_def}) then defines $\langle s(w) \rangle$ as a tempered distribution
on every line of constant $\Im w = \eta$ such that $e^{-\eta k}\langle \tilde{s}(k)\rangle$ is a 
tempered distribution. $\langle s(w) \rangle$ is therefore defined on a strip of the complex 
plane including the real axis, because if $e^{-\eta k}\langle \tilde{s}(k)\rangle$ is a tempered 
distribution for two values, $\eta_1$ and $\eta_2$ of $\eta$, then it is also for any intermediate 
value $\eta' \in [\eta_1, \eta_2]$.\footnote{ 
{\em Proof}: The hypothesies imply that $(e^{-\eta_1 k} + e^{-\eta_2 k})\langle \tilde{s}(k)\rangle$ 
is a tempered distribution and that
\begin{equation}
 e^{-\eta' k}\langle \tilde{s}[\tilde{f}]\rangle 
 = (e^{-\eta_1 k} + e^{-\eta_2 k})\langle \tilde{s}[\tilde{f}']\rangle.
\end{equation}
with $\tilde{f}' = \frac{e^{-\eta' k}}{e^{-\eta_1 k} + e^{-\eta_2 k}}\tilde{f}$. If $\tilde{f}$ is a 
function in the matrix Schwartz space ${\mathfrak{S}}^{2\times 2}$ then so is $\tilde{f}'$, and it 
depends continuously on $\tilde{f}$. $e^{-\eta' k}\langle \tilde{s}[\tilde{f}]\rangle $ is therefore 
continuous in $\tilde{f}$, making it a tempered distribution. \QED}
(See \cite{Hormander} Theorem 7.4.2.)
The strip is symmetric about the real axis because by the reality condition (\ref{reality_real_axis})
$\tilde{s}(-k) = \tilde{s}(k)^*$, and may be 
open, of the form $|\Im w|<\kappa$, or closed, $|\Im w|\leq\kappa$. If $0 < \kappa$, so that the interior of 
the strip is non-empty, then $\langle s(w) \rangle$ is a holomorphic function on this interior. 
(See \cite{Hormander} Theorem 7.4.2.). If $e^{-\eta k}\langle \tilde{s}(k)\rangle$ is a tempered distribution 
also on the boundary $\eta = \pm\kappa$ then $\langle s(w) \rangle$ is defined at $\Im w = \pm \kappa$ and it 
follows from the weak continuity of the Fourier transform on tempered distributions that it is the weak limit 
of $\langle s(w) \rangle$ on the interior of the strip. \footnote{
The value on a given test function $f$ of the {\em weak limit} of a sequence of distributions $T_n$
as $n \rightarrow \infty$ is $T_\infty[f] = \lim_{n \rightarrow \infty} T_n[f]$. That is, it is the limit 
taken holding the test function constant.}
(See \cite{Hormander} remark on p. 194.)
For our purposes we need the extension of $\langle s(w) \rangle$ to cover $|\Im w| \leq \hbar/2$, or a wider strip. 

Our method of extension thus amounts to analytic continuation, but with the restriction that 
$\langle s(w) \rangle$ on the lines of constant $\Im w$ in the strip be tempered distributions. This is a 
non-trivial condition because there exist holomorphic functions, such as $\exp(w/\hbar)$, which do not 
satisfy this requierment. [But it can be made somewhat less restrictive by extending the notions of Fourier 
transform and of distributions, for instance to Fourier hyperfunctions (see \cite{Fourier_hyperfunctions}). 
Furthermore, in the asymptotically flat context many of the states for which the necessary Fourier transforms 
are not defined because $\langle s(w)\rangle$ diverges too rapidly as $\Re w \rightarrow \pm \infty$ are in 
fact excluded by the requirement that the action of $s(w)$ on $\M$ preserves asymptotic flatness. (See section
\ref{classical_Geroch}.)]

Note that we will require not only that $\langle s(w) \rangle$ extends to a strip of the complex 
$w$ plane, but also that the $n$ factor expectation value 
$\langle \overset{1}{s}(w_1)\overset{2}{s}(w_2)... \overset{n}{s}(w_n)\rangle$ extends 
in the same way to $|\Im w_i| \leq \kappa$ with $\kappa \geq \hbar/2$ in each argument 
$w_i$. 

The existence of these extensions of expectation values to complex spectral parameters of course 
requires restrictions on the state. 
In order that $e^{-\eta k}\langle \tilde{s}[\tilde{f}]\rangle \equiv \langle \tilde{s}[e^{-\eta k}\tilde{f}]\rangle$
be a tempered distribution for all $\eta \in (-\kappa,\kappa)$ the expectation value 
$\langle \tilde{s}[\cdot]\rangle$ must admit a continuous extension to a larger space of test 
functions than just ${\mathfrak{S}}^{2 \times 2}$, one that includes $\cosh(\kappa k)\tilde{f}$ 
for all $\tilde{f} \in {\mathfrak{S}}^{2 \times 2}$. More precisely, it must be continuous on 
the image ${\mathfrak{S}}_\kappa^{2 \times 2}$ of ${\mathfrak{S}}^{2 \times 2}$ under the map 
$\tilde{f} \mapsto \cosh(\kappa k)\tilde{f}$, with the open sets of ${\mathfrak{S}}_\kappa^{2 \times 2}$ 
being the images of the open sets of ${\mathfrak{S}}^{2 \times 2}$ under the same map. 
Since ${\mathfrak{S}}^{2 \times 2}$ is dense in ${\mathfrak{S}}_\kappa^{2 \times 2}$ the continuity
of $\langle \tilde{s}[\cdot]\rangle$ implies the existence of a unique continuous extension.
Similar conditions must also hold for the expectation values of products. But note that if one 
state $\varphi_0$ satisfies these conditions then so do the states (expectation values) 
corresponding to a dense subspace of the Hilbert space of the GNS representation defined by 
$\varphi_0$. (See \cite{Khavkine_Morreti}.) 

What we have defined is in fact an extension of the states that satisfy our conditions from
$\A_0$ to a larger algebra, $\A_1$, including as elements $s_\eta[f]$ for all 
$f \in {\mathfrak{S}}^{2 \times 2}$ and $\eta \in [-\kappa,\kappa]$. This algebra provides 
a definition of $s(w)$ for all complex values of $w$ that we will need.
Furthermore, the extension of the exchange relation (\ref{q_Geroch_exchange}) to this range of
$w$ is already built in: Because $\langle \overset{1}{s}(w_1) ... \overset{n}{s}(w_n)\rangle$ 
is holomorphic in $w_i$ for $|\Im w_i| < \kappa$, and the R matrix is holomorphic on all $\C$, 
the fact that the exchange relation (\ref{q_Geroch_exchange}) holds for real $w_i$ implies that 
it also holds for $|\Im w_i| < \kappa$. It then holds for $|\Im w_i| = \kappa$ as well, because 
the expectation value on the boundary of the domain of holomorphicity is the weak limit of that 
in the interior.

An extension of the reality condition (\ref{reality_real_axis}) on $s(w)$ to complex $w$ is 
also built into $\A_1$: 
\begin{equation}\label{reality1}
 [s(w)_a{}^b]^* = s(\bar{w})_a{}^b.
\end{equation}
Any expectation
value of a product that includes a factor of $s(w)$ is holomorphic in $w$ on the strip 
$|\Im w| < \kappa$. Taking complex conjugates, any expectation value including a factor of 
$s(w)^*$ is antiholomorphic in $w$ in the same strip. It follows that an expectation value 
including a factor $s(w) - s(\bar{w})^*$ is holomorphic in the strip, and also vanishes on the 
real axis, by (\ref{reality_real_axis}). It therefore vanishes on the whole strip, implying
(\ref{reality_condition}). (\ref{reality_condition}) at $w = \kappa$ follows by weak continuity.

Note that the exchange relations and other conditions satisfied by $s$ may always be viewed
as conditions on the expectation values $\langle \overset{1}{s}(w_1) ... \overset{n}{s}(w_n)\rangle$ 
of products of $s$, which may be thought of as either the result of applying the state to elements 
of the extended algebra or as analytic continuations of the corresponding expectation values on 
$\A_0$, that is, of the expectation values for real spectral parameters $w_i$.
 
We still have not defined the product (\ref{shifted_product}), 
$\overset{1}{s}(v + i\hbar/2)\overset{2}{s}(v - i\hbar/2)$, either as an algebra element or inside 
expectation values. If $\kappa > \hbar/2$ then the excpectation value 
$\langle \overset{1}{s}(w_1)\overset{2}{s}(w_2)\rangle$  is a holomorphic function at the line 
$(w_1, w_2) = (v + i\hbar/2, v - i\hbar/2), \ v \in \R$ so its
restriction of this line is well defined and satisfies the restricted exchange relation (\ref{ihbar_dif_exchange}). In fact, this is the case also for complex $v$ as long as 
$|\Im v| + \hbar/2 < \kappa$. If $\kappa = \hbar/2$, or $\kappa = |\Im v| + \hbar/2$, then one can try 
to define $\langle \overset{1}{s}(v + i\hbar/2)\overset{2}{s}(v - i\hbar/2)\rangle$ as the weak limit of 
$\langle \overset{1}{s}(v + i\eta)\overset{2}{s}(v - i\eta)\rangle$ as $\eta$ approaches $\hbar/2$ from 
below. This limit might not exist, but if it exists then it is a tempered distribution (\cite{Dieudonne} 22.17.8) which satisfies (\ref{ihbar_dif_exchange}). 
The restriction of the $n$ factor expectation value $\langle \overset{1}{s}(w_1) ... \overset{n}{s}(w_n)\rangle$
to $(w_i, w_j) = (v + i\hbar/2, v - i\hbar/2)$ for any pair of arguments $w_i$ and $w_j$ can 
be defined similarly and if it exists it satisfies the exchange relation.

Suppose now that we limit the set of quantum states once more, to states such that these 
expectation values do exist. The states that satisfy this requierment extend to an algebra $\A_2$ 
which contains $\A_1$ and also $\overset{1}{s}(v + i\hbar/2)\overset{2}{s}(v - i\hbar/2)$, and any 
product $\overset{1}{s}(w_1) ... \overset{n}{s}(w_n)$ with a pair of arguments $w_i$, $w_j$ 
restricted by $w_i = w_j + i\hbar$. In this algebra the restriction (\ref{ihbar_dif_exchange}) of 
the exchange relation holds. Furthermore, the quantum determinant of $s$ (\ref{qdets_def}) is 
defined and the condition that this quantum determinant be $1$ may be imposed. That is, we may 
divide $\A_2$ by the two sided ideal generated by ${\rm qdet}\,s - 1$, yielding a quotient 
algebra $\A$ in which ${\rm qdet}\,s = 1$. 

Notice that the unit quantum determinant condition 
$1 = {\rm qdet}\,s(w) = s(w+i\hbar/2)_1{}^1 s(w - i\hbar/2)_2{}^2 - s(w + i\hbar/2)_2{}^1 s(w - i\hbar/2)_1{}^2$, 
like the other conditions on $s$, needs to be imposed only for $w \in \R$. 
An expectation value containing the product 
$\overset{1}{s}(w + i\hbar/2)\overset{2}{s}(w - i\hbar/2)$ is holomorphic in $w$ as long as 
$|\Im w| + \hbar/2 < \kappa$, so that if ${\rm qdet}\, s(w) = 1$ on the real $w$ axis then it 
is also $1$ in the complex strip $|\Im w| + \hbar/2 < \kappa$. The value at the edge 
$|\Im w| + \hbar/2 = \kappa$ is also $1$ because it is the limit of the value in the interior,
so if (\ref{qdets_is_1}) holds on the real axis it holds wherever ${\rm qdet} s$ is defined.

The algebra $\A$, if it can indeed be constructed as we have outlined, would provide a framework within 
which the formal manipulations of the beginning of this section, and also those of the following 
section, are rigorously justified. (Actually to formulate the ansatz (\ref{ansatzcuantico}) for 
the quantum Geroch group action requires an algebra that extends $\A$ to include products 
$s(w + c_1)_a{}^b s(w + c_2)_c{}^d$ for complex constants $c_1$ and $c_2$ other than $i\hbar/2$ 
and $-i\hbar/2$. But since the ansatz appears to fail already at the formal 
level unless $\Im c_1 = -\Im c_2 = \hbar/2$ this extension does not seem really necessary.)

The main open question in the construction of $\A$ is whether the various restrictions that must be 
placed on the quantum states in order to make possible the extension of $\A_0$ to $\A_2$ leave a 
sufficiently rich set of states so that the entire classical Geroch group can be recovered in a 
suitable classical limit.

We will call the algebra $\A$ the {\em complexified algebra of observables} of the Geroch group, 
or the ``algebra of observables'' for short. Only the $*$ real elements of $\A$ can be observables. 
Recall that we have proposed that the quantization of a polynomial in the components 
$s(w)_a{}^b$ of the classical Geroch group element $s$ be the symmetrized polynomial in 
the corresponding quantum magnitudes $s(w)_a{}^b$ with the same coefficients. The classical 
components $s(w)_a{}^b$ are defined only on $w \in \R$ so only the quantum components 
$s(w)_a{}^b$ for real $w$ will appear in these quantized polynomials. If the classical 
polynomial is real then its coefficients are real and its quantization is $*$ real. This
follows from the fact that $[s(w)_a{}^b]^∗ = s(w)_a{}^b$ on the real axis, and the fact 
that the reordering of products that applying $∗$ entails does not affect the symmetrized 
products. The quantization of a real classical polynomial therefore can be an observable. 
Note, however, that $∗$ reality of an element $a ∈ \A$ only guarantees that any GNS Hilbert 
space representation of $a$ is symmetric, but not that it is essentially self-adjoint 
\cite{Khavkine_Morreti}.

These quantized real polynomials do not exhaust all $∗$ real elements of $\A$, not even $∗$ real
elements having real polynomials as classical limits. The quantum determinant (\ref{qdets_def}) is an
example. Any of these $∗$ real elements are potential observables. Our quantization of the real
polynomials provides one, standard, quantization of each one, but there is no reason that the
quantum observable that is the physical or mathematical analogon in the quantum theory of
a given classical polynomial observable be this standard quantization. It can differ by a term
that vanishes in the classical limit.

One might ask whether there can exist procedures for measuring the observables of the Geroch
group. Here we are using the word ``observable'' first and foremost as mathematical terminology.
The $*$ real mangitudes in $\A$ are observables if the Geroch group is treated as though it were
a physical system. But there is also a sense in which the Geroch group is measurable, just as
the translation group is measurable in a mechanical system, through its generator the linear momentum. 
The action of the Geroch group on the cylindrically symmetric gravitational field defines a 
representation of the Geroch group on the algebra of gravitational observables, and since this 
action is classically Lie-Poisson gravitational observables that generate this action can be 
constructed \cite{Babelon_Bernard}, indeed have been constructed at the classical level \cite{KS}
and partly at the quantum level \cite{Samtleben_thesis}. These generators are presumably measurable.

As has already been mentioned, there is another way to define $s(w)$, borrowed from the $\mathfrak{sl}_2$ 
Yangian, namely to define $s(w)$ to be a formal power series in $1/w$ with order zero term $\One$. 
However, while a series in $1/w$ is natural for the quantization of a loop group element defined on a circle, 
it is not so natural for the quantization of an element of a line group, like the Geroch group: Consider the 
$SL(2,\C)$ loop group defined on the unit circle $|w| = 1$ in the complex $w$ plane, and equipped with the 
Sklyanin bracket {\ref{gerochbracket}). The elements $g(w)$ of this group are parametrized by the coefficients 
of their expansion in Fourier modes in the angle around the circle, that is, by the coefficients in their 
expansions in (positive and negative) powers of $w$. In a natural quantization the coefficients are promoted 
to quantum magnitudes that together define a power series $g(w)$ in $w$ which is a sort of quantum field on 
the circle. Furthermore the exchange relation (\ref{q_Geroch_exchange}) and the quantum determinant condition 
(\ref{qdets_is_1}), both with $\hbar = 1$, are imposed on the quantum group element $g$, quantizing the Sklyanin 
bracket and the unit determinant condition of $SL(2,\C)$ respectively. If, finally, the coefficents of strictly 
positive powers of $w$ are set to zero and the constant term is set to $\One$ then the $\mathfrak{sl}_2$ Yangian 
is obtained \cite{Molev}.

The expectation value of $g(w)$ and its products in a given state should yield a numerical value, at least 
when smeared with a test function, and at least on the unit circle. But convergence is not needed to define 
the algebra generated by $g$. This can be done at the level of formal power series. The product of formal 
power series are well defined formal power series, because each coefficient of the product series is a 
polynomial function of only a finite subset of the coefficients of the factor series. In fact, all the 
operations necessary to define the Yangian algebra are well defined at the level of formal power series in 
$1/w$, including the shift $w \mapsto w + c$ of the argument $w$ by a constant $c$. Indeed, the Yangian 
algebra is often defined in terms of such series \cite{Molev}.

The natural generalization of the Fourier series expansion of the quantized loop group element $g$ to a line 
group is the inverse Fourier-Laplace transform (\ref{inv_Fourier_Laplace}). The role of the Fourier coefficients
of the loop group element $g$ is played by the Fourier transform $\tilde{s}(k)$ of the line group element $s$.
In terms of $\tilde{s}$ the reality condition is $\tilde{s}(k)^* = -\tilde{s}(-k)$, so, prior to the 
imposition of the unit determinant condition, $\tilde{s}(k)$ for $k \geq 0$, like the Fourier 
coefficients of $g$, may be regarded as free data determining the group element.

Nevertheless, postulating that $s(w)$ be a formal 
series of the form $\One + a_1/w + a_2/w^2 + ...$, even though the Geroch group is not a loop group, does 
solve some problems. The product (\ref{shifted_product}) is automatically well defined, at least as a 
formal series (whether its expectation value converges is another question), 
and $s$ automatically satisfies the condition 
$s(w) \rightarrow \One$ as $w \rightarrow \infty$ which ensures that the action of the Geroch group preserves 
a form of asymptotic flatness. It also seems to bring serious problems. If the expectation value of $s(w)$ is 
taken to be the sum of the expectation values of the terms of the series in the conventional sense, i.e. the 
limit of the partial sums, then this imposes strong analyticity conditions on $\langle s(w)\rangle$ which do 
not seem to have any physical motivation. In particular, it must have a singularity at $w = 0$ or else be 
constant. On the other hand, these limitations can be at least partly overcome by suitably generalizing the 
definition of the sum of the power series in the expectation value, for instance by using Mittag-Leffler 
summation \cite{MittagLeffler}. In fact, the model of $s$ as a formal power series with such a generalized 
notion of summation may well be equivalent to the model in terms of tempered distributions that we have outlined. 
The possibility of realizing $s$ as a formal series is therefore not excluded,
but is less natural for the Geroch group than the model based on tempered distributions. 
The quantization of line groups in terms of tempered distributions has not been 
studied before, as far as the authors know, and merits further exploration.

Have we suceeded in giving $\langle s(w)\rangle$ the correct number of degrees of freedom near the classical 
limit? If either of the definitions of $\langle s(w)\rangle$ at complex $w$ that we have considered here is 
adopted then, prior to the imposition of the reality and unit quantum determinant conditions, the independent 
degrees of freedom of the expectation value of $s$ are 4 complex components at each point on the real $w$ axis. 
The reality condition then reduces this to 4 real components. 
The expectation value of the quantum determinant is not a function of the expectation value of $s$, so in 
general it can be varied independently of the components of $\langle s \rangle$ by suitably varying the state. 
However, in the classical limit this is not so. In this limit, as $\hbar$ tends to zero the state $\varphi$ in 
which the expectation value is taken tends to a classical state, which is an algebra homorphism 
$\A \rightarrow \C$, so that $\langle \overset{1}{s}(w_1) \overset{2}{s}(w_2)\rangle = \langle \overset{1}{s}(w_1) \rangle\langle \overset{2}{s}(w_2) \rangle$ in this limit.
Thus, to order zero in $\hbar$, $\langle{\rm qdet} s(w)\rangle$ is just the classical determinant of 
$\langle s(w)\rangle$. Therefore, imposing the unit quantum determinant condition fixes one degree of freedom
of $\langle s(w)\rangle$ in terms of the other three to order zero in $\hbar$. The trace, for instance, is a 
function of the three degrees of freedom of the trace free part of $\langle s(w)\rangle$, plus corrections of 
first and higher orders in $\hbar$. These considerations indicate that the constraints on the classical limit 
of our quantization of the Geroch group are sufficient to restrict the independent degrees of freedom of 
$\langle s(w)\rangle$ to three real numbers at each real value of $w$ which match the degrees of freedom of 
$s(w)$ of the classical Geroch group.

This argument leaves open the possibility that the classical limit of our model is {\em more} constrained than
the classical theory. In fact this does not occur, because the conditions that define the quantum model, 
that is, the exchange relations (\ref{q_Geroch_exchange}), the unit quantum determinant condition
(\ref{qdets_is_1}), and the reality condition (\ref{reality1}), are compatible in a certain sense. This 
compatibility is rather delicate, and fixes the form of these conditions to a great extent.

Consider first the exchange relation (\ref{q_Geroch_exchange}) by itself. Using this relation a product, 
$s(w_1)_{a_1}{}^{b_1} ... s(w_n)_{a_n}{}^{b_n}$, of matrix components of $s$ can be reordered by sucessive 
exchanges of neighbouring factors, that is, it can be expressed as a linear combination of products in all 
of which the arguments $w_1, ... , w_n$ appear in the same permuted order $w_\sg(1), ... , w_\sg(n)$. If there 
are three or more factors then a given final order can be obtained by several distinct sequences of exchanges. 
It is important that all these ways to reach the same final order yield exactly the same result, for 
otherwise there would exist a linear combination of products, all in the same order, which vanishes. Since 
ordered polynomials may be expressed as symmetrized polynomials this would imply that a non-trivial symmetrized 
polynomial vanishes. Dividing this polynomial by the lowest power of $\hbar$ found in the coefficients gives a 
polynomial that is non-trivial in the classical limit and yet vanishes on all states, that is, a constraint on 
the phase space of the classical limit which does not exist in the classical theory that we are quantizing.

Such spurious constraints do not arise in our quantization because the $R$ matrix in the exchange 
relation satisfies the quantum Yang-Baxter equation
\begin{equation}
 \overset{12}{R}(w_1 - w_2)\overset{13}{R}(w_1 - w_3)\overset{23}{R}(w_2 - w_3) 
 = \overset{23}{R}(w_2 - w_3)\overset{13}{R}(w_1 - w_3)\overset{12}{R}(w_1 - w_2).
\end{equation}
(The overset numbers indicate the pair of spaces on which $R$ acts in each case.) This is sufficient 
to ensure that the result of a sequence of interchanges depends only on the final order.
Thus, the algebra $\A_0$ generated by the components $s(w)_a{}^b$ on the real $w$ axis (and the unit element $1$) 
subject to the exchange relations is the algebra of symmetrized polynomials in the components 
of $s(w)$, with a product which is abelian to order zero in $\hbar$, as can be seen from (\ref{explicit_exchange2}). 
More precisely, the product of $\A_0$ is the symmetrized product plus order $\hbar$ and higher corrections (as in
(\ref{symmetrized_ss_product})), where the symmetrized product of two monomials in components $s$ is simply the product 
of all the factors $s(w_i)_{a_i}{}^{b_i}$ of the two monomials averaged over all permutations of the ordering of 
these factors. 

The formal classical limit of this algebra, obtained by letting $\hbar$ go to zero in the coefficients of the 
polynomials and in the definition of the product so that the algebra product reduces to the symmetrized product, is 
the algebra of polynomials in the components of the classical line group element, subject to no constraints. 
The Poisson bracket of this formal classical limit is the $\hbar \rightarrow 0$ limit of the commutator divided by 
$i\hbar$, and reproduces exactly the Sklyanin bracket (\ref{gerochbracket}). The possible $\C$ valued functions 
$s_a{}^b(w)$, which are the values of the elements of the classical limit algebra, are obtained from these algebra
elements by evaluating them on classical states, which are homomorphisms from the classical limit algebra to $\C$.

The formal classical limit is not necessarily the same thing as a physical classical limit, which is a sector of the full
quantum theory, that is, a subset of observables with a corresponding subset of states, that is described by a classical
theory. In the case of the Geroch group we expect the long wavelength sector, consisting of the modes of $s$ with 
wavelength in $w$ much greater than $\hbar$, to be a classical sector in which the quantum Geroch group
reduces to the classical Geroch group. But this issue will be left to future research. Here we will be content to sketch 
an argument showing that the formal classical limit reproduces the classical Geroch group without additional constraints. 

The algebra $\A_0$ also carries a $*$ operation defined by the reality condition (\ref{reality_real_axis}), which 
is compatible with the exchange relation. Taking the $*$ conjugate of the exchange relation (\ref{q_Geroch_exchange}) 
and applying (\ref{reality1}) one obtains the exchange relation (\ref{q_Geroch_exchange2}). Two distinct exchange 
relations would together imply that a non-trivial linear combination of the components of 
$s(w_1)_{a_1}{}^{b_1} s(w_2)_{a_2}{}^{b_2}$ vanishes, leading to a spurious constraint in the classical limit. But 
in fact (\ref{q_Geroch_exchange2}) is equivalent to (\ref{q_Geroch_exchange}), so this does not occur. The equivalence 
of (\ref{q_Geroch_exchange2}) with (\ref{q_Geroch_exchange}) means that if two polynomials are identified
via the exchange relation (\ref{q_Geroch_exchange}) then their $*$ conjugates are also. In other words,
(\ref{reality1}) defines a $*$ operation on the equivalence classes of polynomials which are the elements of $\A_0$.
The consequence of the reality condition (\ref{reality_real_axis}) in the formal classical limit is that
the matrix elements $s_a{}^b(w)$ are real. 

Extending $s(w)$ from the real line by analytic continuation to a strip about the real line in the complex $w$ 
plane extends the algebra $\A_0$ to $\A_1$, and adding the products (\ref{shifted_product}) to the algebra extends 
$\A_1$ to $\A_2$. The compatibility of the exchange relations among themselves, and with the reality condition persists 
in this case. The minimal width of the strip in the complex plane needed to for our formulation of the quantum Geroch 
group is $\hbar$. If this or any width proportional to $\hbar$ is used the extension does not affect the classical 
limit. That is, the classical limit of $\A_2$, like that of $\A_0$, is the commutative algebra of polynomials in the 
components $s(w)_a{}^b \in \R$ of the classical line group element along the real $w$ axis. To reach this conclusion 
we have used the fact that $s(w + i\eta) = e^{i\eta\, d/dw} s(w)$ according to our definition of the extension of $s$ 
off the real axis. 

Now consider the unit quantum determinant condition (\ref{qdets_is_1}). The classical Geroch group, 
the $SL(2,\R)$ line group, is the surface in the space $\mathfrak M_2$ smooth $2 \times 2$ matrix 
valued functions $s:\R \rightarrow \R^{2\times 2}$, defined by the conditions that ${\rm det}\, s = 1$. 
The classical phase space functions that we are quantizing are polynomials in $s(w)_a{}^b$ 
on this constraint surface. Equivalently, they are polynomials on $\mathfrak M_2$, modulo the 
ideal formed by polynomials of the form $({\rm det} s - 1)p$, with $p$ a polynomial on $\mathfrak M_2$.

In the quantum theory imposition of the constraint (\ref{qdets_is_1}) reduces the algebra $\A_2$
of symmetrized polynomials in the quantum matrix components $s(w)_a{}^b$ to the quotient 
of this algebra by the two sided ideal $\cal I$ generated by ${\rm qdet} s(w) - 1$. But, by the exchange relation
(\ref{q_Geroch_exchange}), ${\rm qdet} s(w)$ commutes with all matrix elements $s(v)_a{}^b$ 
(\cite{Samtleben_thesis} section 5.1, \cite{Molev} sections 2.6 and 2.7). The two sided ideal thus 
consists entirely of elements of the form $({\rm qdet} s - 1)p$ (expressed as a 
symmetrized polynomial), where $p$ again is an arbitrary symmetrized polynomial. 

In the formal classical limit ${\rm qdet} s(w)$ reduces to ${\rm det} s(w)$. The two sided ideal 
generated by ${\rm qdet} s(w) - 1$ thus becomes precisely the ideal that is factored 
out in the classical theory, and the set of classical limits of quantum observables is precisely the set of 
classical observables.

Finally, one has to check that the $*$ operation is compatible with the quantum unit determinant condition 
(\ref{qdets_is_1}). In principle the $*$ conjugate of (\ref{qdets_is_1}) could be a new, independent condition,
and even if (\ref{qdets_is_1}) and its $*$ conjugate were classicaly equivalent, differing only to higher order
in $\hbar$, the strict imposition of both conditions would imply a spurious constraint on the classical limit.
But actually there are no further constraints on the classical limit because the two conditions are equivalent: 
By (\ref{reality1}) the $*$ conjugate of the quantum determinant (\ref{qdets_def}) is 
\begin{equation}\label{qdets_alt}
 [{\rm qdet} s(w)]^* = s(\bar{w}+i\hbar/2)_2{}^2 s(\bar{w} - i\hbar/2)_1{}^1 
 - s(\bar{w}+i\hbar/2)_1{}^2 s(\bar{w} - i\hbar/2)_2{}^1. 
\end{equation}
But the restriction (\ref{ihbar_dif_exchange}) of the exchange relation implies that 
$\vareg^{bd} s(\bar{w} + i\hbar/2)_b{}^e s(\bar{w} - i\hbar/2)_d{}^f$ is antisymmetric in the indices $e, f$, 
and hence equal to ${\rm qdet} s(\bar{w}) \vareg^{ef}$. Therefore  
$[{\rm qdet} s(w)]^* = - {\rm qdet} s(\bar{w})\, \vareg^{21} = {\rm qdet} s(\bar{w})$. 
It follows that the condition ${\rm qdet} s(w) = 1$ on a strip symmetric about the real axis in the complex $w$ 
plane is equivalent to the condition $[{\rm qdet} s(w)]^* = 1$ on the same strip. 

Note also that this implies that the ideal $\cal I$ is $*$ invariant and therefore $*$ acts on the equivalence 
classes that are the elements of the quotient $\A = \A_2/{\cal I}$, the algebra of observables. $*$ thus turns
the full algebra of observables $\A$, without constraints, into a $*$ algebra.

We turn now to the question of whether and how this quantum Geroch group is a symmetry of quantum 
cylindrically symmetric vacuum general relativity.

\subsection{The quantum Geroch group as a symmetry of Korotkin and Samtleben's quantization of cylindrically symmetric vacuum gravity}\label{q_Geroch_as_symmetry}

In \cite{KS} Korotkin and Samtleben proposed a quantization of the phase space $\Gamma_\infty$ of 
asymptotically flat cylindrically symmetric vacuum general relativity. They express their model in 
terms of a factorization of $\M$ on the real $w$ line:
\begin{equation}\label{M_factorization}
 \M(w)_{ab} = T_+(w)_a{}^c T_-(w)_b{}^d e_{\infty\,cd}, 
\end{equation}
with $T_+$ defined on the upper half $w$ Riemann sphere, $T_+(\infty) = \One$, and 
$T_-(w)_b{}^d = [T_+(\bar{w})_b{}^d]^*$. $e_\infty$ is the asymptotic conformal metric in spacetime and the 
asymptotic value of $\M$ as $w \rightarrow \pm\infty$. The condition at $w = \infty$ is not discussed in \cite{KS}, 
it is added here because it is necessary for asymptotic flatness. 

$T_+$ and $T_-$ are the quantizations of complex $2\times 2$ matrix valued functions of $w$, also called $T_\pm$, which factorize the classical deformed metric $\M$ according to the same relation (\ref{M_factorization}). The classical $T_+$ has unit determinant, is holomorphic in the open upper half $w$ Riemann sphere and continuous at its boundary, the compactified real axis, and is equal to $\One$ at $w = \infty$. $T_-(w)_b{}^d$ is $\overline{T_+(\bar{w})_b{}^d}$. Such a factorization exist provided $\M$ is a Wiener function of $w$, that is, a function with absolutely integrable Fourier transform plus a constant which is its asymptotic value. 
This follows from theorem 8.2. of \cite{Gohberg_Krein}, and an argument using Lemma \ref{analyticity} to demonstrate 
$\det T_+ = 1$, because $\M$ is symmetric, positive definite and of unit determinant. Recall that
restricting $\M$ to Wiener functions is a natural definition of asymptotic flatness in cylindrically symmetric vacuum gravity, as was discussed in subsection \ref{asymptotic_flatness}. When this hypothesys holds the factor $T_+$ is also a Wiener function, and is uniquely determined by $\M$ among continuous functions up to right multiplication by a constant matrix that is unitary in orthonormal bases of $e_\infty$. 

The Poisson brackets of $T_+$ and $T_-$ can be extracted from (\ref{Monbracket}) by multiplying 
this equation on the left by $\overset{1}{T}{}^{-1}_+(v) \overset{2}{T}{}^{-1}_+(w)$ and on the 
right by $\overset{1}{T}{}^{-1}_-(v) \overset{2}{T}{}^{-1}_-(w)$, and then equating the 
positive(negative) frequency components in $v$ and in $w$ on both sides of the resulting equation. 
The result is 
\begin{eqnarray}
 \{\overset{1}{T}_\pm(v), \overset{2}{T}_\pm(w)\} & = & \frac{1}{v - w}[\Omega, \overset{1}{T}_\pm(v) \overset{2}{T}_\pm(w)],\label{T+T+bracket}\\
 \{\overset{1}{T}_\pm(v), \overset{2}{T}_\mp(w)\} & = & \lim_{\epsilon \rightarrow 0^+}\frac{1}{v - w \pm i \epsilon}\left(\Omega\, \overset{1}{T}_\pm(v) \overset{2}{T}_\mp(w) + \overset{1}{T}_\pm(v) \overset{2}{T}_\mp(w)\, \Omega^t\right),\label{T+T-bracket}
\end{eqnarray}
where the limit is taken in the sense of distributions.\footnote{
Actually Korotkin and Samtleben 
\cite{KS} define $T_\pm$ directly in terms of the geometry of the asymptotically flat spacetimes, 
obtain their Poisson brackets from the action (\ref{action}), and then show that $\M$ defined by 
(\ref{M_factorization}) has the Poisson bracket (\ref{Monbracket}).}

Korotkin and Samtleben postulate the exchange relations 
\begin{equation}\label{T+T+exchange}
R(v-w) \overset{1}{T_\pm}(v) \overset{2}{T_\pm}(w) = \overset{2}{T_\pm}(w) \overset{1}{T_\pm}(v) R(v-w),
\end{equation}
which quantize (\ref{T+T+bracket}) in precisely the same way that (\ref{q_Geroch_exchange}) quantizes 
(\ref{gerochbracket}) ($R$ is defined as in (\ref{q_Geroch_exchange})), and the exchange relation
\begin{equation}\label{T+T-exchange}
\fl R(v-w-i\hbar) \overset{1}{T_-}(v) \overset{2}{T_+}(w) = \overset{2}{T_+}(w) \overset{1}{T_-}(v) 
R'(v-w+i\hbar)\frac{(v-w)(v-w-2i\hbar)}{(v-w-i\hbar)(v-w+i\hbar)}
\end{equation}
between $T_+$ and $T_-$, which quantizes (\ref{T+T-bracket}). $R'$ is the 
``twisted R matrix'', defined by 
\begin{equation}\label{R'}
 R'(u) = (u - i \hbar /2)I + i \hbar {}^t\Omega,
\end{equation}
or equivalently
\beq
R'(u)^a{}_{bc}{}^d = (u-i \hbar)\delta_b^a \delta_c^d + i\hbar \delta_c^a\delta_b^d.
\eeq

These exchange relations imply that the quantum deformed metric $\M$, defined as the product 
(\ref{M_factorization}) of $T_+$ and $T_-$, satisfies the closed exchange relation
\begin{equation}\label{RMR'M}
\fl R(v-w) \overset{1}{\M}(v) R'(w-v+2i\hbar) \overset{2}{\M}(w) 
= \overset{2}{\M}(w) {}^t\! R^{'t} (v-w+2i\hbar) \overset{1}{\M}(v) {}^t\! R^{t}(w-v) \frac{v-w-2i\hbar}{v-w+2i\hbar}. 
\end{equation}
This is easily seen to be a quantization of the Poisson bracket (\ref{Monbracket}).

In the model of Korotkin and Samtleben $\M$ satisfies, in addition to the exchange relation 
(\ref{RMR'M}), quantum versions of all but one of the conditions that define the set of classical 
deformed metrics corresponding to solutions in $\Gamma_\infty$, namely the reality, symmetry, 
positive definiteness, and unit determinant conditions. Only the asymptotic flatness condition is 
not implemented in their quantum model. 
(The requierment that the fields in the classical phase space be smooth is imposed for convencience 
and, as is usual in quantization, is not imposed on the corresponding quantum fields or expectation 
values.)

The quantum reality and symmetry conditions are straightforward: $[\M(w)_{ab}]^* = \M(w)_{ab}$ and 
$\M_{ab} = \M_{ba}$ respectively, just as for the classical $\M$. (Note that $\M$ is defined only 
on real $w$, even in the quantum theory.) The unit determinant condition is quantized by requiring 
that $T_+$ has unit quantum determinant: 
\begin{equation}\label{qdetT+}
 1 = {\rm qdet}\, T_+ = T_+(w+i\hbar)_1{}^1 T_+(w)_2{}^2 - T_+(w+i\hbar)_2{}^1 
 T_+(w)_1{}^2.
\end{equation}
In the classical limit this ensures that ${\rm det} \M = \det T_+ ({\rm det} T_+)^* = 1$.\footnote{
According to the definition (\ref{qdets_def}) of the quantum determinant used in (\ref{qdets_is_1}) this is
the quantum determinant of $T_+$ evaluated at $w + i\hbar/2$, and comparison with \cite{KS} and 
\cite{Samtleben_thesis} eq. 5.23 shows that $w$ in (\ref{qdetT+}) is also shifted by $i\hbar$ with 
respect to the definition of ${\rm qdet} T_+$ used by Korotkin and Samtleben. This difference in the 
definitions does not affect the condition imposed since both in \cite{KS} and here the unit quantum 
determinant condition is imposed at all values of $w$ at which the quantum determinant is defined. 
The definition used in (\ref{qdetT+}) is adopted here so that 
${\rm qdet} T_+(w)$ is defined on the real $w$ axis and the $w$ upper half plane when $T_+$ is.}
The quantum positivity condition requires that the expectation value of $\M_{ab}r^a r^b$ for any real 
2-vector $r$ be positive semi-definite in any state $\varphi$ on the $*$-algebra generated by $T_+$ and 
$T_-$. This condition is a consequence of the existence of the factorization (\ref{M_factorization}):
$\M_{ab}r^a r^b = \sum_c r^a T(w)_{+\,a}{}^c [r^b T(w)_{+\,b}{}^d]^* e_{\infty\,cd}$, which, because 
of the positivity of states, has expectation value $\varphi(\M_{ab}r^a r^b) \geq 0$. This guarantees 
that $\M$ is positive definite in the classical limit because in this limit the condition 
${\rm det}\M = 1$ excludes the possibility $\M_{ab}r^a r^b = 0$.

The asymptotic flatness condition could be incorporated in the model of Korotkin and Samtleben by 
postulating that $T_+$ is a series in non-negative powers of $1/w$, with order zero term $\One$, 
like the generating function of a Yangian. However, Korotkin and Samtleben do not implement asymptotic 
flatness in their quantum model in this way, or any other. In his thesis Samtleben does present 
a power series in $1/w$ which naively converges to $T_+$ in the classical theory, (see eq. 3.82 
of \cite{Samtleben_thesis}) but he points out that it does not do so in general. It seems to us 
that if suitable falloff conditions on the geometry hold at spatial infinity then this series 
might be an asymptotic expansion for $T_+$. Later (p. 61 \cite{Samtleben_thesis}) he rejects a 
fundamental role for an expansion of $T_+$ as a power series in $1/w$ because it is not consistent 
with the analytic structure of $T_+$ in the classical theory, which does not in general allow an 
analytic continuation to the lower half plane. 
(Of course this does not exclude the possibility of an asymptotic series in $1/w$ for $T_+$.)

Asymptotic flatness could also be implemented in the quantum theory by requiering that
the expectation value $\langle T_+\rangle$ of $T_+$ be a Wiener function with constant term $\One$, 
or that $\langle \M \rangle$ be a Wiener function with constant term $e_\infty$. In the classical 
limit these last two conditions are equivalent and correspond to a definition of asymptotic flatness 
discussed in subsection \ref{asymptotic_flatness}. 
The exploration of this proposal will be left to future work.

The question now arises as to whether the action $\M \mapsto \M^G = s\M s^t$ of the classical Geroch 
group can be quantized in such a way that it is an automorphism of the quantum theory of 
cylindrically symmetric gravitational waves of Korotkin and Samtleben. Here we will find a particular 
quantization of the action which appears to satisfy this requierment. To the extent that we are able to check, 
it preserves the structure of Korotkin and Samtleben's theory. Specifically, we show that the action
(\ref{q_Geroch_action}), which is just the classical action with a small ``quantum'' modification, does
indeed preserve the exchange relations (\ref{RMR'M}), as well as the symmetry, reality, and positive 
semi-definiteness of $\M$. We are not be able to check whether the unit determinant conditions is 
preserved, because we have not found an expression for an action of the Geroch group on $T_\pm$ 
corresponding to the action (\ref{q_Geroch_action}) on $\M$. Indeed, we have not been able to verify 
that the action even preserves the factorizabilty of $\M$ into a product of the form (\ref{M_factorization}). 
Regarding asymptotic flatness, we have not checked that the action preserves asymptotic flatness in the sense 
that the expectation value of $\M$ is a Wiener function. On the other hand, asymptotic flatness is preserved 
if one adoptes the, in our opinion less natural, definition that $\M$ is asymptotically flat 
if it is a formal series in $1/w$ with order zero term $e_\infty$, and $s$ is restricted to formal series
with order zero term $\One$.

The difficulty of verifying that the quantum unit determinant condition in terms of $T_+$ and $T_-$ is 
preserved is one reason that it would be interesting to formulate a unit determinant condition directly in
terms of $\M$. Another reason is that such a formulation ought to make it possible to express the quantum 
theory of cylindrically symmetric gravitational waves entirely in terms of $\M$, instead of $T_+$ and $T_-$, 
bringing it into closer correspondence with the classical theory developed in section \ref{cyl_grav}.
It might also allow the theory to be extended to non asymptotically flat fields, which could be useful for 
the quantization of null initial data for vacuum GR in the absence of cylindrical symmetry \cite{ReiAnd}. 
But this project will be left to future investigations.

Instead of postulating from the outset an action of the quantum Geroch group on $\M$ and showing that 
it preserves the symmetry, reality, and positive semi-definiteness of the deformed metric, $\M$, and its
exchange relation, we will make an ansatz for the form of this action, and then show that the action
(\ref{q_Geroch_action}) is essentially the only one among those of the ansatz that preserves these 
properties of $\M$.  
The ansatz for the action is 
\begin{equation}\label{ansatzcuantico}
\M(w)_{ab} \mapsto \M^G(w)_{ab} = s(w + c_1)_a{}^c \M(w)_{cd}\,s(w+c_2)_b{}^d, 
\end{equation}
where $c_1$ and $c_2$ are complex constants to be determined. In addition, since the matrix elements of $s$ 
Poisson commute classically with those of $\M$ it will be assumed that they commute in the quantum theory.

We will find below, in Proposition \ref{preservation_of_exchange_relations}, that this action preserves the exchange 
relation (\ref{RMR'M}) if $c_1 - c_2 = i\hbar$.
The condition $c_1 - c_2 = i\hbar$ also implies that the action (\ref{ansatzcuantico}) preserves the symmetry 
of $\M$: 
Recall that the exchange relations (\ref{q_Geroch_exchange}) reduces to (\ref{ihbar_dif_exchange}) when the two 
spectral parameters differ by $i\hbar$, and that this implies that
$\vareg^{bd} s(v + i\hbar/2)_b{}^e s(v - i\hbar/2)_d{}^f$ is antisymmetric in the indices $e, f$, for any $v$.
This is true in particular if $v = w + c$ with $c = (c_1 + c_2)/2$, so if $\M(w)_{ab}$ is symmetric in $a,b$ then 
$\M^G(w)_{ab}$ is also, because $s(v + i\hbar/2)_{[a}{}^c \M(w)_{cd}s(v-i\hbar/2)_{b]}{}^d$ vanishes.

The action of the Geroch group should also preserve the reality of $\M(w)$ on real $w$. 
If $c_1 - c_2 = i\hbar$ the reality of $\M^G(w)_{ab}$ requires that 
\begin{equation}\label{condition_on_s_ensuring_reality_of_MG_v1}
 \fl s(v + i\hbar/2)_a{}^c \M(w)_{cd}\,s(v-i\hbar/2)_b{}^d = [s(v-i\hbar/2)_b{}^d]^* [\M(w)_{cd}]^* [s(v + i\hbar/2)_a{}^c]^*.
\end{equation}
By the reality conditions on $\M$ and $s$, and the symmetry of both $\M$ and $\M^G$ this reduces to
\begin{equation}\label{condition_on_s_ensuring_reality_of_MG_v2}
 \fl s(v + i\hbar/2)_a{}^c \M(w)_{cd}\,s(v-i\hbar/2)_b{}^d = s(\bar{v} + i\hbar/2)_a{}^c 
 \M(w)_{cd}\,s(\bar{v}-i\hbar/2)_b{}^d,
\end{equation}
which holds identically if $c$ is real, since then $v = w + c$ is real. 

If, on the contrary, $c$ is not real (\ref{condition_on_s_ensuring_reality_of_MG_v1}) requires 
constraints on $\M$ and/or $s$ which are absent in the classical theory, and are thus not fulfilled
by the quantum theory if it matches the classical theory in the classical limit: Condition 
(\ref{condition_on_s_ensuring_reality_of_MG_v1}) is equivalent to the requierment that 
\begin{equation}\label{condition_on_s_ensuring_reality_of_MG_v3}
 (F(v)_a{}^c{}_b{}^d - [F(v)_a{}^c{}_b{}^d]^*) \M(w)_{cd} = 0 
\end{equation}
where
\begin{equation}
 F(v)_a{}^c{}_b{}^d \equiv s(v + i\hbar/2)_a{}^{(c}\,s(v-i\hbar/2)_b{}^{d)} 
\end{equation}
If $c$ is real then, by the same arguments that led to (\ref{condition_on_s_ensuring_reality_of_MG_v2}), 
the components of $F$ are $*$ real, so 
(\ref{condition_on_s_ensuring_reality_of_MG_v3}) holds. If $c$ is not real and the components of $F$ are 
not $*$ real then (\ref{condition_on_s_ensuring_reality_of_MG_v3}) becomes a non-trivial constraint on $\M$. 
Dividing this constriant by the lowest power of $\hbar$ that appears in it defines a constraint which is 
non-trivial in the classical limit. If $c$ is not real, can the components of $F$ be $*$ real anyway? It 
turns out that it can be, but only if $s$ is so constrained that it makes no difference whether $\Im c$ 
is zero or not. Let us assume that $F$ is a tempered distribution. Since the components of $F(v)$ are $*$ 
real when $v$ is real $\tilde{F}$, the Fourier transform of $F$, satisfies 
\begin{equation}
 \tilde{F}(-k)_a{}^c{}_b{}^d = [\tilde{F}(k)_a{}^c{}_b{}^d]^* 
\end{equation}
If we now suppose that the components of $F$ are also $*$ real for
some non-zero constant value of $\Im v = \Im c = \eta$ then $\tilde{F}$ must satisfy the additional equation 
\begin{equation}
 e^{\eta k}\tilde{F}(-k)_a{}^c{}_b{}^d = e^{-\eta k}[\tilde{F}(k)_a{}^c{}_b{}^d]^*. 
\end{equation}
From the two equations it follows that $\tilde{F}(k)$ is proportional to
$\dg(k)$, that is, that $F(v)$ is a constant, independent of $v$. But in this case one might as well
take $c$ to be real, since its value has no effect on the action (\ref{ansatzcuantico}) of the 
Geroch group on $\M$. If the assumption that $F$ is a tempered distribution is dropped then there 
are non-constant solutions to $F(v)_a{}^c{}_b{}^d = [F(v)_ać^c{}_b{}^d]^*$ for 
$\eta = \Im v \neq 0$, such as $F \propto \exp(n\pi v/\eta)$ with $n$ a non-zero integer. But such
$F$ would seem to be excluded, at least near the classical limit, by the requierment that the action of the 
Geroch group preserves the asymptotic flatness of the geometry. In the classical theory preservation 
of asymptotic flatness requires that $s(w)$ for large $w$ approximate an orthogonal matrix of the inner 
product defined by the asymptotic value $e_\infty$ of $\M$, and thus that the components of $F$ be 
bounded as $|w| \rightarrow \infty$.
In sum, the preceding arguments strongly suggest that the action (\ref{ansatzcuantico}) only preserves the 
reality of $\M$ if $c$ is real.

Finally, $c_1 - c_2 = i\hbar$ and $c \in \R$ also ensures that the positive semi definiteness of $\M$ is preserved, because then for any real 2-vector $r$
\begin{equation}
\fl \M^G(w)_{ab}r^a r^b = \sum_d r^a s(w + c + i\hbar/2)_a{}^c T(w)_{+\,c}{}^d [r^b s(w + c + i\hbar/2)_b{}^c T(w)_{+\,c}{}^d]^*,
\end{equation}
which, by (\ref{state_positivity}), has positive semi-definite expectation value on any state. 

Note, however, that it is not obvious that actions of the form (\ref{ansatzcuantico}) preserve the factorizability of $\M$, that is, that $\M^G$ is a product of negative frecuency in $w$ factor, $T^G_+$, and its $*$ conjugate, $T^G_-$. In particular, $T^G_+(w)$ is not $s(w + c + i\hbar/2)T(w)_+$ because this latter expression is not purely negative frequency in $w$. On the other hand it is plausible that, in analogy with the classical theory, positive semidefiniteness, a quantum unit determinant condition on $\M$, and a quantum asymptotic flatness condition suffices to ensure the existence of such a factorization.

When $c$ is real it can actually be set to zero. The classical Geroch group is translation invariant in the sense that if the function $s(w)$ represents an element of the group, then the function $s(w + k)$ also represents an element for any constant $k\in \R$. This also applies to the quantum Geroch group to the extent that we have defined it: $s(w + k)$ satisfies the exchange relation, reality condition and unit determinant condition if $s(w)$ does. But replacing $s(w)$ by $s(w + k)$ everywhere of course does affect the expression for the action of the $s$ on $\M$, adding $k$ to the argument of each factor of $s$ in (\ref{ansatzcuantico}). 

If $c_1 - c_2 = i\hbar$ and $c = (c_1 + c_2)/2$ is real, and is set to cero by shifting the spectral parameter as we have just described, then $c_1 = i \hbar/2$ and $c_2 = -i\hbar/2$, and the action of the quantum Geroch group becomes
\begin{equation}\label{q_Geroch_action}
\M(w) \mapsto \M^G(w) = s(w + i \hbar/2) \M(w) s^t(w-i \hbar/2). 
\end{equation}
Clearly this reduces to the classical Geroch group action (\ref{Geroch_action2}) in the classical limit.

Our claim is that this action preserves the exchange relations of $\M$ as well as the symmetry, reality, and positive semi-definiteness of $\M$.
That it preserves symmetry, reality, and positive semi-definiteness has already been proved. We turn now to the proof of the invariance of the exchange relation (\ref{RMR'M}). We begin again with the ansatz
(\ref{ansatzcuantico}) and show that (\ref{RMR'M}) is preserved provided that 
$c_1 - c_2 = i\hbar$. It will only be proved that this last condition is sufficient, not that it is necessary, although the demonstration suggests that it is.

The following lemma, which restates the two equivalent forms (\ref{q_Geroch_exchange}) and (\ref{q_Geroch_exchange2}) of the
exchange relations for the $s$ in terms of the twisted $R$ matrix $R'$, will be useful. Note that in the following a presuperscript $t$, that is ${}^t\! X$, will once more denote transposition in space $1$, while a postsuperscript $t$ will denote transposition in space $2$. 

\begin{lemma}\label{sR'exchange_lemma}
For any $u, v , w \in \C$
\begin{equation}\label{g1R'g2}
{}^t\overset{1}{s}(v) R'(u) \overset{2}{s}(w) = \overset{2}{s}(w) R'(u) {}^t\overset{1}{s}(v) + 
(u + v - w - i\hbar) [ {}^t\overset{1}{s}(v), \overset{2}{s}(w)]
\end{equation}
and
\begin{equation}\label{g2R'g1}
\overset{1}{s}(v) {}^t\! R'^t (u) \overset{2}{s}{}^t(w) = \overset{2}{s}{}^t(w) {}^t\! R'^t(u) \overset{1}{s}(v) +
(u + w - v -i\hbar) [\overset{1}{s}(v),\overset{2}{s}{}^t(w)].
\end{equation}

\end{lemma}

\noindent {\em Proof}: By (\ref{q_Geroch_exchange}) 
\begin{equation}
R(v-w) \overset{1}{s}(v) \overset{2}{s}(w) - \overset{2}{s}(w) \overset{1}{s}(v) R(v-w) = 0,
\end{equation}
with $R(z) = z I - i \hbar(\Omega + I/2)$ for all $z \in \C$. Therefore
$R(z) = R(v - w) + (z - v + w)I$, and
\begin{equation}\label{lemma_eq1}
R(z) \overset{1}{s}(v) \overset{2}{s}(w) - \overset{2}{s}(w) \overset{1}{s}(v) R(z) 
= (z - v + w)[\overset{1}{s}(v), \overset{2}{s}(w)].
\end{equation}
But, by (\ref{R'}) $R'(u) = -{}^t\! R(-u + i\hbar)$. Thus, setting $z = -u + i\hbar$ and taking the transpose of 
(\ref{lemma_eq1}) in space $1$ one obtains equation (\ref{g1R'g2}).

To demonstrate (\ref{g2R'g1}) recall that the exchange relation (\ref{q_Geroch_exchange2}), which is equivalent to equation (\ref{q_Geroch_exchange}), is obtained from (\ref{q_Geroch_exchange}) by exchanging spaces $1$ and $2$, and the
spectral parameters $v$ and $w$. Thus 
\begin{equation}\label{lemma_eq2}
\overset{1}{s}(v) \overset{2}{s}(w) R(z) - R(z) \overset{2}{s}(w) \overset{1}{s}(v)  
= (z - w + v)[\overset{1}{s}(v), \overset{2}{s}(w)].
\end{equation}
Equation (\ref{g2R'g1}) is obtained from this equation by setting $z = -u + i\hbar$ and transposing in space $2$.
Note that $-R^t(-u + i\hbar) = {}^t\!R'^t(u)$.\hfill\QED

We are now ready to prove the invariance of (\ref{RMR'M}) under the action of the quantum Geroch group.

\begin{prop}\label{preservation_of_exchange_relations}
If $c_1 - c_2 = i\hbar$, then $\M^G$ as defined by (\ref{ansatzcuantico}) satisfies the same exchange relation (\ref{RMR'M})
as $\M$ does.
\end{prop}

\noindent{\em Proof}:
We wish to evaluate the effect of substituting $\M^G$ for $\M$ in the exchange relation (\ref{RMR'M}). To this end 
the left side, $L$, of (\ref{RMR'M}) with $\M^G$ in place of $\M$ will be rearranged until it takes the form of the right
side of (\ref{RMR'M}), with $\M^G$ again in place of $\M$, plus remainder terms which will be seen to vanish when 
$c_1 - c_2 = i\hbar$.

According to (\ref{ansatzcuantico})
\begin{eqnarray}
\fl L & \equiv & R(v-w) \overset{1}{\M}{}^G(v) R'(w-v+2i\hbar) \overset{2}{\M}{}^G(w) \\
\fl & = & R(v-w) \overset{1}{s}(v+c_1)\overset{1}{\M}(v) {}^t\overset{1}{s}(v+c_2) R'(w-v+2i\hbar) \overset{2}{s}(w+c_1) \overset{2}{\M}(w) \overset{2}{s}{}^t(w+c_2).
\end{eqnarray}
As a first step the order of the factors in the middle of the expression is reversed using Lemma \ref{sR'exchange_lemma}, putting $v + c_2$, $w + c_1$, and $w-v+2i\hbar$ in place of $v$, $w$, and $u$ respectively in equation (\ref{g1R'g2}):
\begin{eqnarray}
\fl L = R(v-w) \overset{1}{s}(v+c_1)\overset{1}{\M}(v) \overset{2}{s}(w + c_1) R'(w-v+2i\hbar) {}^t\overset{1}{s}(v + c_2)  \overset{2}{\M}(w) \overset{2}{s}{}^t(w+c_2) \nonumber\\
\fl\qquad + (c_2 - c_1 + i\hbar) R(v-w) \overset{1}{s}(v+c_1)\overset{1}{\M}(v) [ {}^t\overset{1}{s}(v + c_2),
 \overset{2}{s}(w + c_1)] \overset{2}{\M}(w) \overset{2}{s}{}^t(w+c_2).
\end{eqnarray}
Recall that the matrix elements of $s$ commute with those of $\M$. By exchanging such matrix elements the first line of the preceding expression can be written as
\begin{equation}
 \fl R(v-w) \overset{1}{s}(v+c_1) \overset{2}{s}(w+c_1) \overset{1}{\M}(v) R'(w-v+2i\hbar)   \overset{2}{\M}(w) {}^t \overset{1}{s}(v+c_2) \overset{2}{s}{}^t(w+c_2)
\end{equation}
Now the exchange relation (\ref{q_Geroch_exchange}) may be used to reorder the first three factors, and then (\ref{RMR'M}) 
to reorder the four middle factors, yielding
\begin{equation}
 \fl \overset{2}{s}(w+c_1) \overset{1}{s}(v+c_1) \overset{2}{\M}(w) {}^t\! R^{'t} (v-w+2i\hbar) \overset{1}{\M}(v) 
 {}^t\! R^t(w-v) {}^t\overset{1}{s}(v+c_2) \overset{2}{s}{}^t(w+c_2)\frac{v-w-2i\hbar}{v-w+2i\hbar}
\end{equation}
Next, the order of the last three factors is reversed using equation (\ref{q_Geroch_exchange2}) transposed in both space
$1$ and space $2$, and factors of $s$ and $\M$ are interchanged, with the result 
\begin{equation}
\fl \overset{2}{s}(w+c_1) \overset{2}{\M}(w) \overset{1}{s}(v+c_1) {}^t\! R^{'t} (v-w+2i\hbar) \overset{2}{s}{}^t(w+c_2) \overset{1}{\M}(v) {}^t\overset{1}{s}(v+c_2) {}^t\! R^t (w-v)\frac{v-w-2i\hbar}{v-w+2i\hbar}
\end{equation}
Finally, (\ref{g2R'g1}) is used to reverse the order of the middle factors again, producing also another commutator term:
\begin{eqnarray}
\fl L  = \overset{2}{s}(w+c_1) \overset{2}{\M}(w) \overset{2}{s}{}^t(w+c_2) {}^t\! R^{'t} (v-w+2i\hbar) \overset{1}{s}(v+c_1) \overset{1}{\M}(v) {}^t\overset{1}{s}(v+c_2) {}^t\! R^t(w-v) \frac{v-w-2i\hbar}{v-w+2i\hbar}\nonumber\\
\fl\qquad + (c_2 - c_1 + i\hbar)\left\{ R(v-w) \overset{1}{s}(v+c_1)\overset{1}{\M}(v) [{}^t\overset{1}{s}(v + c_2) ,\overset{2}{s}(w + c_1)] \overset{2}{\M}(w) \overset{2}{s}{}^t(w+c_2)\right. \nonumber\\
\fl\qquad + \left. \frac{v-w-2i\hbar}{v-w+2i\hbar}\,\overset{2}{s}(w+c_1) \overset{2}{\M}(w) [\overset{1}{s}(v + c_1),\overset{2}{s}{}^t(w + c_2)]  \overset{1}{\M}(v) {}^t\overset{1}{s}(v+c_2) {}^t\! R^t(w-v)\right\}.
\end{eqnarray}

If $c_2 - c_1 + i \hbar = 0$, then only the first line remains, so
\begin{equation}
\fl L =  \overset{2}{\M}{}^G(w) {}^t\! R^{'t} (v-w+2i\hbar) \overset{1}{\M}{}^G(v) {}^t\! R^t(w-v)\frac{v-w-2i\hbar}{v-w+2i\hbar},
\end{equation}
which is precisely the right side of (\ref{RMR'M}) with $\M^G$ in place of $\M$, proving the claim of the proposition.\hfill\QED

\section{Open problems and possible directions for further work}

The quantization of the Geroch group presented here is by no means complete. Many open problems, large and small, have been indicated 
in the text. Here we summarize the main ones. The first problem to resolve is to complete the verification that 
the action of the Geroch group is indeed an automorphism of the observable algebra of the gravitational theory. If this is not so
there would be little point to resolving the remaining issues. As a minimum one would have to demonstrate that the action
preserves the quantization of the unit determinant condition on the deformed metric $\M$. This could be done either by defining
an action of the quantum Geroch group on $T_+$ and showing that it preserves the quantum determinant of this object, or by 
expressing the quantized unit determinant condition directly in terms of the quantum deformed metric $\M$, and verifying that
that the action on $\M$ preserves this condition.
The issue of asymptotic flatness should also be addressed, but can perhaps be avoided. Since Korotkin and Samtleben's theory 
treats asymptotically flat fields, an automorphism of this theory has to preserve asymptotic flatness. To do this asymptotic 
flatness must first be defined precisely in the quantum theory, perhaps working out one of our proposals.
A way to avoid this issue, which might actually be more interesting, is to extend the theory of quantum cylindrically symmetric 
vacuum GR to non asymptotically flat fields. This could be done either by extending the definition of $T_+$ beyond the asymptotically flat 
context, or by expressing the theory entirely in terms of $\M$.
Most interesting of all would be a quasi-local formulation, in which the theory can be formulated on any compact causal diamond without
regard to what happens at infinity at all. Such a formulation would be relevant to the problem of quantizing null initial data in
full GR without cylindrical symmetry \cite{ReiAnd}. Expressing the quantization of the classical condition $\det \M = 1$ directly
in terms of the quantized $\M$ seems the most direct route to such a formulation.

Another important issue is the connection of the present theory of the quantum Geroch group with the results of Korotkin and Samtleben 
on the generators of the Geroch group in the algebra of gravitational observables \cite{KS3}\cite{KS}\cite{Samtleben_thesis}. In particular, 
one would have to develop the ``infinitesimal'' quantized Geroch group, that is, the quantized Lie algebra corresponding to the quantized Geroch group, and also the action of this quantized Lie algebra on gravitational observables corresponding to the action of the quantum
quantum Geroch group presented in the present work. Then one could attempt to quantize the classical expression given in \cite{KS} for the non-abelian Hamiltonian \cite{Babelon_Bernard} corresponding to a given algebra element so that the action on gravitational observables defined by this non-abelian Hamiltonian coincides with the direct action of the Lie algebra element.  

To complete the quantization of the Geroch group, and of cylindrically symmetric vacuum GR, there remain the related tasks 
of defining rigorously the algebras of observables, possibly by working out in detail the structure outlined in Subsection 
\ref{q-Geroch}, and of finding Hilbert space representations of these algebras.
% The authors are unaware of previous work on a line group with the exchange relations of a Yangian, so this work may lead to
% the definition of a novel class of quantum groups. 
A representation of the gravitational observables has already been proposed in \cite{KS2} but it has not been demonstrated that
the inner product on the state space of this representation is positive definite. An attempt to prove this would be a logical 
first step in the search for Hilbert space representations.

Looking beyond cylindrically symmetric vacuum GR, it would be interesting to extend our results to cylindrical symmetry reductions of
theories of gravity coupled to various forms of matter. %(or other two commuting Killing field reductions of such models).
Korotkin and Samtleben's quantization of cylindrically symmetric vacuum gravity extends to many such models, in particular to maximal supergravity \cite{Koepsell}. We expect that our results could also be extended to this context. 

Finally, one might try to extend the model of quantum cylindrically symmetric vacuum gravity by relaxing some of 
the conditions imposed on the field. The quantization of cylindrically symmetric vacuum GR can serve as a guide 
to the quantization of null initial data in full vacuum GR, without cylindrical symmetry. The Poisson brackets 
between data on a null hypersurface are non-zero only betwen data on the same null geodesic generator of the
hypersurface, because all points lying on distinct generators are spacelike separated. Roughly speaking the Poisson
algebra of initial data breaks up into a sum of Poisson algebras each associated with a generator, and this Poisson
algebra is essentially the same as that of the same data on a radial null geodesic through the axis in a 
cylindrically symmetric spacetime. (See \cite{ReiAnd}) for a detailed discussion.) However, regularity on 
the axis of the cylindrically symmetric solutions treated in the present work fix some of the null initial 
data. In particular, regularity at the axis fixes the conformal factor $\Omega$ in the reduced spacetime metric, 
and also implies orthogonal transitivity which fixes the two twist constants (\cite{Chrusciel} and \cite{Wald} 
Theorem 7.1.1.). It would be interesting to extend our results to a theory in which these degrees of freedom 
have been restored. 

\section{Acknowledgements}

M. R. would like to thank Dmitry Korotkin, Boris Runov, and Hermann Nicolai for helpful discussions, Nicolai Reshetikhin for answering
a query, and Andreas Fuchs for extensive discussions on the inverse scattering method.
J. P. acknowledges the financial support of the Comisi\'on Acad\'emica de Posgrado (CAP) through the graduate fellowship for 
teaching assistants in 2016-2018 while part of this work was carried out.  
M. P. acknowledges the support of ANII grant FCE-1-2017-1-135352.
The authors would also like to acknowledge the detailed critique of an anonymous referee which has motivated them to clarify the quantum part of the present work.
\\

\end{document}